MASTERS THESIS

"How applicable is Python as first computer language for teaching programming in a pre-university educational environment, from a teacher's point of view?"

Fotis Georgatos

June 2002, Amsterdam

AMSTEL Institute
Faculty of Science
Universiteit van Amsterdam

MASTERS THESIS

"How applicable is Python as first computer language for teaching programming in a pre-university educational environment, from a teacher’s point of view?"

June 2002, Amsterdam

AMSTEL Institute
Faculty of Science
Universiteit van Amsterdam

I kindly thank Dr. Ton Ellermeijer for undertaking the task of supervising this research work and accompanying at various cases my introduction to the field of Educational Research.

# Acknowledgements

To the people I learned from.

Each word of this thesis is plagiarized.
I have never generated a single word myself.

# Abstract

This project report attempts to evaluate the educational properties of a relatively new computer language named Python. This is done by examining computer language evolution history, related scientific background work, the existing educational research on computer languages and Python's experimental application in higher secondary education in Greece, during first half of year 2002.

# CONTENTS

# INTRODUCTION

*"**Language** is the only concept that cannot be described without using itself"*

In the year 1999 there was a major reform in the educational system in Greece. Probably it was one of many worldwide during the last decade, while still most educational systems try to adjust to the ever-changing needs of society that becomes increasingly dependent on technology. The reform is being driven by the Pedagogical Institute, a governmental entity set to connect current scientific research in education with the classroom reality. The reform includes the following areas:

- modern curriculum and books
- providing the educational specifications for the ICT infrastructure of schools
- training teachers with new skills, in particular for computer aided or computer aimed teaching.

Computer programming is one of the newly introduced courses and the related book in entitled *Development of Applications in a Programming Environment*[1] [DAPE99]. An important point for this research is that there is no specification or exact definition for a certain computer language in the programming course, and the choice is left up to the teachers.

No consensus exists, on worldwide scale, when it comes to choosing a computer language for an educational setting. Some use PASCAL or BASIC, some use one of C, C++, Java, FORTRAN, others use LOGO. The total result is a fragmentation, a Babel of languages modernized for the 21st century computer-influenced era. A common element, though, of these languages is that they do not cover the age spectrum of learners evenly: some of them are too simple to be practical, others are practical but hard to learn as a first programming skill. Moreover, the most used ones among them are the hardest to teach.

During the last decade, there is a growing interest in Python, a computer language that combines:

- modern design (*Object Oriented* with *Functional Paradigm* extensions)
- practical characteristics (*rapid application development, suitability for prototyping*) and
- educational attributes (easy on the eyes *syntax* and an *interactive environment* named IDLE).

In a few words, Python is a cleanly designed *scripting* language. All these reasons make Python look suitable, even ideal, for use in a classroom setting.

Add to this that, well-known computer experts are quoted extolling Python as a computer language. Among others, stands out the favorable opinion of Eric S. Raymond[2] in the article "Why Python?" found in a technical magazine called Linux Journal [RE00]:

> *[…] I noticed I was generating working code nearly as fast as I could type. When I realized this I was quite startled. When you are writing working code nearly as fast as you can type and your misstep rate is near zero, it generally means you have achieved mastery of the language. But that didn't make sense, because it was still day one and I was regularly pausing to look up new language and library features! This was my first clue that, in Python, I was actually dealing with an exceptionally good design.*

---

[1] The original title will look like Greek to the reader: "Ανάπτυξη Εφαρμογών σε Προγραμματιστικό Περιβάλλον"
[2] Well known programmer of Open Source Software, inventor of the term. Author of many programs, papers, articles, and books. Possibly most quoted by him is "The New Hacker's Dictionary". Also known on the Internet as ESR.

> *[…] this code took me about ninety minutes to write, and it worked correctly the first time I ran it. To say I was astonished would have been positively wallowing in understatement.*
>
> *[…] Even if we stipulate that I am fairly talented hacker, this is an amazing testament to Python's clarity and elegance of design.*

A high school teacher, Jeffrey Elkner, who happens to be a programmer as well, is also a supporter of the Python language. His arguments are of educational nature [EJ00]:

> *[…] The educational objectives at this point in the course are to introduce students to the idea of a programming statement and to get them to make their first program, thereby introducing them to the programming environment. The Python program has exactly what is needed to do these things and nothing more.*
>
> *[…] Using a very high level language like Python allows a teacher to postpone talking about low level details of the machine until students have the background that they need to better make sense of the details. It thus creates the ability to put 'first things first' pedagogically*
>
> *[…] Python's syntax makes programs much easier to read, thereby making it easier to effectively evaluate student work and to identify problems in student programs. This is a big plus for teachers using Python in the classroom.*

The last sentence of this quote, is the motivation for this research. Python is not only a well designed language, it is also a great tool for educators.

The choice made by the Pedagogical Institute of not choosing a certain computer language for the programming course should be considered an elegant and realistic decision. The current situation is that not all teachers do master a single language and on the longer term, education has to build methodological skills of intellectual nature instead of providing product training. The consequence of this way of thinking is that pseudocode, an artificial language, is the focal point of the course.

On the other hand, not teaching an existing language at all can also have drawbacks, because in that case we create a year-long study in isolation with no gains of practical nature; even though, the resources are available to achieve such an aim. After all, programming languages are tools, whose utility lies in the context of applications just as mathematics.

We hope to make good use of this freedom by evaluating Python in the classroom, comparing the results with those of other research and, last but not least, propagating the findings to more teachers and educators as well as raising, eventually, awareness to the educational community at large.

> *For educators, a nagging question is,*
> *How do I learn to use computers today in a way that will not be obsolete in five years?*
>
> *The answer requires a vision of the future.*

Cynthia Solomon, 1986

# AIMS AND BACKGROUND OF THE STUDY

The aim of this research is to evaluate Python in a classroom environment as an educational tool and examine its properties through real application. Python is a computer language that is discussed recently as an easy one to learn and a great one to master, even according to Computer Scientists.

A special aim of this research is to identify if Python can be applied in secondary higher education in Greece and, if yes, under which conditions and circumstances. A side aim is to compare the results with those of previous research and disseminate experience and outcomes.

> ***How applicable is Python as first computer language for teaching programming in a pre-university educational environment, from a teacher's point of view?***

The objective is to use Python in the course "Development of Applications in a Programming Environment" of Lyceum, the equivalent of secondary high school in Greece, as the vehicle language to introduce programming in a few short lessons.

The aims for this study are explored through a set of major questions[3] followed by their respective answers. Colleagues in the AMSTEL Institute, whom I thank for asking in the first place, originally imposed most of the questions, during the course Research Methodology[4].

The background of this research is supplemented with an enumeration of research hypotheses.

[3] "*Computers can only give you answers*". Quoting Pablo Picasso.

[4] As an attribution to their influence, I wish hereby to refer to their names: Dr. Martin Goedhart, Mrs. Mary Beth, Yenni B. Widjaja, Tomasz Greczylo and Michiel van Eijck. I particularly thank Mrs. Mary Beth who kindly carried on correcting my linguistic mistakes in the English language, verbal or written, even up to undertaking the task to review this report as a native speaker, during a very short and unrealistic period of time. Any errors left, are therefore mine.

## *Fundamental Questions*

### What is programming?

*Programming is a human activity that is a great challenge, involving the design of machine behavior that can assist, and at times replace, humans in tasks of intellectual nature.* [PP90, p. 3]

The product of this activity is a 'program', which can be diverse things at different times:

- describing calculations; the *imperative* or *procedural* programming model
- defining and treating objects; the *object oriented* programming model
- defining functions; the *functional* programming model
- defining logical relationships; the *logical* programming model

The programming models are often referenced as *programming paradigms*.

Another definition of what a program is is syntactical: a program is a text constructed according to certain grammar rules. In fact, Noam Chomsky's Generative Theory of Grammars has, probably, influenced Computer Science more than the field of Natural Languages itself [PC90, p. 10]. Any *syntactical* component of programming, though, is accompanied by its *semantical* interpretation.

At first sight, programming seems a straightforward activity. However, programs are always full of errors and debugging them takes time, since it is, more often than not, difficult to track them down and correct them. [PC90, p. 9]

At this point, there is one important thing to remember: there are different ways to write programs, following particular models to describe solutions.

### What are the programming paradigms?

We can think of a paradigm as a modeling technique particularly adjusted to problem solving in a computing environment. There are a handful of major programming paradigms, which include, in order of importance:

- Imperative — where a program is an ordered execution of commands (statements)
- Object-oriented — where a program is a world of objects that communicate to each other
- Functional — where a program is a set of functions; It is a very postulational model
- Logical — where a program is a set of "logical" declarations

These are the generic families in which we categorize computer languages, too.[5] Briefly, most computer languages belong in the Imperative paradigm. An evolution of it is the Object Oriented paradigm. The Functional paradigm is the one closer to mathematical thinking and is suitable for formal program specification. The Logical paradigm is claimed to be closer to human logic. Also, there exist languages that don't fit exactly to these paradigms like special languages for parallel programming or Reverse Polish Notation. The later is adjusted to a computer's internal structures: it assumes a special data structure, a *stack*, that is fed with data and operations performed on them.

[5] This categorization is more than literal: within the same paradigm, languages are "directly" translatable to each other.

## Which programming paradigm to teach then?

In real world, the most commonly used programming model is the imperative one. For this reason, it is the imperative programming that is being taught up to now in most high school level curricula. This is the case with this research as well.

On the other hand, the Object Oriented and Functional programming are gaining ground in importance during the last few years. There is raising concern in teaching other paradigms than the imperative one, but the problem is where to begin from. Related to this, is, that teachers are not computer scientists and should not have to learn a handful of languages to teach a single course in programming.

Ideally, there exists a single language that one can learn incrementally and selectively teach a certain model. Most existing languages are known to fit on a single programming model.

Python suits well this mindset since it is object-oriented, which is a super of imperative, with functional extensions. This means that is possible for someone to use all three paradigms using the same single language. The Python language does not enforce a unique paradigm.

## What is the value of programming in education?

Programming often tends to be viewed as a problem solving rather than a linguistic activity, often ignoring that programming languages are a case of formal languages[6]. Formal languages have the characteristic that their interpretation - their meaning- is unique.

Mathematics, for example, is a formal language: it has a number of symbols, a number of definitions, fixed concepts and relationships between them. In some cases, the basic definitions can change -consider the different kind of geometries that exist-, but even then, the interpretation of the mathematical language is unambiguous. Formal languages allow the postulational model to develop in a deterministic way; and this is true regardless of their application domain.

You should expect a program written in a certain programming language to "run" exactly the same way[7] in any computer system and also **be understood by each and every person reading it, in the very same way**. This attribute of programming languages makes them a stable ground for pedagogical purposes, since there is little space left for ambiguity. The practical advantage of this is that no extra communication is necessary to resolve differences in the ways of comprehension of a certain "sentence"; you may wish to compare this with what happens in natural languages.

We cannot confirm the same property for natural "spoken" languages. Many -and maybe most- natural languages' sentences leave space for diverse interpretation and can mean different things to different people. In the 16$^{th}$ century, Machiavelli wrote "The Prince" considering this observation.

---

[6] For this to happen, surely some criteria have to be met. These conditions relate to Chomsky's language classification, the generative grammar theory, and the so-called BNF notation. More information on this may be found in Appendix E.
[7] Well, even this is not exact always; but it is either due to bugs, or errors in the software that parses the languages; or incomplete definitions and extensions. C has been notable for its different dialects, Basic and Pascal follow closely.

## Why teach programming at secondary high school?

Devices that are in use around us become more and more complex and even more programmable. For example, we can think of a television and a video, a phone -a mobile one particularly- and even modern kitchen apparatuses. Newer generations will eventually need a better grasp in controlling a PDA[8] –this stands for Personal Digital Assistant, the electronic equivalent of an agenda-, or their …bicycle' s electronic monitoring system. In order to do so, they will eventually be in need of a computer language.

Programming is becoming a necessary component of a modern curriculum and the skills developed thereof will be essential for citizens of a technological society. According to the ACM[9] Model High School Computer Science Curriculum *as high school students study the natural sciences to understand the natural world, they need to study computer science to comprehend the social, economic, and cultural environment of the information age*. Regardless of what pupils follow in their future career the sooner they are familiar with what programming is about, the better for them.

One more argument in favor of teaching programming is to support the existing mathematics curriculum at schools. Albeit there is skepticism for the fact that teaching programming to pupils has its extra overhead to reach some minimal and essential level of proficiency, it is recognized by David Johnson[10] that using IT to explore mathematics, discreet in particular, is beneficial [JDC00]:

- *a new way to view or express many mathematical concepts and relationships, e.g., factors and multiples, primeness, solving equations, mapping of a function to a new function, generating values for parameters and variables, evaluating or graphing functions or relationships, transformational geometry, and so on.*
- *a new means for solving problems, e.g., mathematical modeling and simulation become a means for making sense of `real' problems*
- *selected programming environments, e.g., computer `microworlds' or dynamic pictorial environments can be considered to be mathematical domains in themselves, i.e., the computer becomes a medium for expression in that the `thought experiments of mathematicians can be turned into computational experiences for the non-experts'*

We may add to this vision that, so far, teaching programming has failed to permeate the educational system(s) on large and systemic scale. It can be attributed to the claim of time being a scarce resource in modern schools. Schools' plans are devoted to very specific aims, in fact, and little freedom is allowed to the teachers in selecting alternative paths in teaching mathematics, let alone experiment with tools for it. Nowadays, this is changing by having computers installed at most schools. New ways are paved, so, why not take advantage of this opportunity?

Above all, we can view computer languages dissociated from their technological background, as communication tools, exactly as natural languages are. According to Green [GP90, p. 42]:

*We still think too readily of programs as just being for compilation. We should think of them also as being for communication from ourselves to others, and as vehicles for expressing our own thoughts to ourselves. So we should think more about reading versus writing, capture of ideas versus display of ideas.*

[8] This is another technological TLA = Three Letter Abbreviation. TLAs' generation is a plague, which has invaded modern cultures and seems to originate from the Latin. As much as possible, acronyms will be avoided in this research.
[9] Dear reader, ACM stands for Association for Computer Machinery
[10] A vivid supporter of the use of Information Technology for exploring mathematics

## How do students become experts in a computer language?

In principle, someone has to become comfortable with the *syntax* and *semantics* of a language in addition to the *environment* that it is applied in. Also, it is needed to be understood, what would be the results of using certain algorithms and known programming structures:

- Is it going to be an appropriate idea?
- Will it be efficient?
- Fast enough?
- Simple enough?
- Will I get the reply I want?

The fundamental questions regarding acquiring natural languages similarly arise on the acquisition of computer languages. These questions, as described in *Knowledge of language* of N. Chomsky [CN86, p.3], are:

- *What constitutes knowledge of language?*
- *How is knowledge of language acquired?*
- *How is knowledge of language put to use?*

It appears that the principles behind becoming an expert in a computer language are those of a natural one plus the problem-solving component. Proving this and answering these very questions, as far as computer languages are concerned, can well be the scope of a more advanced research than this investigative project with a timeline of a year.

It may be suggested, however, that there are strong indications that the concept of *Universal Grammar* (UG) is also applicable for understanding the acquisition of computer languages. This judgement is based upon the observation that modern computer languages are documented by Context Free Grammars in a special meta-language named BNF. BNF is equivalent to Chomsky's Type-2 grammars, defined in his Generative Grammar Theory. Please read Appendix E on this.

Add to this, that multiple *computer language translators* exist and the related technology is making use of sets of production rules, the same term introduced as *rewrite rules* by Chomsky. An elaborate case of rewrite rules is applied as well by special programs that try heuristically to convert primitive language codes (of lower-level, like assembly) to higher level code, like that of C, Pascal or BASIC[11].

It is worthwhile to note that first languages were not originally described in BNF, like BASIC and FORTRAN. This implies that there exist program texts which when parsed syntactically, can result into two different syntactical trees and, hence, have diverse meanings. The lack of BNF formalism often results in languages that fall in the category of Context Sensitive Grammars; it is quite hard to write a proper parser for a language with Context Free Grammar, correctly, with no aid tools.

As a final note, the *syntactical* aspect of a programming language is only one component of it. The other one is the *semantical*, which defines what kind of behavior is defined for each keyword. The semantical aspect is more complex to describe formally, so a natural language is often employed.

[11] This technique is called reverse software engineering. There are tools for it, although it is, often, prohibited by law.

## What is the course that teaches programming in the Lyceum?

The question can be rephrased as, which is the course that actually takes the students to the computer laboratory, and has the teacher directing to type commands and write programs?

After a thorough investigation of the school textbooks[12], the answer appears to be "Development of Applications in a Programming Environment", hereby referenced as DAPE. The course program lasts 58 hours, which assuming the allocated time of 2 hours per week, corresponds to 29 weeks.

In this research case, the target population is students aged 17-18 with a so-called *practical* or *technological* direction in Greece. This implies students of the last class of higher secondary school level excluding the ones aiming for humanities' sciences.

## What is the course specification according to the course program?

The Course Program, defines the didactical units, their contents and their length:

| Unit | Content | Hours |
|---|---|---|
| 1. Analyzing a problem | • Definition and comprehension of a problem<br>• Problem's structure<br>• Definition of requirements | 12 |
| 2. Designing an algorithm | • Algorithms – Basic concepts<br>• Strategies for designing algorithms<br>• Development of algorithms<br>• Testing of algorithms | 27 |
| 3. Implementation in a programming environment | • Ways, techniques and environments for programming<br>• Principles of structured programming<br>• Principles of modern programming environments<br>• Designing and implementation of a User Interface<br>• Testing and debugging a program | 27 |
| 4. Evaluation & Documentation | • Evaluation, optimization, extension of a program<br>• Documenting a program<br>• Software life cycle | 9 |

Another document, the school textbook *[DAPE99, Teachers handbook, p. 14]* defines the general aim of the course, that pupils must:

- cultivate analytical thinking and synthesizing ability
- develop creativity and imagination in design
- cultivate -and be enabled to practice- the sharpness and clarity of expression
- develop skills of methodological value and not those of software operators
- develop skills of algorithmic approach in solving problems
- be able to solve simple problems by use of basic programming knowledge

---

12 Mr. Kostas Pagonis from Paedagogical Institute has provided the researcher with the complete series of books for teaching Computer Science courses in Greece. His contribution to this research is invaluable, for which I thank him.

## When is the right time to teach a computer language?

The students' textbook is composed of 14 chapters *[DAPE]*. The students do not have to write in or study any particular computer language until chapter 6. At that moment, several languages are discussed and one may be chosen to be used in the laboratory. The chapters from 6 to 10 is where students are expected to implement small programs initially written in *pseudocode* –an artificial language, which is a fusion of a natural and a computer language - and experiment with a real language in the school laboratory. The exact study components, which are instructed in the computer lab, are up to the teachers. Hereby the topics covered in these chapters are listed.

6. Introduction into programming
    - Concept of a program
    - History of programming
    - Natural and Artificial Languages
    - Strategies for designing programs
    - Object Oriented programming
    - Parallel programming
    - Programming Environments
7. Basic concepts of programming
    - Alphabet and data types
    - Constants and variables
    - Operands, functions and expressions
    - Assignment command
    - Input – Output commands
    - Program structure
8. Conditional and loop commands
    - Conditional commands
    - Loop commands
9. Arrays
    - Single dimension arrays
    - When to use arrays
    - Multidimensional arrays
    - Typical array processing
10. Procedures and functions
    - Procedural programming
    - Characteristics of procedures
    - Advantages of procedural programming
    - Parameters
    - Procedures and functions
    - Scope of variables and constants
    - Recursion

The next chapters of the book cover more complicated topics such as object-oriented programming, event driven programming, user interfaces, debugging and documentation.

## What kind of concepts should students learn?

The course "Development of Applications in a Programming Environment" concentrates on the Imperative programming paradigm. Its major ideas, are the priorities to be instructed in the class.

The most emphasized concepts include the three major programming constructs, which according to Green are [GT90, p. 12]:

- sequence: a set of commands that are executed successively
- iteration: a definition for repeating a command or a block of commands
- conditional: a "branching" command, that modifies the flow of a program

Other major concepts that are discussed are data structures, as is the case in imperative languages:

- variables
- constants
- arrays

The selection of these concepts is, of course, a subset of the results of more than 50 years research and experience in programming. This subset is truly indicative and sufficient, though.

## What are the requirements for a programming language used in teaching secondary high school novices?

The requirements for an ideal such language could be defined in criteria such as:

- Easy to learn[13], at least for novice teenagers
- Practical, hopefully already applied in working and popular software
- *General Purpose Language* in scope, as every "computer literate" has to know at least one[14]

Easy to learn are languages that do not obstruct the problem solving process and can express the main ideas clearly and in a few symbols.

Practical languages are the ones, which have an application domain that is not strongly bound to education and can be used with advantages in other –scientific or engineering- fields as well.

*General Purpose Language* means in ICT[15] terminology that a language is designed with no particular application in mind. It can equally be used to define an algorithm for doing a calculation, solving complex mathematical problems, calculating bank interests or controlling a factory process.

The requirements which are specially designed for education can be supplemented by some generic criteria for computer languages. C.A.R. Hoare has written such criteria that are provided later on.

---

[13] As we may discuss later on, what is considered *easy* and *hard* is like what is *big* and *small*: the words have a meaning which is defined in a relative reference context. In this case, *easy* as compared to other existing languages in May 2002

[14] What are you going to do if you want to convert 500 .gif pictures in .png format while you already have a program that can do it for a single picture? The answer is not to find and download a special 2nd program, but (re)write one!

[15] ICT stands for Information and Communication Technology.

## What is *pseudocode*?

A brief answer according to the students’ book [DAPE99, Students handbook, pp. 331] is:

“*A method of describing algorithms by means of predefined* ***natural-language-like*** *keywords*”

*Pseudocode* is a widely applied technique for introducing novices into programming. The idea is that instead of confronting the students with a real computer language, we can use an artificial intermediate that resembles the natural language of the audience (Chinese, English, Greek, etc).

This strategy is quite common when teaching *algorithmics*, as D. C. Johnson reports [JDC00]:
*“It is argued here that over time the teachers and pupils will come to understand and appreciate the use of an agreed code and format for expressing the algorithms in the words and symbols from a programming language, or even some pseudocode which is readily translated into a programming language.”*

The last sentence of this quote describes precisely what is the case with Lyceum’s book [DAPE99]. The pseudocode defined in the textbook resembles, indeed, greatly the computer language ALGOL or the more modern PASCAL. The purpose of this choice is to give the ability to time-pressured teachers –and students!- to generate a working program with word-after-word translation of the pseudocode. As a consequence, it is PASCAL the language most often used in schools, in Greece.

See below, for a few examples of a program written in pseudocodes and existing computer languages. The program is described in the textbook [DAPE99, pp. 42-43] and is the 10th example of chapter 2. What it does, is to find the total of the first hundred of positive integers. Compare the equivalent programs, which are, with clockwise order, in Greek pseudocode, in English pseudocode, in Pascal and in Python languages.

| | | |
|---|---|---|
| **PSEUDOCODE (Greek & English)** | **Αλγόριθμος** Παράδειγμα_10<br>Sum ← 0<br>**Για** i **από** 1 **μέχρι** 100<br>      Sum ← Sum + i<br>**Τέλος_επανάληψης**<br>**Εκτύπωσε** Sum<br>**Τέλος** Παράδειγμα_10 | **Algorithm** Example_10<br>Sum ← 0<br>**For** i **from** 1 **to** 100<br>      Sum ← Sum + i<br>**End_iteration**<br>**Print** Sum<br>**End** Example_10 |
| **COMP. LANGUAGES (Python & Pascal)** | Sum=0<br>**for** i in **range**(1,101):<br>      Sum = Sum+i<br>**print** Sum | **program** Example_10<br>**var**<br>   Sum, i: integer;<br><br>**begin**<br>Sum:=0;<br>**for** i:=1 **to** 100 **do**<br>  Sum:=Sum+i;<br>writeln(Sum);<br>**end**. |

## Why prefer a scripting language in a school?

First, a definition is appropriate: *scripting programming language* is one which is meant to be used for rapid problem solving at the moment that a need arises, without requiring a top-down design phase. For this reason, scripting is synonym concept with "rapid applications development" or "gluing applications" as it is often called in Computer Science and Software Engineering. Well known scripting languages, nowadays, are Perl, Python, Rexx, Tcl and the *UNIX shells*.

Scripting languages are easier to learn and easier to write. John K. Ousterhout, the author of Tcl, comments in the paper '*Scripting: Higher Level Programming for the 21st Century*' as seen in *IEEE Computer magazine, March 1998* [OJK98]:

> *Scripting languages are easier to learn because they have simpler syntax than system programming languages.*
> *[…] Casual programmers are not willing to spend months learning a system programming language, but they can often learn enough about a scripting language in a few hours to write useful programs.*

Execution time is usually doubled in a scripting language, than a compiled one, which means that it takes twice as much time for a program to finish. On the other hand, it takes roughly half as much time to implement the same idea, even for professionals. Quoting Lutz Prechelt in a report comparing seven programming languages in a certain problem-solving scenario, involving the calculation of possible mnemonic names for phones[16] [PL00]:

> *In every case the scripting version required less code and development time than the system programming version; the difference varied from a factor of 2 to a factor of 60.*
> *[…] Designing and writing the program in Perl, Python, Rexx, or Tcl takes only about half as much time as writing it in C, C++, or Java and the resulting program is only half as long.*

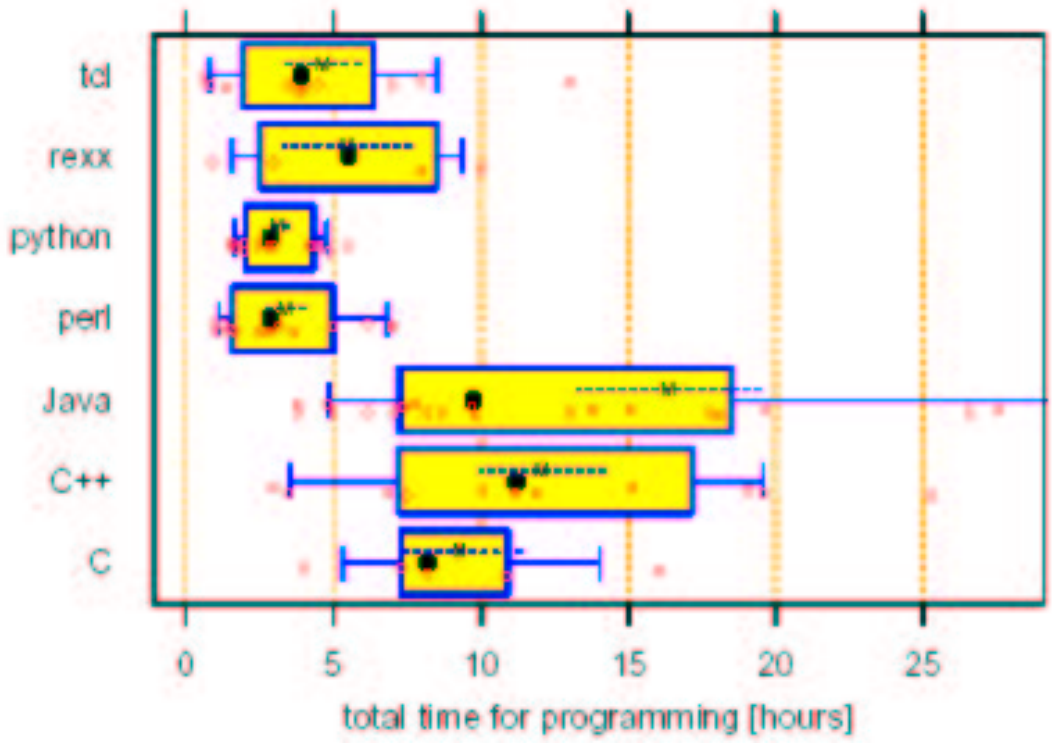

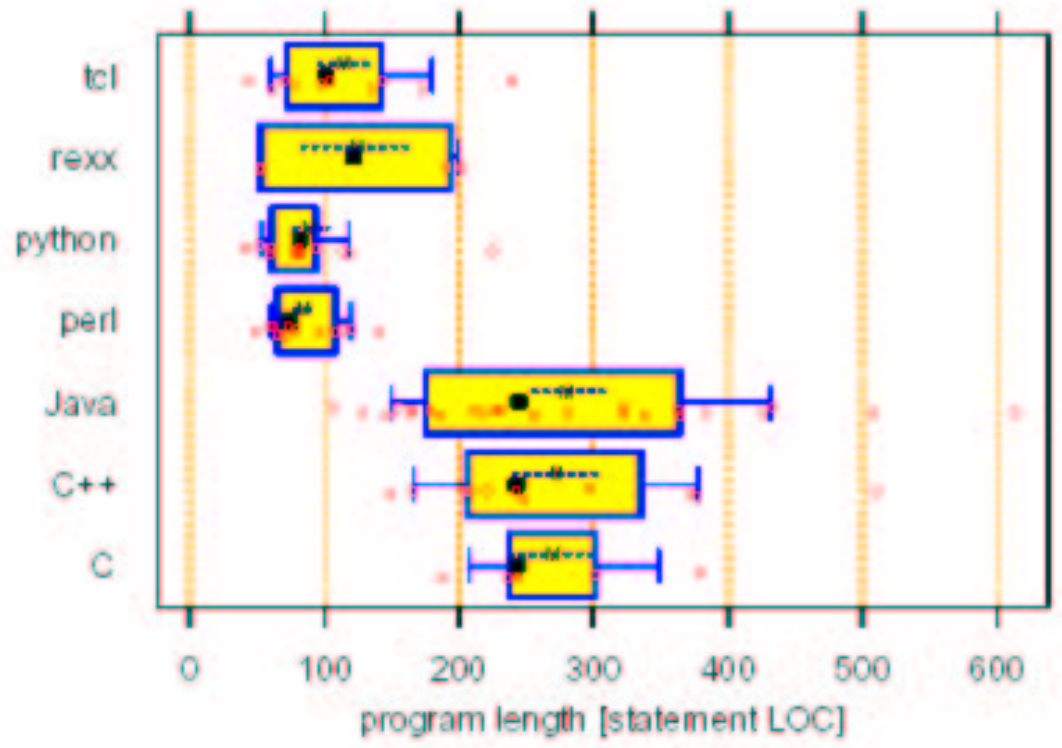

| Total time for programming [hours] | Program Length [statement LOC] |
|---|---|
| Total working time for realizing the program. *Script group: times as measured and reported by the programmers. Non-script group: times as measured by the experimenter.* | Program length, measured in number of non-comment source lines of code. |

[16] For example, typing AMSTEL on a phone's keyboard corresponds to the number 267835. The requested program involved the opposite operation: given a list of numbers and a supplied dictionary, convert to all possible strings.

In the same report, there are comparisons in efficiency and resulting performance, which can be of high interest to Computer Scientists:

> *"The results indicate that, for the given programming problem, scripting languages (Perl, Python, Rexx, Tcl) are more productive than conventional languages (C, C++, Java). In terms of run time and memory consumption, they often turn out better than Java and not much worse than C or C++."*

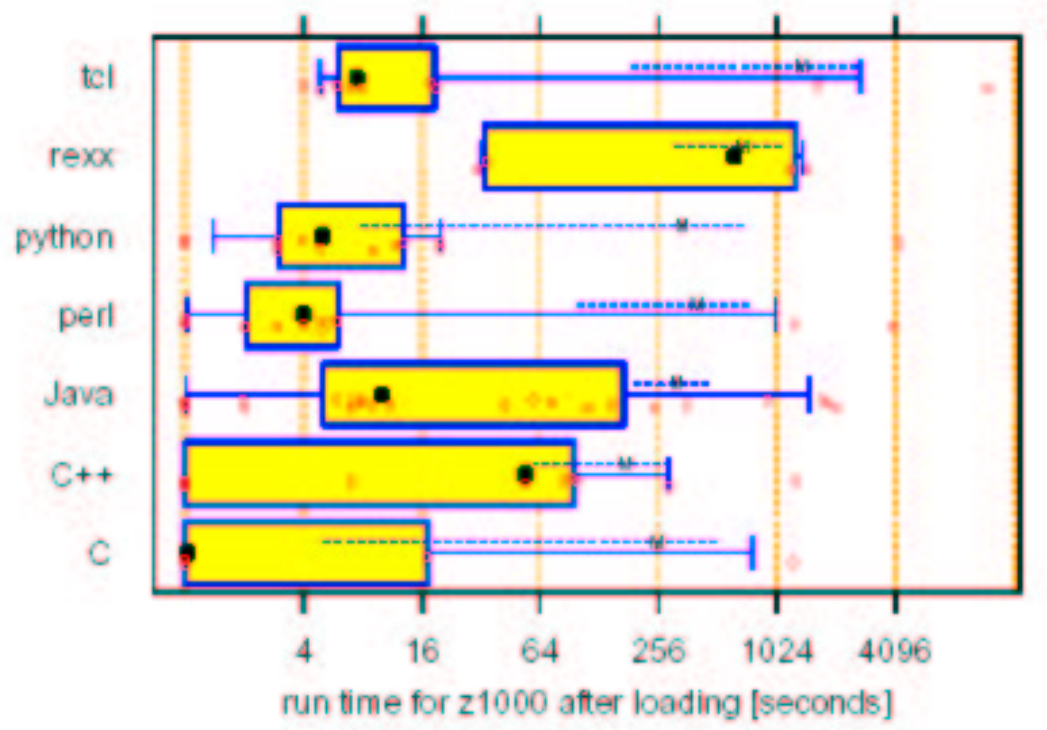

If scripting is so much simpler and efficient for expert programmers, then maybe this is a subtle hint for languages in educational settings as well.

## Why Python?

Python is a General Purpose Language. It is easy to learn *[CP4E, p.1]* and widely used in science in the fields of numeric programming, artificial intelligence, image processing, biology and others. It is known to be easy and can express with clarity complex ideas. Such complex ideas easily resulted to errors in other languages for solutions to the same problems [SF98] [RE00]. Python's syntax is making use of *indentation* to denote *syntactical structure*. Below is what the teacher Jeffrey Elkner has to say, based on his experience with Python [EJ01]:

> *I am a high school teacher who has been using Python to teach our intro course for two years now. I have never seen a better language for that purpose.*
>
> *[…] Python is good as a teaching language for the very same reasons that you are now using it to program, it permits you to focus on the problem at hand and doesn't get in your way.*

Python has a development environment, called IDLE, which is particularly suited for editing in it. The language does not require or enforce its usage in any way. IDLE is part of the standard distribution of the language *[IDLE]*, offers syntax highlighting, online help and interactive editing facilities. These features make it advantageous for schools' computer centers and laboratories. The trend to use such environments in programming has also been noted by Green [GT90]:

> *Low-level test-based environments are still commonest but structure-based editors, usually built on syntactic structure are becoming more common.*

What Python is not the first choice language for, is performance, which is attributed largely to the fact that it is an *interpreted* and not a *compiled* language.

An advantage of interpreted languages over compiled languages is that it is much easier to make rapid changes and test those changes. The disadvantage of interpreted languages compared to compiled languages is, in fact, performance. Since computer hardware is increasing in speed and decreasing in price rapidly, the performance factor is less and less important. In addition to this, we are considering here an educational environment and performance of computer systems is unimportant next to performance of people, teachers and students.

## What does Python code look like?

Here is a typical example of a program in Python. The lines that start with symbol # are comments.

```
# This is equivalent of the C code, for solving the linear equation ax+b=0,
# as this is described in page 125 of the Greek students' handbook, named:
#
# "Ανάπτυξη Εφαρμογών σε Προγραμματιστικό Περιβάλλον",
# (Development of Applications in a Programming Environment)
# Pedagogical Institute, Athens 1999, ISBN 960-7251-23-7
#
# Copyleft by Fotis Georgatos <gef@ceid.upatras.gr>, 1-5-2002
# This code has been placed in the public domain.

A = input('A= ')
B = input('B= ')
if A==0:
    if B==0:
        print "AORISTH" # Undefined
    else:
        print "ADYNATH" # Impossible
else:
    print "X = ", -B/A # solution of a linear system
```

## What is the trick with indentation in Python?

Python uses spaces and tabs, which are regarded as "whitespace noise" by other languages, to denote a program's block structure of statements. Most languages' syntactical specifications use special *tokens (eg. {})* or a *delimiter ;* in order to denote such statement blocks. Python needs none:

| Pascal (**BEGIN**, **END**, **;** ) | C, C++ or Java ( **{**, **}**, **;** ) | Python |
|---|---|---|
| D:=b*b-4*a*c;<br>**if** D>0 **then**<br>    **begin**<br>    rt1:=(-b+**sqrt**(D))/(2*a);<br>    rt2:=(-b+**sqrt**(D))/(2*a)<br>    **end**; | D=b*b-4*a*c;<br>**if** (D>0) **{**<br>    rt1=(-b+**sqrt**(D))/(2*a);<br>    rt2=(-b+**sqrt**(D))/(2*a)<br>**}** | D=b*b-4*a*c<br>**if** D>0**:**<br>    rt1=(-b+**sqrt**(D))/(2*a)<br>    rt2=(-b+**sqrt**(D))/(2*a) |

The sequential block structure is fundamental in imperative programming and plays a major role in the meaning of a program. What is promoted as the standard style in other languages is, to write the code such, that visual layout matches the syntactical[17], because this is not enforced by language[18].

The total effect of this, is that Python is explicitly spartan in its description, promotes readability for the eyes and skips a whole range of problems that occur when someone omits by error tokens or delimiters. Even the most experienced programmers fall frequently in these kinds of traps.

---

[17] "Recommended C style and Coding Standards": *Indentation and spacing should reflect the block structure of the code; […] Make proper use of white space so that the structure of the program is evident from the layout of the code.*
[18] "Linux kernel coding style", paragraph #1: *"The whole idea behind indentation is to clearly define where a block of control starts and ends. Especially when you've been looking at your screen for 20 straight hours, you'll find it a lot easier to see how the indentation works if you have large indentations.*

## What is a problem and what is an algorithm?

Psychologist J.M. Hoc has studied the role of mental representation in problem solving, as far as learning programming is concerned [HJM77]:

> A situation can be said to be a **problem** to the subject when he represents to himself a pair of states – the initial state and a final state, that is, the objective that he wishes to attain- and a procedure which leads from one to the other using one or several *calculation representations* that he has at his disposal, -that need not be necessarily algebraic but can also be rules and actions of a *language*- providing such a procedure is not translatable, word-by-word, into already known rules or actions.
>
> The subject who can find a specific solution for each of the specific cases of a class of problems is said to have an algorithm. [...] When a person is able, without difficulty, to give a solution for every case, he is showing an algorithmic behavior; but this does not mean that he is capable of exteriorizing the algorithm concerned.

The very process of exteriorizing an algorithm by expressing it in an artificial language, which is suitable for a computer system, is what programming is all about.

## What if you were a teacher?

If I were a teacher I would like to have a language that is compatible with the theoretical content of the course and attracts the minimum of errors. The syntactical and semantical aspects of a language are known to contribute greatly in the amount and nature of programming errors - commonly known as bugs- that are generated when programming. Novices are particularly vulnerable in such design influences and this can play a major role in a class: The errors promoted by language will happen not in one or two students but usually the majority, while trying the type a program. This can affect greatly the course schedule, because the same error has to be corrected for many students. Explaining it as formalism takes a lot of time.

Maximizing clarity is also important when correcting students' transcripts. Furthermore, a language that does this is beneficial over one that allows too much freedom to students in the way of laying out code. A teacher that has to correct tens of (manu-)scripts will appreciate a lot to find properly indented code that denotes the structure of the program. Python is advantageous, because proper indentation is required.

Another aspect which is important is that the language should not be obstructing or mystifying the abstract problem solving process and still be motivating: the students should constantly feel that there is more to learn by using their newly acquired tool and that they can do it with minimal effort.

## *Research Hypotheses*

### Teachers tend to teach in the language in which they learned programming

In the current lack of scientific ground upon the selection of a language for teaching programming, this is not a bad practice in itself. Furthermore, it is the most logical choice since a person usually teaches best in the language that he feels more comfortable with, a language, which is well known. It is hoped that this research will be one more small and effective contribution to the scientific knowledge about learning computer languages, an asset of the field Didactics of Informatics.

### Some languages are designed considering the human factor

Few languages consider seriously the learning phase and the educational aspect. Languages that present themselves as such are LOGO, BASIC, PASCAL, ABC, Python and there are others. Most of the languages, while being developed, are inspired by a problem at hand and little attention has been given to the human factor or their reusability as communication means.

### The documentation and help provided by the language are important

Instructing a computer language means teaching programming concepts, the respective keywords' *semantics*, while applying them with a particular *syntax*. This kind of information is typically found in a computer language's manual. It is not assumed that the materials that the teachers are going to use are the same. In fact, they can be others than the school's textbook if the teachers wish to do so.

### Python will sometimes be advantageous and sometimes not

While instructing, teachers will come up with easy and hard moments by using the language. There is a set of questions to be developed or analyzed, exploring the experience, by a teacher's point of view, regarding the applicability of Python in their educational setting.

### Python is a case, worthwhile to investigate

Final hypothesis, last but not least:

**It is assumed that there is content -and that a conclusion can be derived- when comparing Python with other computer languages used in education**.

This must not be taken for granted.

# HISTORY OF PROGRAMMING

## *The first steps*

This introduction is based on the book " *Fundamentals of Programming Languages*" by Ellis Horowitz, and is, in an extent, back translated from a Greek translation of the original [HE84].

The first known algorithms[19] were discovered by archaeologists to be written on clay tablets. Those tablets are dated between 3000-1500BC and were found in an area close to Babylon in Mesopotamia, which is located near to present times Baghdad, Iraq.

Babylonians are known to have had a peculiar arithmetic system based in the number 60[20]. On top of this they had invented *floating-point* numbers, the latter is the way that numbers with decimals are often called in computer science. To illustrate an example the number 8, 50, 36 can correspond to 31836 or 530.6 or generally to 31836*60^k, [kEQ].

Babylonians did not only write computational tables to facilitate arithmetic operations, they also could solve algebraic expressions using an algorithm that would compute it. The algorithm was a generic method of solving a particular category of problems following various steps and ending with a note like "This is the procedure". In fact, this has a lot of similarities with the way most programs are written nowadays. There is an article by Donald Knuth [KD72] where some of these algorithms are described. Albeit Babylonians had invented an elaborate, for those times, symbolic system, not much evolution happened in computational methods since then, until the 19th and 20th century.

We may find algorithmic procedures being introduced later, most known being the Euclid' s' one for computing the Greatest Common Divisor[21] and Eratosthenes' Sieve which is a technique to find the prime numbers up to a certain limit with least computational effort. These cases are dated 1500 years after the Babylonians and the most important fact is that they were isolated examples expressed in natural language, which made no significant advancements towards a formal symbolic representation of algorithms.

The next conscious effort in programming was triggered by Charles Babbage (1792-1871) who designed two mechanical computational machines during the years 1820-1850: the *Difference Machine* which was based on finite differences theory and the *Analytical Machine* which had a lot of similarities with a modern digital computer.

[19] The word is coming from the study of the Persian mathematician Abu Ja'far Mohammed ibn Musa al Khowarizmi, which is dated close to 825AC. Five centuries later his study was translated in Latin and started with a phrase like "Algoritmi dixit...". This was the first complete report in algebra, another word that comes from the Arabic Al-jabr, meaning restoration, since one of the aims in algebra is the restoration of equality in an expression. [DAPE-S, p.25]

[20] Actually, that is the very reason that we nowadays count hours, minutes, seconds in this way and divide the circle in 360 degrees. 60 is a great number to have as a radix of an arithmetic system, since it is easy to do multiplication and division; it has many divisors, some of them being prime numbers, and even the number ten, which is commonly used elsewhere…

[21] Also known as GCD. It is a reference algorithm in many computer languages because it is relatively simple and uses a special technique called *recursion*. In *DAPE's Students handbook* the algorithm is presented in page 71. You may find the Euclid' original statements in ancient Greek together with a translation in English, in APPENDIX G.

Ada Augusta, Countess of Lovelace and daughter of famous poet Lord Byron, was the programmer of Babbage' s analytical machine. She is often recognized as the first computeprogrammer ever[22].

Babbage and Ada had identified the great capabilities of their inventions -early forms of hardware and software- and in an effort which paved the way of the research field of "algorithmic analysis", commented:

> *In every computation there is a wide spectrum of possibilities for task succession, and multiple investigations must influence the choice of the applicable method for a certain computational machine. It is important that the method chosen will tend to minimize the computational time.*

The next steps towards formalizing a symbolic system for programming happened in the 30s and 40s. These efforts where mostly theoretical with no particular computer in mind:

- Plan Calculus by Zuse
- The Turing Machine by Turing; this is still used as reference model for computers of today
- Lambda Calculus by Church; the principles of which, have influenced the functional style
- Mark I by Aiken
- Flow diagrams by Von Neumann

There were more scientists who did similar efforts; Donald Knuth and Trabb-Pardo did a great report of this work in their book "The early development of programming languages".

Unfortunately, a great part of this research was unknown to the people that were among the pioneers of the modern digital computers and those who contributed for the first languages; so the early computer languages' research findings were not taken full advantage of.

[22] A language with her name was designed in the 80s with triggering and funding by US ministry of Defense.

## *The revolution of modern computer languages*

Here is how it all started…

The first computers, could only understand machine language, which is actually a set of 0s and 1s:

```
01000010
11010100
00010001
00010100
10001001
```

For example, this could be a valid program adding numbers 212 and 20, supplied in binary format. The way this was implemented was by setting a row of switches, set them up and down, press another button to proceed to the next command, repeat for each command, then press the "execute button" to retrieve results on a row of lamps or a printed paper. Albeit tedious, this has been the way first computers worked and the first programs were written[23].

If someone had to calculate something more complicated than an addition or a multiplication, like a division[24], he would probably have to break down the work into discrete steps and code these on the computer. This should be done each and every time a new need occurred and, indeed, it happened. Of course, it was necessary to automate this process a little more and be able to generate the necessary code simply by stating a mathematical expression. There was a need for **language compilers**, in order to have some abstraction over the machine and focus on the problem at hand rather than the peculiarities of the system.

The first language compilers[25] were developed for use in scientific and commercial applications during the 50s and 60s. The vast majority of those were written in FORTRAN and COBOL, the former being the standard in academics and the latter being the standard in business applications. Soon after, there was ALGOL, PL/1 and BASIC. Then, there was ALGOL68, PASCAL, MODULA, C and others. All these were imperative languages, a programming style based on the principle that everything is a sequence of commands, which are executed one after the other:

```
step 1: do this
step 2: do that
step 3: goto step 2 until date>30/6/2002
step 4: end program
```

This is how the first computer languages were born and within a few years there were more than 200 of them. A revision of history would help to place this research in a historical context.

[23] And this is the way programming within a CPU happens, a skill which is required for chip designers and still is practitioned today in University Departments of Computer Engineering in a course entitled Microprogramming
[24] Division is probably the first algorithm that children learn at school. Albeit primitive, it is still an algorithm!
[25] The first compiler ever, was named A-0 and was written in 1951 by the Late Rear Admiral Grace Hopper, who later led the committee for designing the COBOL language.

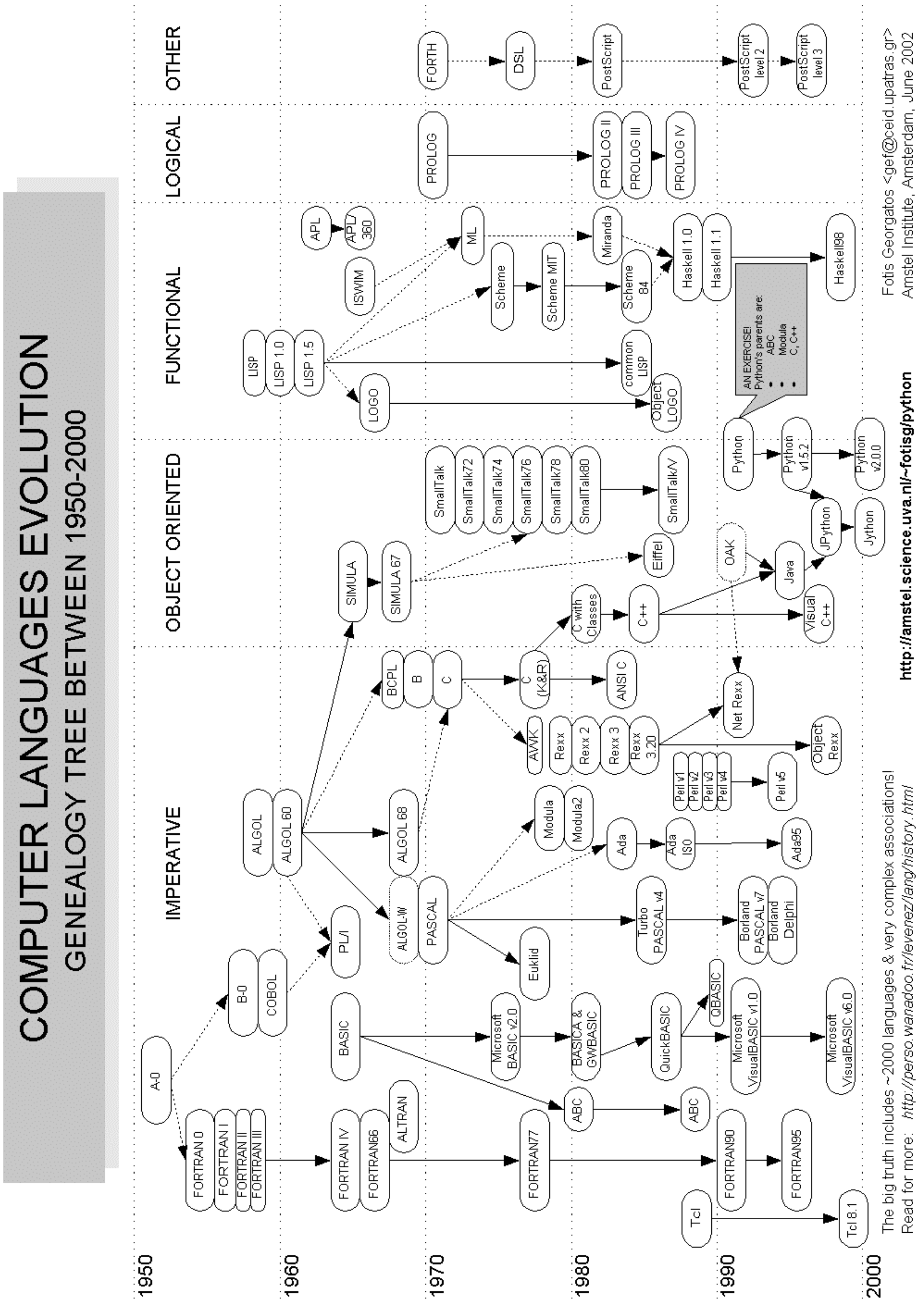
COMPUTER LANGUAGES EVOLUTION
GENEALOGY TREE BETWEEN 1950-2000
IMPERATIVE
OBJECT ORIENTED
FUNCTIONAL
LOGICAL
OTHER
1950
1960
1970
1980
1990
2000
A-0
FORTRAN 0
FORTRAN I
FORTRAN II
FORTRAN III
B-0
COBOL
FORTRAN IV
FORTRAN66
ALTRAN
BASIC
PL/I
ALGOL
ALGOL 60
ALGOL 68
ALGOL-W
PASCAL
BCPL
B
C
SIMULA
SIMULA 67
LISP
LISP 1.0
LISP 1.5
LOGO
APL
APL 360
ISWIM
ML
Scheme
Scheme MIT
PROLOG
FORTH
DSL
SmallTalk
SmallTalk72
SmallTalk74
SmallTalk76
SmallTalk78
SmallTalk80
FORTRAN77
Microsoft BASIC v2.0
BASICA & GWBASIC
ABC
Euklid
Modula
Modula2
Ada
AWK
Rexx
Rexx 2
Rexx 3
Rexx 3.20
C (K&R)
ANSI C
C with Classes
C++
Eiffel
SmallTalk/V
common LISP
Object LOGO
Scheme 84
Miranda
Haskell 1.0
Haskell 1.1
PROLOG II
PROLOG III
PROLOG IV
PostScript
Tcl
FORTRAN90
QuickBASIC
QBASIC
Microsoft VisualBASIC v1.0
Turbo PASCAL v4
Borland PASCAL v7
Borland Delphi
Ada ISO
Ada95
Perl v1
Perl v2
Perl v3
Perl v4
Perl v5
Net Rexx
Object Rexx
OAK
Java
Visual C++
Python
Python v1.5.2
Python v2.0.0
JPython
Jython
AN EXERCISE! Python's parents are:
ABC
Modula
C, C++
Haskell98
PostScript level 2
PostScript level 3
FORTRAN95
Tcl 8.1
Microsoft VisualBASIC v6.0
The big truth includes ~2000 languages & very complex associations!
Read for more: http://perso.wanadoo.fr/levenez/lang/history.html
http://amstel.science.uva.nl/~fotisg/python
Fotis Georgatos <gef@ceid.upatras.gr>
Amstel Institute, Amsterdam, June 2002

## FORTRAN

FORTRAN stands for FORmula TRANslation and was the result of a research effort of an IBM team of the 50s. John Backus was the head of this team, whose aim was to create a high level language that would produce efficient code from algebraic expressions. FORTRAN's design started in 1954. The development of the computer program that did the translation –a primitive *compiler*- took two and a half years and was written in machine code. Nothing else was possible at the time, other than writing in 0s and 1s. In 1957, when it was released it became readily very popular, in great part due to IBM that offered it for free; so much popular, that, in the future, it would be hard to direct programmers to newer and, possibly, better languages developed soon after. FORTRAN has been in wide use since then, having received countless revisions, each one trying to overcome initial limitations or introduce implementations of newer concepts like functions and procedures, structured design, less restricting syntax and so on. It is still very common among academics, because many of the current professors learned programming with it. In engineering sciences, it is preferred due to the excellent libraries written in it for matrix computations[26].

## COBOL

In 1959, US Ministry of Defense assigned a committee the task of designing a computer language for general application in government, industry and the private sector. COmmon Business Oriented Language compilers were still being developed almost a decade later, most known written by RCA and Remington-Rand-Univac. The cooperation between US government and industry led to an abundance of COBOL code and made it very popular during the 60s and 70s. Nowadays, COBOL is declining in use albeit there are still many applications written in it[27].

## BASIC

BASIC stands for Beginner's All-purpose Symbolic Instruction Code and was designed by professors John Kenemy and Thomas Kurtz at Dartmouth College as a modification of FORTRAN for beginners, with an interactive environment for ease in experimentation. Computer users welcomed BASIC and its use was magnified with the introduction of microcomputers in 1975: Dr Wong developed TinyBASIC[28], which takes 2Kbytes of memory and is loaded from paper tape. Bill Gates and Paul Allen were the first to embed the language in a ROM, thus, making it usable by the moment a computer is switched on. Microsoft was founded later for the commercial exploitation of their interpreter[29] and by 1981 they presented BASICA & GWBASIC for the first IBM PC. This was instrumental in the propagation of the language. BASIC has been later redesigned under the names QuickBASIC and VisualBASIC and numerous applications were written in it. The development of BASIC is still continuing in personal computers, and it is the programming environment of choice for implementing macros under Microsoft Office, the most widespread office suite available today.

---

[26] In particular the FORTRAN libraries for matrix computations according to the specifications LAPACK, BLAS1,2,3 are fundamental building blocks for software such as Matlab or its freeware clone Octave and several simulation tools.

[27] But, some ideas which where introduced in COBOL influenced other commercial-oriented languages like dBase and Clipper and resulted to what we now know as database systems. Nowadays, typical such applications are the Relational DataBase Management Systems (RDBMS) and the language used is SQL, standing for Structured Query Language.

[28] The first known freeware program. The text strings *All Wrongs Reserved* and *Copyleft* appeared in this program.

[29] MS Basic was the first program sold by Microsoft Corporation and also the first major case of software piracy – It was copied widely before Microsoft made it available. Bill Gates lost a copy on paper tape during a computer show.

## *LISP*

LISP stands for LISt Processor and was designed by John McCarthy at MIT as the first functional language. LISP is a breakthrough in the way languages are designed because of the introduction of the functional paradigm, an idea that was adopted later in other languages, most important of them being Smalltalk, APL[30], Scheme – a LISP dialect by G.L. Steele and G.J. Sussman –and, recently, Haskell. It brings the notion that everything in a program is a function; whose parameters may be other functions and this concept can apply recursively until the parameters of the final functions are well known, i.e. we reach the data given by the problem definition or language primitives. A functional program is a single expression, which is executed by evaluating the expression. LISP is a language that has influenced a lot the Artificial Intelligence community and has been applied in expert systems, game theory, natural language processing and other disciplines. It is often distinguished by the use of a lot of parentheses to denote the program structure[31].

## *LOGO*

LOGO is a high-level programming language developed by Wallace Feurzeig and Paul Wexelblat for use by learners, including children. Its history is rooted in artificial intelligence and LISP. It is also rooted in Piaget's research into how children develop thinking skills [HB84], since its educational consultant, Papert Seymour, has been a close collaborator of Piaget. Papert promoted LOGO as a learning language, not for a specific branch of mathematics but for problem solving behavior. LOGO has been the first language offering a graphical interface. Learners are expected to drive a turtle through an interactive environment and discover the virtues of programming by experimentation, applying commands like: go forward 50 (steps), right 90 (degrees) 32. The power of turtle geometry is that movement is not described in terms of absolute position -Cartesian coordinates-, but relative to the position and direction of the turtle, a conceptual animal that moves around the screen[33], which makes the turtle "an object to think with".

## *PROLOG*

Prolog demonstrates *logical programming*, which is a popular style in Artificial Intelligence. The dominant idea in logical programming is that a program is a set of logical definitions, that when combined with the initial data, various computable conclusions can be derived of. Prolog is based on a set of a deduction system and rules. Although the theoretical component of logical programming is promising, current computer systems' architectures do not favor towards interesting results in terms of performance and scope in applying this style of programming in practice, so its role is limited. Even more attempts to use Prolog in education have not been so successful [MP90]:

> Prolog has been a troublesome language and fewer interesting research results have come out of teaching children Prolog.

---

[30] APL was designed by Kenneth Iverson to teach mathematics at high school level without intention of actually implementing it on a computer, so it used symbols, which did not exist on keyboards. APL is very powerful: Many computations that require iterative loops and auxiliary variables in other languages are stated in a single line in APL.
[31] LISP: "Lots of Irritating Superfluous Parenthesis", by unknown on the Internet
[32] Try this in a LOGO environment: REPEAT 360 [FORWARD 90 RIGHT 156]
[33] Or …the floor as the LOGO research team suggested, using a robot-like turtle with wheels and a pen.

## ALGOL

ALGOL has never been a language that was widespread in practice, but has influenced greatly almost all subsequent ones. It introduced the concept of *procedural* and *structured* programming, as early as in 1960. A few Europeans designed it and was the first one to have a formal grammar defined in BNF[34], a formalism introduced by Backus and revised by Naur [NP60]. The use of a single command, GOTO -often seen in FORTRAN, COBOL and, later, in BASIC code- had resulted in programs that were difficult to develop and even more difficult to maintain. It is now recognized that the use of GOTO is playing a role in the number of errors that are encountered in programs and that it makes their correction a complicated task. Structured programming is based on the use of only three logical constructs and their combinations: *sequence*, *condition* and *repetition*. GOTO should be avoided and with structured programming it didn't have to be used almost in any case. This, phenomenally simple, contribution of ALGOL was adopted by all subsequent languages and this is what makes it a milestone. The main successor of ALGOL is considered to be PASCAL, but it has truly influenced all languages thereafter, including C and any derivatives thereof.

## PASCAL

Two of the members of the team working on ALGOL were Niklaus Wirth and C.A.R. Hoare, who did not agree on the upcoming ALGOL68 definition and found it too bloated and unsuitable for further development. In 1968, Wirth started designing a variation of ALGOL, named after the French mathematician and philosopher of the 17th century Blaise Pascal, for use in teaching institutions and academic environments. Two years later, PASCAL had combined the best features of the languages in use at the time: COBOL, FORTRAN and ALGOL. It was a general-purpose language that promoted boldly the structured programming style and could be used in a range of applications without difficulty, which is why PASCAL is still so popular. PASCAL influenced later the development of ADA and MODULA, also under Niklaus Wirth' direction.

## HASKELL

Functional programming is appropriate for exploring formal algorithms and mathematics. The language Haskell[35] is considered, as the best vehicle for someone to investigate the possibilities and features of this style of programming, in particular the version Haskell98. What is most noticeable in Haskell is that it incorporates much of the existing research in the field of functional languages and that it takes mathematics symbolism to a higher level. Please refer to appendix F, for a discussion that relates to the definitions of $\Pi$ and $\Sigma$ symbols from the standard mathematics and a case demonstration explaining why this language can become important in the future.

---

[34] Backus Normal Form, later renamed Backus-Naur Form at the suggestion of Donald Knuth; a formalism to denote what Chomsky independently had also called context-free grammars (CFG) or Type-2 languages in his theory of grammars, in 1959. The biggest difference between Chomksy's CFG and BNF is notational; the two symbolisms have been proven to be equivalent. A grammar is called to be context-free if and only if it can be defined by a set of *productions*, whose left-hand-side component has one and only one *non-terminal* symbol. See appendix D for more.

[35] Thanks to the coordination done by Dr. Peter Uylings and Dr. Peter van Emde Boas I had the opportunity to teach this language to a small group of students at University of Amsterdam, on behalf of Dr. Jan van Eijck for the first courses of the laboratory work. Haskell98 is not only a great language to learn, but it is also a great language to teach!

## *C; the one after B*

Dennis Ritchie developed the language C in years 1969-1973, while developing the UNIX system at Bell Laboratories in New Jersey together with Ken Thompson [RD83]. In his own words:

> *The C programming language was devised in the early 1970s as a system implementation language for the nascent Unix operating system. Derived from the typeless language BCPL, it evolved a type structure; created on a tiny machine as a tool to improve a meager programming environment, it has become one of the dominant languages of today.*

The research that existed over structured programming with ALGOL 60 influenced the design of C, as well as the later work done by the committee of ALGOL 68. The predecessors of C were in sequence: CPL (Combined Programming Language), BCPL (Basic CPL), B (a…subset of BCPL).

C is particularly oriented towards system programming and its first priorities have been simplicity, efficiency and flexibility. Because of this, C has been a successful language, dearly used for system programming as well as application programming, for more than 30 years. C combines both the elegance and efficiency of machine language and the readability and maintenance of high-level language[36]. C is still considered best for system-level programming and it is the one in which many known operating systems are written, including those in use on the majority of current computer systems: UNIX, Windows, MacOS, Linux[37]. Major other languages are written in C as well.

In 1978, Dennis Ritchie together with Brian Kernighan wrote a book on C [RD78] and this served as a reference material in the computer industry for many years to come, until the moment that a formal standard was adopted[38]. This book is also known as K&R and included a definition of the C language in BNF[39], which was instrumental in removing syntax ambiguities that existed between different versions of the language by then. There existed C compilers that would produce different programs, due to differing interpretation of the language's syntax[40].

As a bottom line: *C, the most influential language of all in system software engineering, has had an evolutionary development and passed numerous revisions over long periods of time before being a stable tool with a clearly defined syntax and semantics. There is a moral here to remember when programming!*

---

[36] According to others: "C combines both the elegance and efficiency of machine language and the readability and maintenance of machine language". Well, C can be indeed difficult to comprehend in many cases.

[37] Linus Torvalds chose C for writing Linux in 1990. The Linux kernel contains now a little more than a million lines of C code. In order to make this large software project readable and, thus, maintainable by many people, Linus has introduced "Linux kernel programming guide".

[38] *By 1982, it was clear that C needed formal standardization. A year later, American National Standards for Information Systems (ANSI) established a committee with the goal of producing a C standard. This report was published at the end of 1989 and was accepted by ISO as well […]. C language escaped nearly unscathed from the standardization process, and the standard emerged more as a better codification than a new invention.* [RD83, p.11]

[39] For the technical reader: it was actually a BNF-like input file for *yacc*, the standard tool used to generate computer language parsers in C, even up to now. Yucca' s original author is Steve Johnson who later produced also *lint*, a tool to remark on dubious constructions within C source code. For your information, Python is also written in C and has a BNF-like syntactical definition to be used by yacc and relevant tools.

[40] Even then, the K&R reference BNF definition was insufficiently precise on many details of the language that had to do with semantics. For the technical reader, here is an example of semantical ambiguity: a[i]=a[i++];

## *FORTH & POSTSCRIPT*

Forth[41] was developed to regulate movement of astronomical observation equipment. Since the control equipment was driven by primitive microprocessors, the overheads of conventional high-level languages would have been unacceptable. Forth was the compromise between the virtues of assembly code and the virtues of high-level languages [GT90, p. 33]. Many scientific machines used FORTH as their control language, including telescopes like Kitt Peaks. A program written in FORTH controlled the submersible sled that located the wreck of Titanic in the summer of 1985.

Forth is a very special language, because it designed for what is called a stack-based machine. The main idea here, is that all information to be processed, is stored in a single major data structure, the **stack** and is followed by the operations to perform on it, a technique also known in symbolism as Reverse Polish Notation[42]. The main idea behind RPN is that the operations follow the operands:

*AX+B* becomes *AX*B+*

*N(N+1)/2* becomes *N 1 + N * 2 / or N N 1 + * 2 /*

The expressions are interpreted from left to right, which for a small system means that no complete language parser is necessary, which is complex to build and would require more memory.

Green raises questions about the use of FORTH and this programming paradigm [GP90, p. 34]:

- Is it readable?
- Can we parse for structure?
- Is it error-prone?
- Is it easy to learn because of having fewer concepts?

The answers to these questions make FORTH look comparable to machine language in complexity, since it trades speed of execution for risk of errors.

A descendant of FORTH is PostScript[43], a device-independent page-description language, which is used to control laser printers, nowadays. In fact, PostScript is the programming language that is being executed at the microprocessor chip of laser printers or similar output devices.

This is from a problematic printout from a PostScript printer, as seen on the 28th of May 2002, at the 1st floor of Euclides Building, University of Amsterdam:

```
ERROR: undefined
OFFENDING COMMAND:

STACK:

-mark-
```

What probably happened, is that the **stack** was empty, most probably due to some programming error, either in the .ps document or the microprocessor's code. Do you want to debug this further?

---

[41] Forth's creator, Charles Moore, regarded it as a fourth generation's language, hence the name. Well, the machine that he used could only handle filenames up to five characters long, that's how Fourth became Forth.

[42] Also known as postfix notation. There is also prefix and infix, the latter is the one commonly used in mathematics etc.

[43] In between, there is a language named Design System Language, forerunner to Postscript

## *A few words on Object Oriented Programming; technicalities included…*

Object Oriented Programming is one of the most important innovations in programming languages. It was introduced in the 70s, particularly with the languages Simula[44] and Smalltalk. The idea behind Object Oriented Programming (OOP, or simply OO) is an observation taken from real life:

*We manipulate or speak about objects, things or living beings, and have a less-or-more clear insight in the properties, actions and reactions of that object. Why not do the same with computers?*

To have a first view of what object oriented programming is like, we can think of mathematical objects that are related in a hierarchical manner: **Quadrilateral => Rectangle => Square**.
The properties of a *parent* object are *inherited* to the child object. Inheritance can be defined as an action of refinement of a notion without repeating everything that can characterize a former notion. Note that a square can as well be defined as **Quadrilateral => Rhombus => Square**, so a square's properties are both those of a rectangle and a rhombus. This is exactly what *multiple inheritance* in OOP is all about!

In the object-oriented programming paradigm, the principal idea is that all software components are object-like entities with attributes and inheritance relationships defined between them. To connect this with the experience that most people already have with computer systems, let's provide an example for defining an "YesButton" of a graphical interface:

- code is first written by the programmer who defines what is a "Button" object and how it looks like: it is rectangular, it can be pressed etc.
- from that, an "YesButton" object is inherited, assigning its *text* property to contain "Yes"
- using this new object "YesButton" within another window, will have the obvious result on the screen.

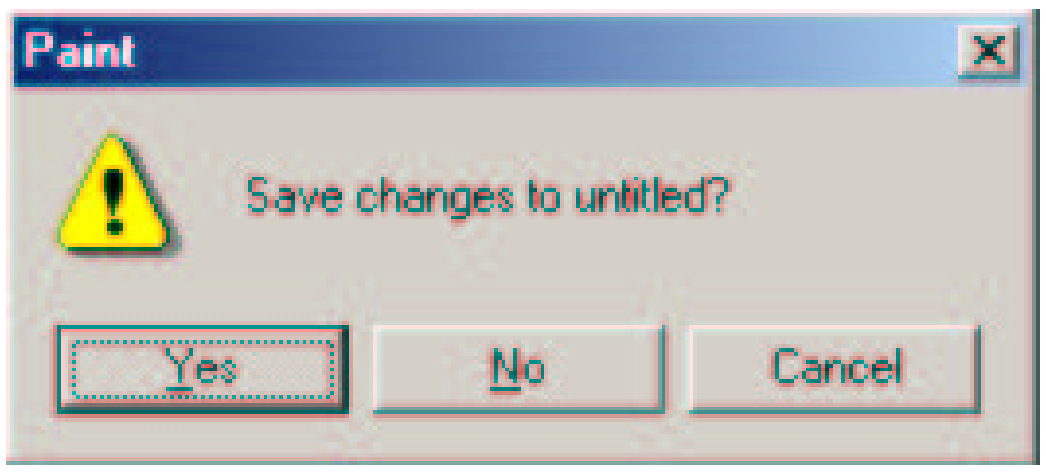

The concept that everything is an object was nourished by the Graphical User Interfaces (GUIs): In the late 80s, the advancements in graphical capabilities of modern computer systems triggered a massive reform in programming methodology. Many existing languages had been rewritten[45] to adapt this important innovation. The following table demonstrates this evolution:

| Language | Object Oriented Version | Visual Programming, for GUI Environments |
|---|---|---|
| Basic | | Visual Basic |
| Pascal | Borland Pascal 6.0 | Delphi |
| C | C++ | Visual C++ |

[44] Simula was created in Oslo, Norway. It was based on Algol60 and originally was meant for simulation tasks.
[45] Their syntax included extra constructs that made possible newer concepts to be defined. For example, C++ syntax was a modified C that added now the keyword *class*, while in Pascal's syntax the keyword *object* has been added, of course among other related things. The keywords *class* & *object* have an equivalent purpose: they define an object.

BASIC turned out as Visual Basic, an effort done by Microsoft. Pascal adopted its Object Oriented capabilities in Borland Pascal v6.0 and Delphi, which is graphical environment meant for building applications. C has been redesigned under the names: C with Classes, Objective C, and C++. The later was designed by Bjarne Stroustroup and is pronounced C-plus-plus; it' s the one that prevailed.

C++'s next evolution is the well-known language **Java**, whose most advertised attribute is the ability to run within web browsers and gain true independence from the underlying platforms[46]. The convenience of running Java applets directly from the web, without having to know anything about what is a compiler or interpreter –because it is embedded within the web browser-, made it very popular on the Internet. It is often considered the "Internet language". A Sun Microsystems' team is leading the evolution of Java, which is one of the primary company's projects.

FORTRAN?

Well, that was too difficult and clumsy to rewrite while adopting the OOP model!

In fact, even BASIC, PASCAL and C did not manage to follow the wave of OO painlessly: the syntactical structure of these languages had to be readjusted in such a way that existing programs would continue to work and the new OO functionality would be added. This resulted in these three cases, in languages that needed very verbose OO code and a behavior not intuitively predicted.

Concluding, let's remind that Object Oriented programming is often taken advantage of, when combined with event driven and visual programming. These are mutually orthogonal though: the one does not require or enforce the other but, usually, they are implemented next to each other.

What is most important with Object Oriented programming is that it becomes increasingly useful nowadays. It is apparent that the instructional language of the future must contain object-oriented features. On the other hand, having novices in a programming course with such a refined topic, could be quite difficult and make programming look complicated, even to the brightest students. Using initially an imperative and then an object oriented language has its disadvantages, mainly that there would be two languages to focus on[47]. Furthermore, it should not be misunderstood by the students that Object Oriented Programming need not be relating to a Graphical User Interface. It is an abstract model, but the facilities to work with GUIs should be available there, too.

This means that serious consideration has to be taken when considering which language to use for teaching, because it should allow discussion of the OOP model, at any level, without enforcing it.

---

[46] But that was the reason computer languages were introduced in the first place. Many people praise Java for the fact that it is independent of the system being run, but this should be a property of other languages as well. Why this did not happen has to do with the way companies exploited the other computer languages.

[47] This is indeed an unsolved problem with the case of DAPE. The book begins with plain Pascal and BASIC to discuss the main Imperative constructs. When the moment of Object Oriented programming comes, there is some mentioning of C++, Delphi, VisualBasic, Java and Smalltalk. The books' pictures are from VisualBasic. This may give the false impression to students, that Object Oriented programming is about building Graphical User Interfaces, which is not always the case. Python can work as an Object Oriented environment, without enforcing it, with relatively little effort.

## *First scripting languages: AWK, TCL, REXX*

In 1978, Alfred V. Aho, Peter J. Weinberger and Brian W. Kernighan realized that there was a growing need for a simple text processing scripting tool and implemented a tiny processing language with advanced features named after their initials, AWK. They wrote:

> *AWK is a convenient and expressive programming language that can be applied to a wide variety of computing and data-manipulation tasks*[48]

Until that moment, people were writing in UNIX' *shells*, primitive programming environments designed for interactive use; they were facing problems of features and portability[49].

Rexx was developed by Michael Colishaw in 1979 and was circulated within IBM's network. A few years later it became the standard language of IBM's OS/2, albeit it can run on any platform. It is an easy and clean procedural language oriented towards manipulation of words and numbers. During the 90s, an Object Oriented and a Java version appeared, called Object Rexx and NetRexx.

Tcl has been developed by John Ousterhout and followed a similar path. It is particularly popular among UNIX systems, where it is combined with the high level graphical toolkit Tk to build easy to use interfaces for command-line tools. It is adopted and supported by Sun Microsystems nowadays.

## *PERL*

Perl[50] was written by the linguist Larry Wall in the early 90s, for the purpose of text parsing and manipulation and became very popular in just a few years. It was initially written as a quick reporting language that would combine the facilities of UNIX's sed and awk commands.

Perl language faced successive waves of expansion during the early 90s and then its syntax was cluttered with too many new ideas that were not originally provisioned. This has an effect upon, particularly larger, programs that become unreasonably difficult to read and grasp as a whole, after only a few days absence. For one thing, Perl is not possible to be described in a BNF-like formalism and its interpreter is too obviously a collection of evolutionary developments.

In any case, Perl is currently the scripting language, which is in widest deployment. It is commonly estimated to be the one behind more than 80% of the 'live' content of the Internet [RE00], which includes forms, mail robots and anything automated.

---

[48] Anyone familiar with the field of seismology will easily observe that scientists in this discipline do a lot of work with awk: abundance of digitally recorded data in the 70s forced seismologists towards the first well written tool available…
[49] Most known of them are: sh, ash, bash, csh, tcsh, ksh, zsh. The portability problems meant that programs would not run when changing environment, either different shell or even the same shell on different systems. Nowadays, they have converged in features and syntax, and are nearly finished general purpose languages.
[50] Depending on whom you ask, PERL stands either for *Practical Extraction and Reporting Language* or *Pathologically Eclectic Rubbish Lister*.

# PYTHON

Research in programming languages didn' t stop and in the 90s more languages had appeared, each having a different priority in mind. One of them by that moment was ABC, an evolved and indentation-structured form of BASIC, mainly designed for use in education. Guido van Rossum was working at this time at the Amoeba Distributed Operating Systems group for the Centrum voor Wiskunde en Informatica, and contemplated an extensible scripting language with ABC's easy on the eyes syntax, while having good facilities for calling and cooperating with other programs. He considered a few ideas found in C, C++, Icon and Modula-3, and started writing Python.

Python is a Very High Level Language and an Object Oriented Dynamic Language. It concentrates the experience of multiple years of work in Computer Science and incorporates ideas found in imperative, object-oriented and functional programming paradigms; it does include *exceptions*, *modules* and *classes*. Python's object model is easy to use[51]. It is interpreted and sacrifices some performance in order to maximize development speed. It can call other programs and can be embedded in other programs, which makes it ideal as a scripting language within bigger packages.

The first release of Python in the public domain happened in 1991. The version of the language that is mostly found in various systems is v1.5.2. At the moment, Python is at version 2.2.1

Python, contrary to LOGO and ABC which are also educational languages, is not considered at all a toy language: it is not only designed with educational requirements in mind, it also has properties that make it stand high next to professionally widespread languages like C and Perl.

Last but not least, it is available for free. Anyone can download it from the Internet, install it, give to a friend, a student, a teacher and this is perfectly legal. The language is supplied together with source code and documentation, which means that it is possible to see what are the internals of it, if you are interested and even contribute in its development or steer its direction. There are also plenty of Python programs on Internet sites, available for download.

What more could someone ask for a language to be useful in primary and secondary education?

What more, rather than a language that can continue to be useful for the years to come?

[51] Without this implying any deficiency in multiple inheritance, encapsulation, polymorphism or operator overloading.

# REVIEW OF THE LITERATURE

## *Psychological studies of programming*

The history of reviewing programming from a psychological point of view, begins probably in 1967 with the work of Rouanet and Gateau, in seek of finding patterns in business programmers' behavior[52]. For the interested reader, their conclusions have been that the programmers lacked a methodology of using abstract representations of information and control flow, consistently. In other words they were taking no advantage of higher level language' facilities and their code included low-level machine language constraints. [PP90, p. 3]

Computer scientists, who developed a "normative" approach, to what they considered to be the most powerful programming concepts, did most subsequent studies. These concepts included in those early days ideas like using *flowcharts* and *structured* programming.

The second generation of programming begins with the work of G. Weinberg in 1971, entitled "The Psychology of Computer Programming", in which it is supported that a psychological point of view must influence the way programming is conducted. This was in contrast with other studies of the times that were, often, rapid assessment of tools that led, equally often, to contradictory results. For instance, improvements and lack of effect have both been reported while studying the use of flowcharts, by Wright and Shneiderman respectively. Later, J. Brooke and D. Gilmore found that there are certain circumstances that can favor the use of flowcharts. [PP90, p. 4]

The third generation of programming' studies happened on firmer ground, initiated with a debate on the theoretical and methodological contexts of this kind of research, by Hoc, Moher, Schneider. This era is marked both by psychologists studying programming, like Green, as well as computer scientists in the cognitive science field like Soloway.

## *Conditions for successful learning of a computer language*

Du Boulay et al [BD81] have set conditions, that they consider necessary for the successful learning of a computer language, as the case is with novices:

- a conceptually simple notational machine
- some processes of this machine should be visible to the user
- the system should be interactive
- the teaching materials employed should be harmonious with the particular implementation of language, e.g. they should use the same terms;
- the commentary should be at the appropriate level of detail for the task given to novices, and for their conceptual grasp, e.g. error messages

---

[52] I thank Dr. Panagiotis Politis, one of the authors of "Development of Applications in a Programming Environment" textbook, for helping me with suggestions on literature and providing study materials in printed and electronic format.

## *Programming in the school curriculum*

The most commented benefit of the application of computer languages in education is the ability to do reflection while learning in them [JDC00]:

> *...to acknowledge the potential of the range of software tools now available, this paper takes the position that a programming language provides an essential tool/environment for learning, and thus warrants the time and effort necessary for developing the ability to use and then apply the power of the software in the study of mathematics throughout primary and secondary school. [...]*
>
> *In treating a big idea as a procedure one gains in two ways. The procedure provides a dynamic description, as contrasted with a static definition, and further the procedure, whether it is executed manually or electronically - by hand or computer - now becomes a powerful means i.e. an object or basis for the construction of a new 'object', for further exploration and investigation in mathematics itself.*
>
> *[...] it is also the case that there is a critical dialectic between the constructs available in the language and the pupil's thought processes - one tends to think in terms of the constructs and capabilities of the languages which is to be used in the implementation - and as such this also impacts the way pupils learn.*

Johnson shares similar views with Cynthia Solomon [SC86, p. 8]:

> *Thus computer environments can offer children opportunities to develop their intellectual abilities by making personal discoveries through a continuous process of building on what they already know.*

Moreover, programming is confirmed by Suppes to conform to the postulational model of thinking:

> *Mathematics - or any other subject matter - can be broken down into individual facts; the relationship among the individual facts or elements can be organized hierarchically. In this analysis subject matter is composed of local knowledge: One fact leads to another fact that exists higher in the logical structure.*

## *The role of pseudocode*

In most cases of teaching a programming language, a teaching technique of using an intermediate language, named *pseudocode*, is deployed. Pseudocode fulfills an educational role: bridge the gap between the natural language's way of thinking and the statement order, which is expected by a computer system. The technique is well established both in academic Computer Science courses, as well as in the cases that computer programming is taught in compulsory education, high school level included. Even expert programmers make use of a private form of pseudocode to devise problem solutions. According to Petre:

> There was evidence that they (expert programmers) solve problems, not in the target programming language, but in a orivate, pseudo-language that is a collage of convenient notations from various disciplines, both formal and informal.

Kernighan and Plauser endorsed as a principle for good design, first to capture a solution in an abstract private pseudo-language and then proceed with writing the real code [KB74]:

> Write first in an easy-to-understand pseudo-language; then translate into whatever language you have to use

## The importance of multiple paradigms

It seems as if new needs are developing in the methodology of teaching programming, which involve more than one paradigm; which practically means to deviate from the imperative model. According to Pair's view [PC90, p.18]:

> *Formerly one used to say that the language in which a program was written was not very important: all languages used were indeed once similar, all were algorithmic*[53] *languages (except Lisp which was not well known). This is no longer the case nowadays.*
> *[…] The development of programming languages and methods, and the teaching of them, have up to now hardly been linked to a psychological study of the activity of programming, and this can account for certain failures. To be of any use however psychology must go beyond the procedural aspect of programming; it must take into account those other styles*[54] *which, even if they are not new, are becoming more and more important nowadays due to the variety of applications and the training that programmers receive.*

Green stipulates that there are problems, whose solutions are a better expressed in a combination of procedural –imperative- and declarative –functional- paradigm [GT90, p.18]:

> *Many problems seem to be peculiarly intractable to any single programming paradigm. If approached procedurally, it becomes clear that part of the problem is best approached declaratively and vice versa. Why not use a mixed paradigm, and treat each aspect of the problem on its merits?*

## The opportunistic planning effect

Unlike what someone could originally think, programming is not a top-down strategy, although the total result has such a structure. According to Pennigton and Grabowski [PP90, p.50]:

> *The dominant view of planning discussed in the psychology and artificial intelligence literatures is one of step-wise refinement, in which the primary process is one of top-down, breadth-first decomposition. In this method, a complex problem is decomposed into a collection of (ideally) non-overlapping sub-problems. The subproblems are decomposed into further subproblems and this is repeated until the subproblems are simple enough to be solved by retrieving or specifying a known plan for solution (N. Wirth, "On the composition of well structured programs", 1974).*
> *[...] However, as a view of what the design subtask actually involves, design by step-wise refinement presents an overly simple view. First, there is evidence to suggest that the design process is not as orderly as that required by step-wise refinement. Miller and Goldstein (1977) found that in many instances their computer coach needed a mechanism to alter the coach's approved (orderly) expansion. Other data also suggest that there is some amount of alternation between levels of planning as early decisions have implications for later steps and later steps may call into question some aspects of earlier decompositions. Some are even more pessimistic, suggesting that good programmers "leap intuitively ahead, from stepping stone to stepping stone, following a vision of the final program; and then they solidify, check, and construct a proper path.*

Green suggests that opportunistic planning is what people really do when programming, not what they might do if they were perfect and followed the "approved waterfall" top-down style of development [PP90, p. 119]:

> *Many designers accept the evolutionary style of programming, in which the activity of program design is one of repeated modification – frequently starting from a seed, which was an already existing program.*

[53] The original author probably wants to say *imperative*, not algorithmic; since all computer languages are algorithmic
[54] Among others: procedural, object-oriented, functional paradigms

## *The discussion on modern computer languages*

**To begin with, there is no such thing as a perfect language; no language is a panacea. It is usual that some languages excel in one way, while other languages excel in another. Having said that, it is better to have a bad, but working, language than having nothing at all.**

Not all computer languages have the same characteristics, nor are they emphasizing by their design in similar topics. Such characteristics can be

- imperative paradigm (FORTRAN, COBOL)
- structured programming (ALGOL, PASCAL, C)
- educational orientation (BASIC, LOGO, ABC)
- object orientated paradigm (C++, Java, Smalltalk)
- scripting & prototyping (Perl, Tcl, Rexx)
- functional paradigm (Haskell, LISP, Scheme)

These characteristics are sometimes orthogonal, which means that a computer language can have both. For example, LOGO is also an *applicative* –synonym for functional- language. BASIC, ABC, C++, Java, Perl, Tcl, Rexx are all imperative languages but they emphasize on different aspects. Haskell is also good in education. The modern dialects of BASIC and PASCAL are object oriented, as well as newer versions of Perl. Almost any language that has been rewritten in the 90s and then, adopted the object oriented paradigm's ideas…

### FORTRAN

**FORTRAN** was designed in the 50s and being one of the very first languages, does not follow a particular paradigm other than that of imperative programming. It is still a widely used language, particularly among engineers [CMJ82]. Here follows an example program, written by Knuth and Trabb-Pardo, in FORTRAN looks like. This program will fill an array named **A** with 11 numbers supplied by the user, calculate the function $\sqrt{|T|+5*t^3}$ for each of them, then print all the results which are smaller than number 400, in reverse order, and their exact value. Note the use of GO TO statements.

```
C       An example program in FORTRAN
C       By Knuth and Trabb-Pardo, 1978
        FUNF(T)=SQRTF(ABSF(T)) + 5.0*T**3
        DIMENSION A(11)
1       FORMAT(6F12.4)
        READ 1,A
        DO 10 J = 1,11
        I = 11-J
        Y = FUNF(A(I+1))
        IF(400.0-Y)4,8,8
4       PRINT 5,I
5       FORMAT GO TO 10 (I10, 10H TOO LARGE)
8       PRINT 9,I,Y
9       FORMAT(I10, F12.7)
10      CONTINUE
        STOP
```

Most people that are using other languages cannot avoid criticizing FORTRAN and mention its limitations. Hereby is provided a humorous and slightly sarcastic quote over FORTRAN as written by Ken Thompson[55], in the paper "*Reflections of trusting trust*": [TK84, pp. 761-763]

> *In college, before video games, we would amuse ourselves by posing programming exercises. One of the favorites was to write the shortest self-reproducing program. Since this is an exercise divorced from reality, the usual vehicle was FORTRAN. Actually, FORTRAN was the language of choice for the same reason that three-legged races are popular.*

## BASIC

**BASIC** was designed in the 60s. It is known to be excellent in the prototyping aspect due to its interactive environment, although in its first years lacked design and structured syntax almost just like FORTRAN. Easy problems are very easy to solve in BASIC, but hard problems are close to impossible [HB84]. Why? The reason is that the early versions of the language were not designed around the concept of *code blocks* with *procedures* and *functions*. Any large BASIC program is bound to be an unreadable maze of GOTOs. GOTO was and is still unavoidably used very often in FORTRAN and BASIC programs, because this is what the language encourages. Students come to rely on it, even though it is a poor mechanism [LM88, p. 78]. Even now, that newer versions support *structured* code, programmers' inertia resists in the old habits. The practice of using the GOTO command has been criticized, at first by Edsger W. Dijkstra in a paper well-known in Computer Science world entitled "*Go To Statement Considered Harmful*" [DWE68, pp. 147-148].

Another problem that programmers in BASIC faced is the different dialects of the language. There are tens, probably more than a hundred different versions, that have been created since there was no single authority on designing it and taking care of its continuous evolution. For more on this topic, there is book by the original authors of BASIC, Kenemy and Kurtz, entitled "**Back to BASIC: The History, Corruption, and Future of the Language**" [KT85].

## PROLOG

PROLOG had attracted initially a lot of attention in educational research, because it diverged a lot from the imperative (execution-like) programming model. According to Green [GT90, p. 25]:

> One school of thought supports logic-based programming, on the grounds that our natural mental model is supposed to concentrate on logical relationships, rather than on the order of executing functions. In reality, many difficulties arise in Prolog, some of which reflect the simple fact that logic may be logical, but it is not natural. Empirical research indicates that Prolog novices find execution-based models of computation easier than logic-based models and that even experts use both types of model, not just logic.

Not only imperative than logical programming seems more intuitive to novices, but Prolog also requires an intimate knowledge of the core internals of the language, in order to be understood:

> The need to learn the operating rules of the Prolog machine is already shown to be a necessary condition to designing complex programs in this kind of language. [TJ87]

[55] A well-known pioneer in software engineering and one of the two initial designers of the UNIX operating system. Creator of B language, which in turn influenced greatly **BCPL** of Martin Richards and famous **C** of Dennis M. Ritchie.

## PASCAL

PASCAL keeps the user away of the machine's design internals and protects against as many mistakes as possible. Its syntax is unambiguous and defined precisely be a context free grammar [GT90, p.22]. What PASCAL is really good at, is for composing hierarchical -block structured- programs. Hierarchy is the primary tool for partitioning software complexity and making programming problems tractable. The foundations of PASCAL were built while the language ALGOL was still developing: In 1966, Bohm and Jacopini presented in a conference the theoretical aspects of block-structured programming [BC66]. Their findings did not became publicly known until two years later, when the professor Edsger W. Dijkstra published the paper entitled "GOTO statement considered harmful"[56] [DWE68].

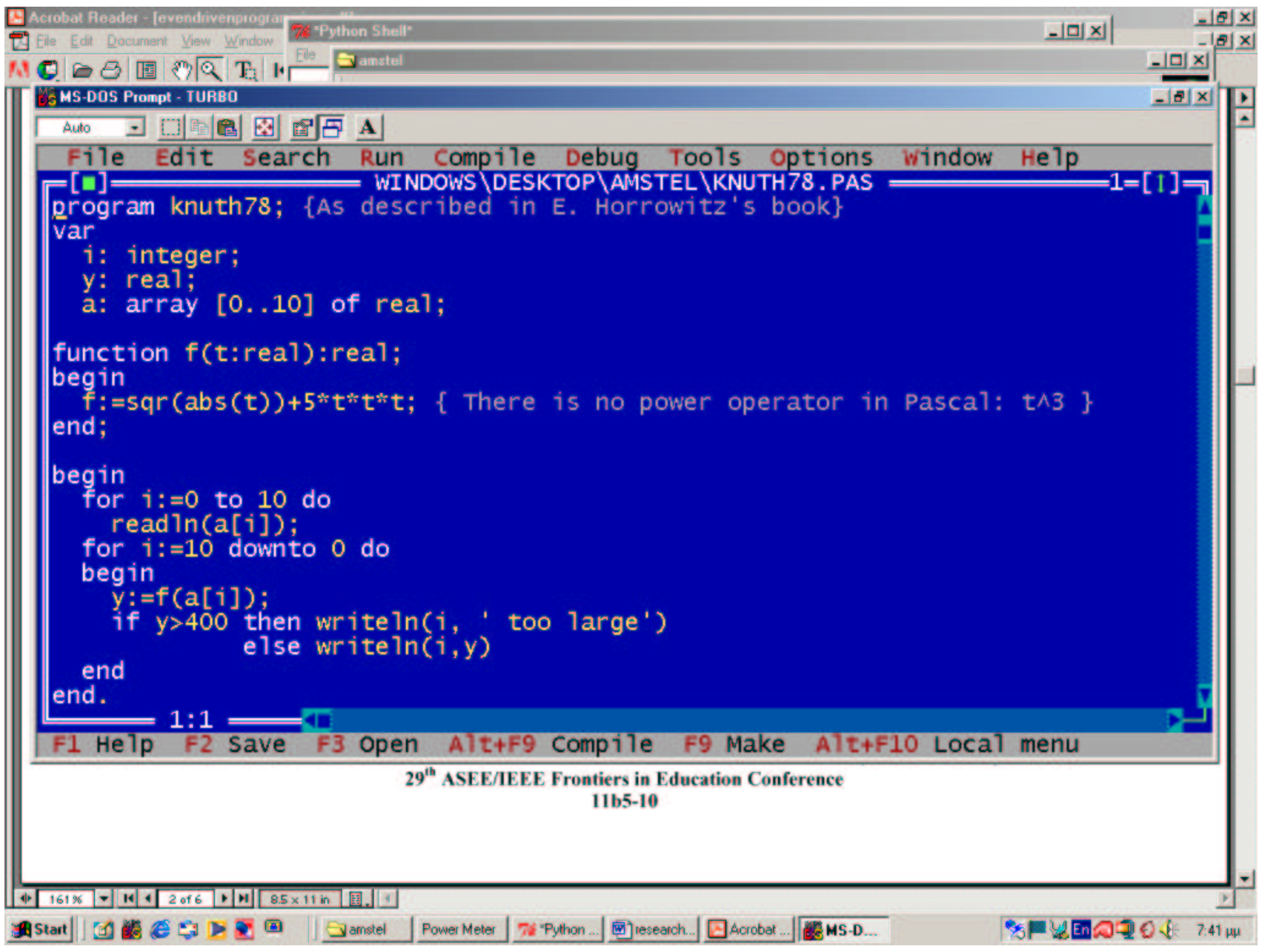

This attitude of promoting structure, which is by design, encapsulated in PASCAL, made it also very "beaurocratic". PASCAL can really teach structured programming, but it takes an effort to write and understand the programs only due to the length of code. It has been called *"a toy language, suitable for teaching but not for real programming"* and "*a wordier and less expressive language*" by Brian W. Kernighan [KBW81], one of the most influencing people in the field of software engineering[57]. He comments further on:

> *It is suitable only for small, self-contained programs which have only trivial interactions with their environment and that make no use of any software written by anyone else.*

In PASCAL indeed, you may have to write verbose code even for the simplest things: A lot of code means for students that they have to use more symbols per line, so the likeliness of an error is higher.

PASCAL is still considered great for Didactics of Informatics, promoting readable and well-structured programs, but it can also demonstrate drawbacks in a school class or in real applications.

---

56. This, in turn, triggered an opposition with the paper ' "GOTO statement considered harmful" considered harmful'. This was answered with a paper "GOTO statement (considered harmful)$^{n}$, n is odd" and led to further discussion and evolution of the *exception handling* mechanisms of the existing programming languages.

57. One of the people that influenced the language C and the design of UNIX systems; a creator of the AWK language.

## LOGO, Papert & Piaget

Papert has a background, which includes mathematics, cognitive psychology and artificial intelligence. Saymour Papert's early work with Artificial Intelligence caused him to think about what human intelligence is, which eventually led him to Geneva where he studied with Piaget from 1958 to 1963. Piaget described stages of mental maturation through which, children go with age and experience. Piaget's ideas had a strong influence on Papert, which is boldly demonstrated in his two important and often cited books referred as *Mindstorms* [PS80] and *The Children's Machine*. LOGO has been used extensively in educational research, not only by Papert but also by many others, including the professors Harold Abelson and Andrea diSessa. The two most often quoted books over LOGO are the *Mindstorms* of Papert and the one by Abelson and diSessa entitled: *"Turtle geometry: the computer as a medium for exploring mathematics", 1981, MIT Press*

Papert asserts that by providing the Logo Turtle as an object to think with, children between 8 to 14 years old, who are concrete thinkers, can overcome their way of thinking, which at that age is only about objects, and think about thinking in itself. In other words, it allows *metacognition* to happen, which is believed to be a formal operational process, too abstract for concrete stage thinkers. The program's code that pupils write themselves is available for further experimentation and reflection.

> *The advantage is pedagogic rather than computational: turtle geometry is compatible with a learner's own experience in the world; it is "body syntonic* [LOGO-F]

LOGO compared to other languages has the advantage that promotes an abstract thinking and does not develop bad habits to the people learning it, thanks to its emphasis in symbolic computation. According to the guide "Preparing to study Computer Science at Cambridge", regarding advance practical preparation, as of June 2001:

> *Teaching yourself to program can lead to your picking up bad habits that will hinder your progress later. In particular, you should avoid languages like Basic, C and Fortran. Pascal is passable but your best bet would be to experiment with Logo.*

LOGO does teach structured programming, but it is not an imperative language: it belongs to the functional paradigm, just like LISP. This fact has been little recognized and is probably a reason why teaching more than the basics in programming with it, ignoring its LISP inheritance, has been a complicated adventure for educators and children. Researcher R. D. Pea reports: [PRD83]

> *LOGO is cognitively complex beyond its early steps, and quite difficult to learn without instructional guidance, even if students are intellectually engaged with that learning. While the semantics and syntax of LOGO are readily learned, the pragmatics – how to arrange lines of legal programming code to achieve specific ends – is a great challenge.*

Even researchers that did use and commented on LOGO did not realize this property of it and give LOGO characterizations like *procedural* or *recursive* [HB84].
The research effort done with LOGO by Saymour Papert, Alan Kay and Cynthia Solomon is great, but LOGO is, nowadays, considered outdated and is not actively maintained as in the past[58].

---

[58] In the meantime, languages evolved. Haskell is the most advanced from the family of functional languages. Python includes the components of functional languages that are useful for practical and educational purposes.

## C, C++ and JAVA

**C,** which is distinguished for brevity and dense symbolism, is unnecessarily complicated for novices. C was built to be fast and powerful, at the expense of being hard to read. It is a good language in terms of technical design and can be very structured. This is not always advantageous for beginner level educational purposes: more often than not it is not easy on the eyes and the same applies for its object-oriented derivatives **C++** and **Java** (there is an example in the next page).

According to Green's opinion about C [GT90, p.22]:

- *Error messages are short and often hard to interpret*
- *C frequently gives up when a runtime error occurs, leaving the user to work out the mistake*
- *C has a very terse syntax which is not completely unambiguous*[59]

Let' s stress this out a little more; C can be so complicated for the human reader that there are even annual contests on that, under the name IOCCC: "*International Obfuscated C Code Contest*"

```
                                                   extern int
                                                       errno
                                                         ;char
                                                            grrr
                             ;main(                           r,
  argv, argc )            int    argc                           ,
   r        ;           char *argv[];{int                     P( );
#define x  int i,       j,cc[4];printf("      choo choo\n"     ) ;
x  ;if    (P(  !        i              )     |  cc[  !      j ]
&  P(j    )>2  ?        j              :     i  ){*  argv[i++ +!-i]
;              for    (i=              0;;    i++                   );
_exit(argv[argc- 2    / cc[1*argc]|-1<<4 ]    ) ;printf("%d",P(""));}}
  P  (    a  )   char a   ;  {    a  ;   while(    a  >      "  B   "
  /* -    by E            ricM    arsh             all-      */);    }
```

This example is a valid C program written by Eric Marshall in 1986, printing on screen "choo choo".
It is not easily understandable even for an experienced programmer albeit it is fine for a computer.
For more examples like this, please visit the original source *http://www.ioccc.org*

Even after laying out the code in another way, C can be distinguished as a language heavily challenging the psychological limit of seven points of interest in a single contextual area. This is due to the case that its syntax allows heavy nesting of *expressions* and *operations*. The human cognitive processing capability is known to be able to track up to seven items in short term memory. This may serve as an explanation in why it is hard to cope with valid C statements like:

**for (i=0;++i>>8;) { printf "i is %d:", i };**

This limit on cognitive skills was first introduced in a paper of the psychologist George A. Miller [MGA56, pp. 81-97]. Computer programmers adopt this constraint[60] and it is known to researchers in the field of Human Cognition and Programming [OT90, p. 69] as well.

---

[59] C was not originally defined by a formal grammar; read about BNF or N. Chomsky's Type-2 languages for more.

[60] Sometimes without realizing how much it does happen without explicit directive. Linus Torvalds, inspirator of Linux, suggests in the document "Linux Kernel Coding Style", to avoid using more than 7 variables in a function for this very reason. Also, the fact that there is a "style definition" document for this huge software project in C, is a subtle hint that there is more to "style" than just style: indentation and code formatting can be important in programming.

## PYTHON

So far, there has been no extensive research in Python, with the exception of the preliminary work done by the teacher Jeffrey Elkner [JE00] [JE01]. Most papers published over Python are either studies in application use of the language or personal perspectives on its suitability for teaching. In some of these papers, it is easy to realize that there is great potential in deploying this language in education. This thesis is trying to do just that.

Hereby, there is an attempt to intrigue the interest of the reader, by providing an example.

Next examples are taken from a report comparing Python and Java on technical features [LG00], which includes eight very small examples of programs which are demonstrative of the languages' expressiveness. Here we have the two cases of a program…

a) doing nothing

| Python | Java |
|---|---|
| | `Public class NoTest { public static void main(String[] args){} }` |

b) writing the first million of positive integers on the screen:

<table>
<tr><th>Python</th><th>Java</th></tr>
<tr><td><pre>for x in xrange(1000000):
          print x</pre></td><td><pre>public class ConsoleTest {
  public static void main(String[] args) {
     for (int i = 0; i < 1000000; i++) {
        System.out.println(i);
     }
  }
}</pre></td></tr>
</table>

The first Java program, doing nothing, is non-empty and requires a lot of explanation.

The second one in Java is writing the first million integers. Java requires knowing the typing model of the language; if not the whole object oriented methodology and its libraries. Note that Java is similar to C++, which is an evolution of C, and the respective programs are also similar. It has become almost impossible to program well in Java or C++ without an intimate understanding of their internals, which is ironic given that one of their primary design aims is to allow the programmer to work at a greater level of abstraction.

Python seems to allow an incremental learning of the language without enforcing to dive deeply in the internals of computer science in order to write the first proper programs.

Hopefully the two last code fragments were indicative enough for the non-technical readers!

## *Common errors encountered in programs*

This chapter is based on the book of E. Horowitz "*Fundamentals of programming languages*", particularly chapter 3.4 on *Syntax and Reliability of programs* [HE84], which is a collection of research done by other scientists on common software errors, what is publicly known as *bugs*.

### Unexpected assignments due to syntax

The language has to be designed such that its characteristics are easily analyzed and correctly interpreted by both human and machine. Look at this example, written in FORTRAN:

```
C       This is a FORTRAN program; this is a comment line
        DO10I=1.5
        A(I)=X+B(I)
10      CONTINUE
```

A quick look would make as think this is a DO loop of FORTRAN, in which the variable I, will successively take values 1,2,3,4,5. A closer look reveals that there is a dot where a comma should be. A well-designed syntax should mention this as a problem. In FORTRAN, this is a valid assignment command that will assign the value 1.5 to a variable named DO10I, to the great embarrassment of the programmer or the ones making use of it[61].

Another problem occurs when the syntax of the language is ambiguous in its interpretation, which can occur when a symbol or command has two different meanings. Take, for example, PL/1's command:

```
A = B = C
```

A person would think that this is assigning the value of variable C to variables A and B. What is indeed happening, is A = ( B = C ), which means B gets the value of C, and A is true or false depending on the equality of B and C, because "B=C" is a conditional expression, as well. In other words, the expression B=C is treated both as comparison and assignment in the same expression.

In a research done By H. Morgan [MH70], it was found that 80% of typing errors in a typical program has to do with the replacement, insertion or deletion of a single character, or the swapping of two. In languages that variables are not declared, these kinds of errors within variable names result in indirect declarations. This is a very common kind of error.  For this reason, it has been supported that *strongly typed* languages are better for writing correct programs. A next chapter discusses the typed versus typeless languages debate.

A similar case with the previous example, is the most typical single error of novices in C:

```
if (a=0) {a=10};
```

This statement will **always** be executed; because it is actually an assignment and not a comparison, and after this command variable a will always have the value 10. Subtle[62] or not?

---

[61] It is heard that an unmanned spaceship to Venus was lost due to such software error.
[62] A combination of these two subtleties in this page, is an exercise for the reader, written in C: if (0<a<5) {a=10};

## Syntax errors probabilities, in Pascal programs

In 1978, Ripley and Druseikis examined the Pascal compiler of a CDC 6000 machine [RG78], collecting information regarding the nature and frequency of syntax errors in 589 programs. In 41% of the errors, only a single keyword or symbol of the language is missing.

Here are their findings:

| Number of errors | % | What is missing |
|---|---|---|
| 83 | 48.5 | ; |
| 18 | 10.5 | **END** |
| 13 | 7.6 | **BEGIN** |
| 8 | 4.7 | identifier |
| 7 | 4.1 | : |
| 7 | 4.1 | = |
| 6 | 3.5 | ) |
| 6 | 3.5 | # |
| 5 | 2.9 | ( |
| 5 | 2.9 | **DO** |
| <4 | 7.7 | other |

What is observable is that errors relating to the ";" terminator in Pascal are contributing a 14%. Another source of errors is the **BEGIN**, **END** and **DO** identifiers, which do contribute another 6%.

At those times, the programmers that wrote these programs were most probably academic students, researchers or professors. The equivalent error scores for novice low-aged programmers would be much higher for these very kind of errors, since their existing experience with what programming language syntax is, is close to negligible.

Is it at all possible to avoid such errors?

If we assume that BEGIN, END and DO are not required, as well as the ; terminator and assuming that this pattern is independent of era and ignoring people' s background, this would account for a 20% reduction in human time spent debugging **any** PASCAL program's syntax **ever, at the least**. It would be a 25% improvement, for any first round of error correction[63].

## Changing the precedence of mathematical expressions

Further research by Gannon and Horning in 1975, over the languages TOPPS and TOPPS II, shows that calculating expressions in left-to-right order with operation priority is producing fewer errors than right-to-left with no priority at all[64].

The moral? Never ignore the human factor.

---

63 Python by using indentation, avoids such errors

64 The first case is what people know from a mathematics class, the second is what was probably convenient for building the original implementation of the TOPPS compiler.

## The dangling else problem

Another kind of error due to ambiguity is the *dangling else* problem [HE84, p. 84]: Assume the conditional statement:

**if** condition **then** S1 [ **else** S2 ]

The **else** component can be omitted. When nesting another **if** statement where S1 is, we have:

**if** condition1 **then if** condition2 **then** S1 **else** S2

which can be interpreted in one of two distinct ways, depending to which **if** the **else** refers to:

**if** condition1 **then** (**if** condition2 **then** S1) **else** S2
**if** condition1 **then** (**if** condition2 **then** S1 **else** S2)

It is interesting that someone may write a program that is understood in a specific way by a human, due to its indentation, but the computer will do something different depending on the subtle rules adopted by a language's syntax, that resolve the *dangling if* question[65].

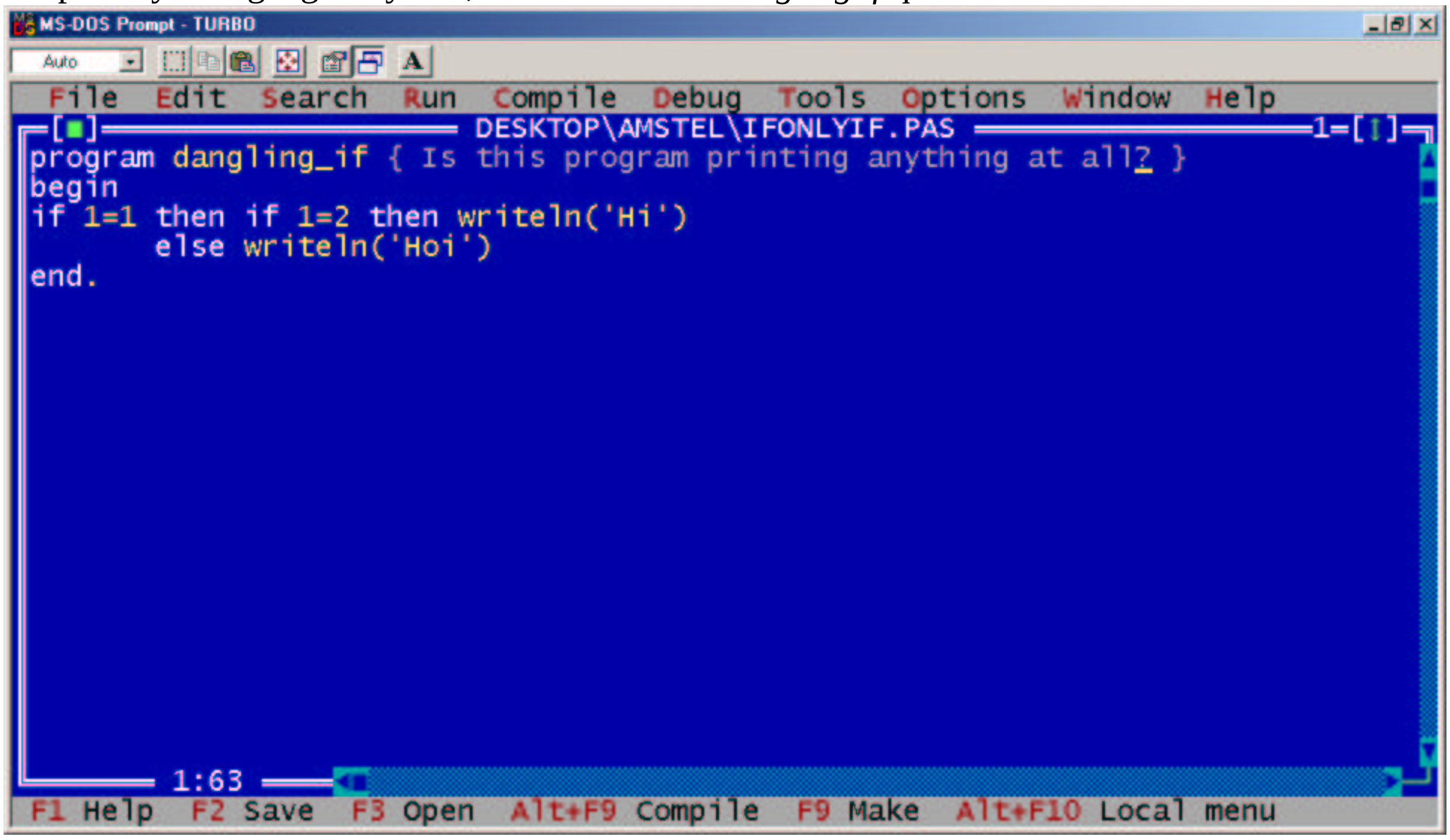

Python solves the dangling else problem with its indentation-implied syntax. According to the Python' s online documentation:

```
[...] thus there are no ambiguities (the `dangling else' problem
is solved in Python by requiring nested if statements to be indented).

[...] the following is illegal, mostly because it wouldn't be clear
to which if clause a following else clause would belong:

if test1: if test2: print x
```

[65] In Python this cannot happen, since indentation **is** defining the syntax of the language.

## *Typed versus typeless languages debate*

Imperative computer languages use the notion of *variables* in order to store the program's state at each moment. For example, when we store a vector for further manipulation, it is treated as a set of variables, as far as the computer is concerned. The programmer has to adjust to this concept.

Variables are a little different, than what we are used to in standard mathematics, in the sense that they are entities within a computer system's memory and not mere algebraic abstractions. For instance, a code that will do a matrix calculation, might use variables i, j, k and look like this:

```
(* matrix multiplication *)
var
    i,j,k: integer;
begin
for i:=1 to m do
  for j:=1 to n do
    for k:=1 to k do
      C[i,j]=C[i,j]+A[i,k]*B[k,j]
End
```

In the previous example, you may notice that the variables are *declared* in advance. Some languages do not require this and the **for** statement in an indirect *declaration*. So there exists a respective discrimination of languages: typed[66] and typeless.

The distinction of languages in *typed* and *typeless* divides programmers as well as those who study the programming techniques and researchers in Didactics of Informatics, into two opposite campuses, depending for which one, there are more arguments at hand:

- typed languages require the programmer to think and "design" the algorithm, well ahead
- typeless languages allow to leapfrog in providing a solution and circumvent formal planning

Two things to note:

- A lot of errors relate with the incorrect use of variables and erroneous passing of values among them. The typed languages are better at catching this kind of errors while writing the program.
- Typeless languages facilitate prototyping by "documenting" only the basic idea of the program, and not what its internals should work like. This is great for "evolutionary" programming.

The *variable typing* in languages appeared initially for purely technical reasons, relating to the construction of *compilers*. Typed languages promote a top-down strategy because you have to know what the algorithm is like, when declaring the variables. Harvey Brian comments [HB84]:

> *Some language designers have taken the position that variable typing is a good thing, apart from implementation issues, because it disciplines the programmer to use variables for only one purpose.*

A problem with typing is that it does not conform with the opportunistic way of developing programs, where the programmer does not know in advance all the details of the solution:

---

[66] Those are further split into two categories, strongly and weakly typed depending on what the compiler or interpreter knows about them at compile time, and usually are referenced as such: "Strongly typed", "Weakly typed".

> *According to studies in psychology of programming, even expert programmers are not able to conform themselves to the top-down strategy, but show "opportunistic" strategies which mix top-down and bottom-up components* [PP90, p.3].

So the chance is, that a programmer will come back and fill in the "missing variables" at a later stage. All in all, the idiosyncrasy of the language to declare the variables in advance does not mean that someone is going to make use of it while writing the program.

Teaching a typeless language to pupils is simpler: there is no need to explain what are the types of the language like a string and a Boolean; or that a number is either integer or floating point. The typing mechanism of languages is not an easy topic to explain. For the students this means frustration with the language until they thoroughly understand the internals of the typical mechanisms. For the teacher it means considerable time, because he may have to explain even to the weakest student in the class how the internals of the computer are working like.

## *Assignment using the symbol =*

It seems that use of equality symbol ' =' in assignments in many computer languages is unjust: it has led to confusion many generations of novice programmers. According to Harvey [HB84]:

> Since the beginning of time (in 1954), programming students have been getting confused about common programming statements such as X=X+1, a frequently used assignment construct that seems to go against one' s algebraic intuition.

Pascal tried to correct this with adopting an expression like "X:=X+1". It is seems that this has been equally problematic within Computer Science, with people coming from other languages. Also, many other language designers went the other way round, by redefining equality. In Python the equality comparison between two expressions is accomplished with the operator "==".

## *The errors in TEX*

Donald Knuth started in 1977 to work intensively on a software system for typesetting, which later resulted in the well-known TeX and LaTeX systems, widely used for scientific publications. What is particularly interesting with Knuth's work, is that he kept a very detailed journal of all errors and tried to organize them in categories according to their nature [DK84, pp. 243-291]:

| | | |
|---|---|---|
| Algorithm | Forgotten | Portability |
| Blunder | Generalization | Quality |
| Cleanup | Interaction | Robustness |
| Data | Language | Surprise |
| Efficiency | Mismatch | Typo |

We may note that the accumulation of errors over time in two distinct efforts in developing TeX, in 1978 and 1982, seem to evolve in a similar pattern [DK84, p. 272]. This maybe is an indication that errors are likely to happen in any project, regardless of preknowledge over the problem domain.

## *Why using an interactive language is important*

An interactive language is one that does not require a special preparation phase, called *compilation*, in order to be executed. The commands are supplied directly and executed without requiring from the user any extra knowledge rather than that of the language's vocabulary and syntax.

Harvey Brian has written a very detailed explanation on this topic [HB84]:

> *Whether or not a language is interactive has an effect on its efficiency. In brief, program development is generally faster with an interactive language, but already written programs generally run faster in a language that is not interactive. The difference has to do with the mechanism by which the computer understands your program.*
>
> *Every computer is built to understand one particular language. This machine language is different for each type of computer. Since machine-language instructions are represented as numbers, they are not easy for people to read. For example, the number 23.147.265 might mean 'add the number in memory location number 147 to the number in memory location 265'. Programs written in a high-level language […] must be translated into machine language before the computer can carry them out. This translation is done another computer program that comes in one of two flavors: compiler or interpreter.*
>
> *A PASCAL compiler for example, takes a program written in PASCAL and translates (compiles) it into the machine language of whatever computer you are using. The translated program is permanently saved as machine language (probably as a file on your floppy disk). Thereafter, the machine-language program can be executed directly. The compiling process takes a long time. But once it's finished, running the compiled program is very fast because it need never be compiled again.*
>
> *A LOGO interpreter, on the other hand, does not create a permanent machine-language version of your program. Instead, each LOGO statement is translated and executed every time the statement is supposed to be executed. The interpreter does not produce a machine-language representation of your program but simply carries out the machine-language steps itself. If a LOGO statement is to be executed six times, it's translated six times. […]*
>
> *Interpreted language can be interactive. Suppose you want to find the value of 2+2 in PASCAL. First, you must use the text-editor part of your PASCAL compiler, which will translate the program into machine language. Finally, you run the compiled program and your computer types out 4. In an interpreted language like LOGO, you can simply type PRINT 2+2 to see the same result.*
>
> *The situation in which interaction is most important is program development. If you are writing a complicated program, it probably won't work right the first time you try it. You will have to try it, see what goes wrong, change the program and try again. […]*
>
> *The flexibility and ease of use of an interactive language is particularly valuable in an educational setting. For a student of programming, there often is no production phase – the program is of interest only as long as it doesn't work. When it does work the student goes on to the next problem. In that sort of environment, the speed advantage of the compiler never materializes. In a business environment, on the other hand, the actual production use of a program is likely to be more important, which makes a compiler more desirable.*
>
> *Some languages use mixed schemes. BASIC (normally an interpreted language) has compilers that allow the user to give up interaction for efficiency. Some LISP compilers can coexist with interpreters, so that some procedures can be compiled while others are being debugged interactively. Some versions of PASCAL are compiled into an intermediate language called P-CODE, which is then interpreted. FORTH uses a similar system of partial compilation, but the compiler is part of the run-time environment, so single statements can be compiled and run interactively.*

## *Hoare's criteria on computer languages*

Hoare is one of the established theoreticians of programming languages and inventor of the popular sorting algorithm QuickSort. He followed closely the birth, development and maturation of major computer languages and set criteria in deciding what is a good computer language design. A subset of his criteria list, which excludes the ones that are quite technical and less relevant for education like fast compilation, portability, efficient object code production etc, are [HE84] [PM90]:

### The need for a Formal Syntactical and Semantical Specification

This criterion describes the necessity for a formal definition of a computer language. The *syntactical specification* is achieved by using either BNF notation or syntactical diagrams. The use of such notations, in fact, facilitates also the design of languages, because it is possible to use certain tools, to automatically generate a program that can "understand" the syntax of programs written in that language. These tools are called parser generators and make simpler the next step, which is the semantical description of the language.

*Semantical specification* is much more complicated than the syntactical since it describes what the keywords will *mean* in this language. The proposed techniques for its definition are: Interpretive by Wegner, Axiomatic by Hoare & Wirth and Denotational by Scott & Strachey [HE84, p. 50]. The semantical description formalisms are relatively orthogonal and can be complementary applied for the definition of a computer language. What is most interesting of all, is that, up to now, using plain natural language is the most safe and common way of describing the expected behavior of statements. The reason for doing so is that a formal definition using the techniques described above is exhaustive and requires extensive review, which cannot be done by humans in reasonable time.

Because it is so difficult to formally describe the semantical aspects of the language, Hoare did not fail to mention "formal semantics helpful but also need well-written manual". So, documentation should be considered as an aim in language design and not as an afterthought. After all, it is humans that are going to use the programming language.

### The need for Security

No program that violates the definition of the language should be able to escape detection without explicit user-friendly notification.

This relates with the ability of the language's interpreter or compiler to:

- accept the syntactically correct programs, and only them, and produce predictable results
- any error messages should be comprehensible in terms of the source language program

No program should cause the computer to run wild [PM90, p.105].

The definition seems overly simplistic, but security is critical for novices in programming since they are hardly ever experienced with computer software at the level that they can discriminate a valid machine's reply from a bug or, interpret an "esoteric system message".

## The need for Simplicity

A definition of simplicity as given by Hoare is:

> *A small range of instructions with uniform format, each having a simple effect that can be described and understood independently of other instructions. Terseness or simplicity of syntax is a common goal, one supported by studies of notations.*

According to Horowitz, a computer language is by its nature, a complicated tool to design. The source of this complexity is that it has to satisfy two different needs [HE84, p.48]:

> *On the first hand, there is a need for notation that can be easily used by the programmer. He should be able to describe real problems with ease. Because it is impossible to hope that the notation will explicitly address all application needs, it has to be generalized. Moreover, the facilities set of the language has to be a short list; otherwise the language will be hard to learn*[67]*.*
>
> *On the other hand, the language should allow efficient programs. This means that from any possible use of its notation the language will be interpreted fast, or compiled to efficient machine code. These two arguments are contradicting and lead to problems while designing a language, thus turning the creation of a language an interesting challenge that only few managed to succeed in.*

The simplicity of a programming language's syntax can have a dramatic effect in the capability of programmers to use it properly, as well as in the reliability of programs.

## The need for Readability and Clarity of structure; use of indentation

According to Hoare, *the readability of programs is immeasurably more important than their writeability*. The argument is, simply, that programs are written for and read by people and programming languages should encourage clarity. Language designers should aspire to express the intended solution structure visibly. [PM90, p.106]

The readability of a program is greatly helped by proper *indentation*. In fact, all people involved in programming know that indentation is far more than a stylistic component. Over time, it became recognized that using indentation is a simple and practical means of denoting the code's structure, rather better than using special characters, called *separators,* such as ; : , . etc. Indentation is used in practice, but is not enforced by the majority of languages today, because they were designed with the separator technology. As early as in 1974, Donald Knuth noted once on the use of indentation:

> *We will perhaps eventually be writing only small modules, which are identified by name as they are used to build larger ones, so that devices like indentation, rather than delimiters, might become feasible for expressing local structure in the source language.*

Recent languages adopted indentation style with known examples being ABC, Python and Haskell.

---

[67] Which is related to the need for orthogonality, only a few characteristics that can be combined without side effects. C.A.R. Hoare also comments: "*A requirement for the success of a good language is the simplest possible design. Without simplicity, even its designer cannot predict the consequences of his decisions. Without simplicity, the compiler writer cannot achieve reliability, compactness, speed and efficiency. The one that benefits the most though, is the user of the language.[…] The real master is the one that fully understands his tools and the same applies for programmers.*"

## The need for Orthogonality

Orthogonality is the notion that all language components are mutually independent and that there should be no more than one way of expressing any action in the language. A small set of basic facilities may be combined using systematic rules, without arbitrary restrictions, and this may result in arbitrary complexity [PM90, p.104]. Orthogonality is related to the previous term of simplicity. Too many keywords (=commands) in a language is an indication of poor orthogonality.

## The need for Generality

The language should allow all concepts to be the combination of a few primitive ideas. One typical such example, is ability to use a single command to denote both procedure and functions, by considering the procedures as functions with no return value. This has a lot of advantages which range from technical ones, while building the language, to very important while using it or instructing it: less complexity is observed by the person who tries to "think" in this language.

## The need for Subsets organization

This criterion refers to the ability for incremental learning of the language. This is the concept, that it is not necessary to understand or refer to the total design of the language, when only a subset of it is needed to solve a particular problem. A language, which is good in this, for example, would allow you to easily calculate a mathematical expression, without requiring knowing everything about declarations, initializations, objects, libraries, etc.

The idea behind subsets is key in an educational setting, where the language should be as simple as possible during the first steps, yet be powerful enough for more complex constructs later on.

Such an example of a language effort, that very consciously tried to optimize on this aspect, has been the SP/k series, where k denotes an incremental subset of the language PL/1, used in the 60s.

# METHODOLOGY

This chapter covers the initial consideration of methodology and why particular methods and techniques were employed.

The research is conducted because it is claimed, during last few years, that Python constitutes an easy to learn or use language which is combined with exceptional technical design, incorporating several years of research in the fields of Computer Science and Didactics of Informatics.

According to Green [PP90], there are three different ways to study computer languages and programming:

- Languages as Notations
- Programming and Natural Languages
- Programming as Problem Solving

This work attempts to evaluate Python mainly for its notational aspects and as a problem-solving environment, as that could be of concern for a certain school level and for a given purpose, in Greece. It is not an attempt of studying problem solving per se, rather a "language as a tool" investigation. A tool not only meant for the programmer but also for the student and the teacher.

The idea to compare Python against other existing languages, like BASIC and PASCAL, has been tempting. Then again, the research would narrow down to context-dependent evaluation and would probably miss further important conclusions. So, this is a more generic enterprise trying to find the true applicability of the language, with an eye –or two- looking at the future.

The road ahead, lays long: this time not only it is required to teach to future citizens this new subject of programming, but also most current teachers may need training, as well. The work that has to be done should not be taken lightly: it is equivalent of trying to teach thinking in a new language, like what happens with the natural ones: new symbols, new concepts, new contexts. This time, the teachers are not necessarily at a very advanced position compared to the students.

*For the aforementioned reason, the teachers' opinions are the subject of this study.*

Moreover, the language is considered to be the very first one that students learn, for two reasons:

- It is well known from other research studies, that the first programming language is very influential in the way people think and that it can have an effect on learning more languages.
- A student that knows already another language may start translating in a "mechanical" way from the one that he knows to Python. This is probably a special case of the previous one.

The last few paragraphs explain also the direction for an evaluative character of inquiry, using non-performance measures such as open questionnaires or interviews, in order to measure as much as possible. The structure and content of these questionnaires is discussed later on. We rejected the idea of quantitative analysis in a field so new to both teachers and students while considering a scale that is appropriate for a few months' work.

The proposed schema of questionnaire based interview is a discussion which is led by the researcher. This allows enough freedom to collect information that cannot be foreseen by the researcher in advance and enough space for answer development. Being able to trigger a more detailed description of an answer can produce extra information on a particular direction of interest. This could help when interrelating with the other interviews' outcomes and may ensure more converging results and better justified conclusions.

Because Python is, still, a relatively new language and its environment is not tested widely in class settings, any experiment in it should rather be accompanied with observations. Observational techniques have recently become popular in Didactics of Informatics, due to their ability to capture the more complex aspects of programming [GD90, p. 95] and support findings of other techniques.

So, the primary research instruments are the questionnaire-based interviews to be answered by the teachers. It would also be beneficial to have the interviews taped, so that all information will be transcribed and, thus, make it possible to have it cross-checked at later stages. There is space for observational research, though. What will really happen depends upon teachers' requests and allowance. Consensus exists on allowing maximum participation of researcher in class.

## *Research instruments & development thereof*

The final questionnaire included 20 questions, with the most critical being open-ended. This is what was deemed appropriate in an interview-like approach of data collection. The original questionnaire was written such as to be available to work with it interactively, while taking an interview. The open-ended questions are intended to be rephrasable, at interview time. It is assumed that as stated such, the answers evolving can focus on a conclusion and can end up with information that is possible to be correlated later on.

The structure of the questionnaire, which is shown in appendix A, has as follows:

- **General information about the teacher**. Information like name, and age
- **Educational experience & background.** This includes data like major and minors, specialization courses, professional experience and opinion over school and students.
- **Before Python**. This section refers to what is the situation before Python enters the class
- **After Python**. This is the situation after having applied Python.

The first two sets of questions are meant to provide a ground for the rest of the questionnaire, as well as uniquely identify the participants; not only in name but also in qualities.

The last two sets of questions try to extract information on what is the situation at this moment and where Python can have a role in; how easy is that; what considerations should be taken care of.[68]

[68] I thank Dr. Panagiotis Politis for reviewing the questionnaire, from the point of view of an expert in the Didactics of Informatics field, and giving me practical advice on its content. Albeit brief, I consider his contribution to assign a higher weight in this research work.

## *Setup of the experiment*

The size of the sample was selected to be small, for practical reasons. Teaching a programming language which is not only first for the students, but also for the teachers is a major undertaking. The complexities involved are hard to perceive in advance and the technical ones are probably the easiest: what is the most hard is proposing to inject in the existing and intensively working curriculum a tool that almost no one has ever heard of. This is the reason that up to five teachers are expected to participate in this research.

## *Teaching materials*

The existing course [DAPE99] provides these materials for teaching:

- A textbook for the students; the main reference material
- A notebook for the students
- A textbook for the teacher
- A set of two disks with programs in PASCAL and BASIC

The course syllabus suggests the use of a computer language in the laboratory, without defining which one and specific materials: Because the formal course focuses on the pseudocode, no attempt is made to finger-point to certain programming language manuals. In fact, there are many manuals for BASIC and PASCAL, depending on the specific dialect discussed and it is hard to choose: wouldn't this imply that the textbook [DAPE99] is insufficient? Teachers are expected to be able to find the language's reference manual themselves, learn it, and be prepared for students' answers.

Having realized that **Python's Tutorial** is the only necessary manual needed for learning the language, a **Python CD** for the purpose of this research was created, which contained on 26th February 2002:

- The Python language; with interpreters for DOS, Windows & Linux operating systems[69]
- **Python's complete documentation**, as found on the website; please read below for more
- The then current version of the research proposal and the questionnaires in .rtf format
- A mirror of the project's website at http://amstel.science.uva.nl/~fotisg/python
- A mirror of Python's website at http://www.python.org
- Relevant papers found on the Internet in .pdf or .ps formats
- Example libraries for Graphical User Interfaces; anticipating future needs of the DAPE course

The CD was replicated about a dozen times and was distributed to the teachers and a few students that asked for it.

**Note that the Python CD contained material, which is freely available and could be copied with no legal restrictions, infinitely.** This would not be possible with, say, Pascal.

[69] The Greek Ministry of Education has a provision for two operating platforms, Windows and Linux

## Python's documentation

Python does not suffer from the problem of multiple dialects and its documentation is coming together with the interpreter from a unique place, its Internet site. Because of the language's development model, that anyone can contribute and enhance its development, this will probably remain the same in the future, too.

At **http://www.python.org/doc/** these documents and their explanations may be found:

- Tutorial (start here)
- Global Module Index (for quick access to all modules)
- Library Reference (keep this under your pillow)
- Macintosh Library Modules (this too, if you use a Macintosh)
- Language Reference (for language lawyers)
- Extending and Embedding (tutorial for C/C++ programmers)
- Documenting Python (information for documentation authors)

There are many more documents and references for teaching material, even whole books, from the website. The most important one is simply the **Python Tutorial,** which is available in many formats, including HTML, PostScript, PDF and LaTeX.

## Method of teaching

Students have to be prepared with the theoretical background well in advance, as this is part of learning the theory of the course, and will be invited to write programs in a computer laboratory.

The initial setting of this research, suggested two important axes:

- Learning by doing and
- Using the constructivist approach

Teaching won't be limited to a teacher centered approach only, but will put the students in the role of a programmer, in order to write a program that can be interpreted and executed by the computer. The importance for the students of actively *applying* the information given them is also referred in the dissertation work of P. Kamsteeg: "Teaching Problem Solving by Computer" [KPA94, p.33-34]

> *Educational strategies that use little direct expository instruction, but expect learners to acquire most of their knowledge from their actual activities and experiences, are known as 'learning by doing' approaches. The educational approach known as constructivism (Papert 1980; Harel & Papert, 1990; Perkins, 1986) is particularly ardent in its insistence on learner activity.*
>
> *[…] According to Papert (1980), this results in a better retention and a wider transfer of what is learned than other educational means can provide.*
>
> *[…] As applied to teaching problem solving, the aspect of learning by doing vs. by being told boils down to giving the learner practice problems to solve, versus providing him/her with worked examples to study.*

## Expected learning activity

Another topic that has been a focal point of interest, is constructivism. Constructivist learning is based on students' active participation in problem solving and engagement in critical thinking regarding the learning activity. They are "constructing" their own knowledge by testing ideas and approaches based on their prior knowledge and experience, applying these to a new situation, and integrating the new knowledge gained with pre-existing intellectual constructs.

> *The constructivist approach should be the norm followed by the teachers, but, then, it has to be verified that all teachers follow constructivist approach*[70].

This is agreeable, but alas, can only be fully confirmed once the courses are over. Making sure that constructivism is followed, would require very close observations over several weeks of courses: it requires examining teaching strategies and activities used with each particular group of teacher and students. The effort required to do this properly should not be underestimated. If it is measurable, can this really influence this research?

Constructivism compliance is not a priority parameter to be tested, because it does not directly relate with the applicability of this certain computer language in a certain case environment. It may serve though as a note for the researcher, while observing the teaching methods.

## Reliability & Validity

> *Reliability is the extent to which a test or procedure produces similar results under constant conditions on all occasions.* [BJ99, p. 103]

The research instrument in this work, has been the questionnaire. Albeit the questionnaire was used at radically different circumstances – taped versus hand-written answer to the questions-, there seems to be a comparable result in all three schools. There are very strong indications that the same will happen with other schools as well, that may try the exact same methodology:

- Preparing the students in pseudocode about the *sequence, iteration, conditional* constructs
- Nearly a monthly preparation of the teacher with the language Python
- Brief but intense application of the language in these constructs; the complexity of the problems has not been very high, like adding the numbers of an array, finding the maximum or minimum element etc.

> *Validity tells whether an item measures or describes what it is supposed to measure or describe.* [BJ99, p.104].

The original aim of this research has been to measure the applicability of Python within the Greek Lyceum. The proposed investigation suggested an investigation in the 3rd class of the higher secondary education. The fieldwork included one group from a second class. Albeit deviated from the proposed investigation, it still remained within the research subject title, which means that it did reply another part of the research question for which there was no initial provision.

---

[70] This advice was given by the expert in the field of Didactics of Informatics, Mrs. Kordaki, University of Patras.

## *Research schedule and progress*

This is the research schedule and what has been accomplished according to it:

| Envisioned progress | Finished by: |
|---|---|
| ✓ Finalize research proposal.<br>✓ Set questionnaire for teachers' language evaluation: The interest is to spot areas in which the language or the course itself didn't progress as expected and maybe difficulties arose, as well as collect the experiences of the teachers for the general applicability of Python in education. | JULY 2001<br>(The Netherlands) |
| ✓ Research on the Internet and identify similar or related projects. There is a special interest group and a mailing list named EDU-SIG (Python in Education). Subscribed and monitoring activity. *[EDUSIG]*<br>✓ Find what ICT and computer language topics are covered by the existing curriculum. The expectation is to find at least 1 computer language being instructed, most likely that will be one of BASIC, Pascal, Java or C. **Indeed, this is what happens in practice; Ministry of Education *[YPEPTH]* leaves space for choice of language by the teachers themselves. In practice it is either BASIC or PASCAL.**<br>✓ A few contacts, including teachers in schools, will be asked for interest to participate in the project. Pedagogical Institute in Athens will be asked for even more directions on who and how can help. The aim is to find up to five potential teachers willing to join. If there are teachers that have experience in a diverse computer language they will be preferred, in order to do a comparative study with least bias. | SEPTEMBER 2001<br>(The Netherlands) |
| ✓ Bibliographical research on teaching computer languages and building programming skills. This includes research published in journals, books as well as opinions of experts on the World Wide Web.<br>✓ Concepts' order in learnig a computer language is very important: Either write a course outline, or find a suitable book for Python. This has to be appropriate for the proposed teaching period. **(Python's tutorial can be used as course material and plenty of resources are on the Internet)**<br>✓ Identify the technical components to be used. Python is available both for Windows and UNIX operating systems. The development environment for each case should be investigated. **OK: IDLE is the appropriate working environment; it has been used and proved successful.***[IDLE]*<br>✓ Prepare a web site for easy information dissemination and documentation exchange; focus on sharing the project state, rather than having a colorful site. `http://amstel.science.uva.nl/~fotisg/python/`<br>✓ There is an EDU-SIG mailing list at Python's central site for items of public interest. Setup a local mailing list for the project (Well, the small number of participants, all speaking the same Indo-European dialect of Greek implies that no particular list is necessary at the moment.). | NOVEMBER 2001<br>(The Netherlands) |
| ✓ Write the answers of exercises that will be worked by the students and use as examples other languages' code from [DAPE]'s two floppies se<br>Or simply just prepare the particular exercises in the relevant chapters. | DECEMBER 2001<br>(Greece) |
| ✓ Communicate with the teachers and try to tune their efforts in learning Python.<br>✓ Teachers should arrive to an experienced level before training time and be able to solve the exercises with no difficulty.<br>✓ Setup experimental mini-Labs in prototype school(s); yes, this requires schools that have already equipment, but that should not be a problem because there are quite a few nowadays.<br>✓ The researcher will provide technical support. It will take some time to install Python in all the computers. It will require emails, phone calls and transport to make sure that the IDLE environment is installed. | JANUARY 2002 - FEBRUARY 2002 (Greece) |
| ✓ Let courses begin in parallel during month March, as much as teachers' schedule permits. Hard to predict what will be needed by that very moment but teachers' feedback will be the driving force. | MARCH 2002 (Greece) |
| ✓ Concentrate experience: Ask teachers for information through questionnaire-based interviews.<br>✓ Ask students' comments from their experience; limited observations | APRIL 2002 (Greece) |
| ✓ Analyze data and summarize. | MAY 2002 (The Netherlands) |
| ✓ Write thesis and integrate latest knowledge, possibly also from other researchers in the field.<br>✓ Move the thesis from a draft text to a finalized document.<br>✓ Present thesis. | JUNE 2002 (The Netherlands) |

# EMPIRICAL RESEARCH

This chapter reports the evidence collection process and how research was carried out in the field[71].

The research took place in 3 schools, all located in Athens, prefecture of Attica. The teachers that participated[72] in the research team are:

- **Y**iannis Voyiagis
- **N**ikos Vagenas
- **K**ostas Skentos

The selection of schools in the capital city introduces a bias that can be considered representative of a little more than the third of the current population in Greece[73]. The schools are distributed within the city of Athens with no particular geographical correlation and their respective names are:

- Varvakios Scholi (Y school)
- Evanggeliki Scholi, also known as 3rd Lyceum of Nea Smirni (N school)
- Third Lyceum of Argiroupolis (K school)

All three schools have in common that they were Lycea and the course *Development of Applications in a Programming Environment* is instructed as part of the standard curriculum in the third class of the higher secondary education institutions. Hereby, the teachers and respective schools are referred with letters Y, N, K for brevity reasons. [OBS] stands for observation data.

## *Realized research settings*

What was the primary correlation with all schools, is the component of the textbook [DAPE99] that they have been applying Python in. The language instruction related with the chapters 6 to 9, which include mainly the concepts of *sequence, iteration, conditionals* and *variables, arrays, constants.* This is a subset of the standard curriculum practice in the 3rd class of Lyceum.

The range of problems that the students handled did not require elaborate problem solving, rather an application of their existing knowledge with pseudocode, as it would happen for other languages like PASCAL and BASIC. Students were introduced to the environment incrementally, first by learning mathematical expressions, then the three imperative constructs and combining them.

As one could expect, there has been convergence as well as deviations when comparisons are made with the original research design. In more detail, this is what happened in schools.

---

[71] I thank professors Dimitris Kontogiannis, Georgios Papadopoulos and Adam Aggelis of the Paedagogical Institute, for their support in this effort and their sustained contributions in making this happen.
[72] Who I thank dearly for the interest shown and the prompt answer to the initial invitation at a moment that was critical for the evolution of this research. Our combined work has results far greater than our individual achievements.
[73] According to the most recent results of demographics as announced on April 2002, the total population in Greece, as of 18th March 2001, was counted at 10.946.080, with a number of 3.761.810 (34.37%) living in the capital and its suburbs. Another 1.060.000 was found to live in the second biggest city and its suburbs named Thessaloniki.

## School Y

In February, a brief experiment occurred in school Y: the researcher presented Python to three volunteering students[74] in order to record reactions. The three students were very enthusiastic about the language and came up with a lot of questions. Some of the questions were very elaborate and technical and it was immensely clear that this level of proficiency probably does not relate to the average school reality, or the generic nature of this research. So, the first finding from the "field" was: *Make sure you are aware of the students who have programming pre-knowledge.*

The school Y has followed the theoretical preparation plan with the pseudocode during months October to March and applied the Python language within a period of 4 weeks within March and April. The technological direction of the 3rd Lyceum class of this school included 15 students.

Observation in this school, revealed first an important difference between the languages Pascal and Python, in the keyword **range**() which is used instead of **for**, when implementing iteration blocks. This difference can obstruct the transition from pseudocode to a program, since it requires an understanding of the internals of the range() command in order to write proper code.

The teacher Y adhered to a handwritten questionnaire, which was found to be more "natural"; after having tried a voice recorded interview; even a combination of a researcher-written interview and a computer-typed reply was not deemed appropriate.

## School N

The school N lacked a complete laboratory to apply the language in, due to some ongoing work in it. It was possible, though, to make use of the facilities in its library, which included a few PCs and a computer with a projector. It was clear that the latter was the best way to go, in terms of class management. As a result, the second class of Lyceum was chosen to participate in the introduction of the Python language, since the 3rd class should focus on the upcoming examinations. In order to bring the set of students up to the level of being able to learn programming, teacher N prepared the students with pseudocode examples from the textbook [DAPE] for a period of 5 weeks starting from late February. Then the Python language was demonstrated to the pupils and several examples were discussed in a period of 3 weeks. The language features were presented on a screen. Students were asked to solve small problems in it, after each educational unit was instructed. It is surprising how quickly students were able to write correct programs, even at their first attempt.

Teacher N, also, adhered to a handwritten questionnaire and follow up questions.

## School K

School K was probably the closest to the model school for this research, since the envisioned methodology was adhered to with great detail and only the timeframe had to be moved forward in time. The teaching period was followed by a planned long school excursion and this, in the opinion of the researcher, made the students look and behave free from the weight of examinations. The interview with the teacher afterwards was recorded on a magnetic tape.

---

[74] Who belonged also to the 3rd class, but in a different group than the one with which the teacher used Python.

## *Data from observation*

### School Y

This school, was the first one to reveal a difference between the languages Python and Pascal and, hence, the pseudocode. The difference lies in the commands for implementing *iteration*, what is commonly known as looping:

In Pascal, students would write

```
    for i:=1 to 10 do
        <statement-block>
```

while in Python this is stated as

```
    for i in range(1, 11):
        <statement-block>
```

There are a few things that can confuse a student in this case, which happened also at school N. These are discussed later on, in a special chapter, because this is a common programming construct.

### School N

The observational evidence adds, for this school, two important notes:

- First, a student that had no previous programming experience and was not even "actively" participating in the class, wrote a working program[75], which run the very first time, at once, with no help from anyone. She was only chosen to receive instruction on how to do it and she had never before programmed in any computer language or environment.
- Second, the language easily attracted the interest of the non-technological direction students who were also present. At one point that the course was over, one of them came to the researcher to ask how to finish a program she was working on, with quite clueful questions; the researcher was surprised to find out that this was a "theoretical direction" student, aiming to study the law. What is typical with this kind of students at this period of school, is that they are very focused on their field and usually have reduced interest and attention to other courses, the ones of exact sciences in particular.

These both were pleasant surprises and someone may note that in both observations the subjects were girls. This can be a dim signal for whomever wants to investigate paths for shortening the gap between boys and girls in science education in general.

### School K

The most important thing observed with school K is that everything went "as designed" but maybe not as expected after the experience with the first two schools: There were no problems, even with the for…range() construct; only a month's delay in starting the courses, which happened in April. At that time, the students in school K had nearly finished their intensive preparation for the examinations, had little pressure in coming courses and only the upcoming vacations to think of. It is the researcher' s opinion that this played a role in their easy-going relationship with learning the language and experimenting and explains why they felt familiar with the language quite early.

[75] One of finding the maximal element from an array; after initializing it with a few numbers.

## *Data from the questionnaires*

This chapter attempts to synthesise a "bigger picture" from the qualitative data. The way this is attempted is by identifying key concepts within the teachers' replies, interrelating with each others' comments, noting them down and referencing for easy cross-checking.

Identification of key concepts is somewhat subjective but the references to the original answers can serve to validate personal opinions; they should be considered as anchor points of the statements.

The references to the questionnaire materials –which can be found in appendix B- is done with indexes like: [K1]. K1 stands for teacher K, answer in question 1.

### Teachers' background

The background of the teachers is as follows:
- They are all aged 33-34 years old. [K1] [N1] [Y1]
- Their formal training in teaching spans both possible options: One teacher, K, is a graduate in Physics with further postgraduate training in Computer Science at the discipline of ATM Networks. K had a formal training in teaching during his studies, composed of courses *Didactical Methods* and *Education I* and *II*. [K3] Two of the teachers have graduated in Computer Science and followed with educational specialisation in SELETE, a special institute for potential teachers with no formal teaching preparation. Depending on background, this training lasts either 6 or 12 months. [N3] [Y3]
- Their teaching experience ranges from 6 up to 10 years, in general, and at least 4 years in the field of Computer Science. One teacher, N, has taught exclusively Computer Science for 9 years. [K4] [N4] [Y4] [K5] [N5] [Y5]
- All teachers are instructing Computer Science in all three classes of Lyceum. [K8] [N8] [Y8]
- All teachers have had professional experience in programming, which means that they have written programs as part of a job that they held in the past. They all have working knowledge of Pascal and C as a minimum; they were found to have some experience with LOGO, FORTRAN and Visual Basic as well, depending on the teacher. [K11] [K13] [N13] [Y13]
- Before this research, they knew Python, at best, only by name. [K14] [N14] [Y14]

The teachers have many similarities in their Computer Science Education experience. For this reason, the information they provide in the questionnaire answers will be treated as a unit and it is considered that the conclusions derived will be representative for teachers with such background.

### Teachers' expectations from a computer language

Teachers' expectations from a computer language are:
- Easy to learn the grammar and syntax [K10] [N10] [Y10]
- Easy error recognition, comprehensible by the students [N10] [Y10]
- Efficiency, ability to rapid program development [N10] [Y10]

It is worthwhile to note that the answers of teachers N and Y correlate in content, only their exact phrasing is different.

## Language in use up to now

Currently, all teachers use Pascal [K11] [N11] [Y11] for the following reasons:

- Its commands resemble words of natural language, which allows students explain "intuitively" the meaning of a program without knowing a formal description of the language. This is contrary to what happens in C, which would be hard to teach [K10] [K11]
- The school textbook promotes the use of pseudocode, which has a Pascal-like[76] syntax and keywords [K11] [K12] [N12]
- The school textbook examples are either in Pascal or BASIC [K11] [K12] [DAPE99]. The two accompanying floppy disks of the schoolbook also have the programs in Turbo PASCAL and Quick BASIC. [DAPE99]
- All teachers have previous experience in Pascal [N12] [Y12] [K7]
- Documentation [Y12], friendly environment [Y12] and language tools [N12]
- Pascal places the student at a position to write good code directly without having too many things to consider [K12].

The only trouble faced is the required initial declarations [K12] [Y13]
The conclusion made out of all this, is that a language that is going to replace Pascal has to be equivalent or better in all these arguments and, at least, ease the problem of variable declarations.

## Teaching materials

The materials, on which the teachers based their teaching on, included the standard school textbook of the course *Development of Applications in a Programming Environment*, which formed the basis for the instructed concepts. [K16] [N16] [Y16]. Teachers took advantage of the Python Tutorial for learning Python themselves [N16] [Y17], but did not use any other of the available materials, like presentations referenced from the language's website, for application in the classroom. This is the standard practice followed up to now with Pascal, as well.

## Concepts applied

The concepts taught covered the three major constructs of imperative programming [K17] [N17] [Y17]:

- *Sequence* (of commands)
- *Iteration* (commands **for**…**in range()** )
- *Conditionals* (**if** <cond>:, **then:** <sequence1>, **else:** <sequence2>)

*Variables* and *arrays* were covered also[77] [K17] [Y17], next to the principal aforementioned constructs.

What was not taught, is what was not examined at the end of the year: *lists*, *stacks*, *queues* and everything else from then on [K17].

[76] The English reader could think of it as a Pascal translation in the Greek language. Pascal is an ALGOL-like language, so the pseudocode is, then, an ALGOL-like language as well.
[77] In the questionnaires, variables are not mentioned at all and teacher N does not explicitly mention arrays. In any case, both were taught at all three schools. Variables and arrays are very fundamental in imperative programming [K17] and are hard to miss: even the simplest program needs them to store its data somewhere.

# ANALYSIS AND DISCUSSION

## *The effect of the examinations*

Currently, the examinations drive the content of the course. The official examinations examine *pseudocode*, which is an artificial language, especially made for the needs of this course in Lyceum with keywords in Greek.

Pseudocode is commonly used in academia and elsewhere to tune novices' minds towards an imperative model of problem solving, but it is common that it only serves for "bootstrapping" into programming. The proposed pseudocode is, indeed, a well-thought proposal for the needs of Greek students in approaching programming techniques and there are reasons to keep it being so. [K20]

On the case of Lyceum, pseudocode is not used only as an introductory vehicle in learning a programming language, which is usual in programming courses, but rather as the major language to learn and for this reason it is called *pseudolanguage* [K16]. Because pseudocode is the examined language at the end of the year, all students' and teachers' attention is oriented towards it, which distracts them from using a real programming language. [K20]

Python was the language vehicle for demonstrating real programming in the laboratory, so its application was only a slice of the formal curriculum. Teachers liked Python but were skeptical; realizing that Python cannot be a priority to teach, unless this condition is raised. [K20]

## *Method of teaching*

According to my observations, all teachers employed a method of "teaching by example": they usually demonstrated a program or command in pseudocode, then in Python, then asked for a problem solution on a slightly different but related case or topic. This seemed to be the case both for the laboratory, where Python coding took place, as well as in the classroom's blackboard.[OBS]

## *Differences between formal and actual curricula*

The differences between formal and actual curricula are sharply defined by the examinations:

- The formal curriculum supports the teaching of development of applications in a programming environment. A custom-made DAPE programming environment would require a great amount of time and effort; a fair alternative is to use an existing programming language [OBS]. But,
- Actual curriculum is the pseudocode; this is examined at the end of the year. The teachers who will try to teach real programming have a practical constraint of selecting an existing, suitable programming language. This trims down to the choices of PASCAL and BASIC because these are included in the textbook's examples and disks. PASCAL is promoted by the pseudocode and BASIC is usually regarded as inadequate due to the GOTO promoting and unstructured nature, which is probably an artefact of its history, so the list trims down to PASCAL. Even then, there is the practical problem that the examiner should be able to read the language used by a student, [K20] that favours focusing around pseudocode instead of a true programming language[78].

---

[78] Examining in pseudocode ensures fairness for all involved, with or without a preknowledge of a computer language.

## Advantages in applying Python

Here are the positive comments that Python received from the teachers:

- code is smaller [N18] [K18], less chance for errors [Y18] [N18]
- easier to write, faster development [N18]
- resembles a lot the pseudocode, pupils can write code immediately [K18]
- does not need initial declarations, which helps greatly to get to the point [K18]
- students often write code that runs correctly "at once", the first time [K18]
- attracts students attention, students are enthusiastic about it [K19]
- suitable for use in education [N20] [Y20], a great tool [K20]

Most positive criticism was expected in advance, and has been mentioned during the setting of the research. An exception to this is the initial declarations parameter. In fact, Python is less demanding than, say Pascal in this aspect and allows rapid application development with the least friction with a language's typing system.

## Disadvantages in applying Python

The major drawbacks found by the teachers while applying the language are listed here:

- New syntax which, albeit similar, still is different from the pseudocode [N18] [Y18] [N20]. On the other hand, students can pick it up easily [K20]
- Time needed to explain the internals of the **for**...**in range()** construct [Y17] [Y18]. The third teacher, K, had no problems while applying this construct; students adopted it at once.
- Reality check: students are now learning pseudocode and spending nearly all their time on that, because this is what they are examined on [N19] [K20].

The greatest weight of all is in the last observation. That the students are currently learning pseudocode and not a true programming language, because this is what they are examined, should not pass unnoticed. In order for Python to be applicable as curriculum component in the last class of the Greek Lyceum, the course has to change shape and focus to other things than what the examinations demand [K20].

The first disadvantage is somewhat expected, because the pseudocode is very much Pascal-like. It resembles that language so much, that any other language, which is not Pascal or a derivative of it, would generate such a comment.

The second disadvantage has to do with the difference between Python and, actually, many or most other languages in implementing the **for** command. It is probably more powerful and more generic, but as long as people are not used it, it will be necessary to address to this topic specifically in teaching.

## *Encountered problems*

Using Python, all problem-solving cases were an improvement, with a notable exception, the iteration statements.
The way of implementing the *iteration* construct is just different in Python.

Students would write in pseudocode or Pascal an expression like
**for** i:=1 **to** 10 **do**
<block of statements>
and rightly expect that the variable **i** will take values 1 to 10.

However, to achieve the same effect in Python, students must write:
**for** i **in range(**1, 11**):**
<block of statements>

The differences, are twofold, syntactical and semantical:

- Syntactical: Pascal has the construct **for** <assignment>:=<firstitem> **to|downto** <lastitem> **do**, where Python would use **for** <variable> **in** <set>, and <set> is given by **range**(<start>,<stop>).
- Semantical: The last item in the keyword **range**() is the <stop> operand, hence it is never "executed" by **for**. This seems to have surprised students [N17] [Y17] and needed time for explanation by teachers why it worked this particular way. In fact, this surprised the researcher and the teachers as well, the very first time of using the language. Is there any research or other formal foundation that supports this language design decision? Is there a practical benefit?

There are advantages in the syntactical design, though. Because of its syntactical definition, Pascal can only do iteration enumerating a linear set of numbers up or down, a single step at a time, and only that. Python allows an arbitrary *browsing of a set instead of being a linear scale* [Y17]. There are practical and pedagogical advantages to this, explained after the examples in Python and Pascal:

Pascal

```
Var
      A:array[1..4] of integer;
      Max, I: integer;
Begin
A[1]:=3;
A[2]:=7;
A[3]:=1;
A[4]:=4;
Max:=0;
For I:=1 to 4 do
      If Max<A[I] then
            Max:=A[I];
Writeln(Max);
End.
# By the way: I forgot 5 ";"s
# when first wrote this, despite the fact
# I am mostly experienced in using PASCAL!
```

Python

```
A=[3,7,1,4]
Max=0
for I in A:
      if Max<I:
            Max=I
print Max

# It ran correctly the first time.
```

Python’s syntax is more expressive and can allow a variable to “run” in a set without imposing as a need to iterate a special *index* variable for accessing an array' s elements. In Pascal, it is unavoidable to use a number based addressing scheme, when browsing the contents of a matrix. If you use the referenced value a lot of times in the code, you need one more variable, which also has to be added in the declarations. There is just no other way to write it. *From either a Computer Scientist's point of view or an Educator’s, Pascal is closer to the machine's needs rather than the programmer's.*

Python’s syntax is also more powerful, because it allows to refine the step of counting up or down. In Pascal the step is always implied to be a unit in any case; anything more complex requires either calling a function, using an array or doing extra calculations originating from the linear scale. In Pascal, the problem can become acute if the value of the enumerating variable is used multiple times within the loop. In Python it is possible to pass as parameters arbitrary sets of lists ( with data like floating point numbers), functions, tree structures, etc.

## *How to avoid the problem with for...range() in the future*

There is some ongoing discussion on how to overcome Python’s peculiarity with the <stop> point by adding an ad-hoc function named *series()* with the more predictable behavior. This can be automatically loaded in the Python environment during start-up of the language:

```
# This is the function series(<first>,<last> [,<step>])
# It returns a set, overcoming Python’s peculiarity with the range() keyword
#
# Copyleft by Fotis Georgatos, AMSTEL Institute, UvA, June 2002
#
# EXAMPLE OF USE
# >>> series(10,-10,-2)
# [10, 8, 6, 4, 2, 0, -2, -4, -6, -8, -10]
#
def series(first,last,step=1):
      if last>=first:
        return [x for x in range(first,last+step,step) if x<=last]
      else:
        return [x for x in range(first,last+step,step) if x>=last]
# For better performance and generality, use xrange instead of range.
# See Reference Manual for details on this
```

On the next page, you may observe how the function looks within the specially modified environment of Python; the only true difference is this additional function and the window title!

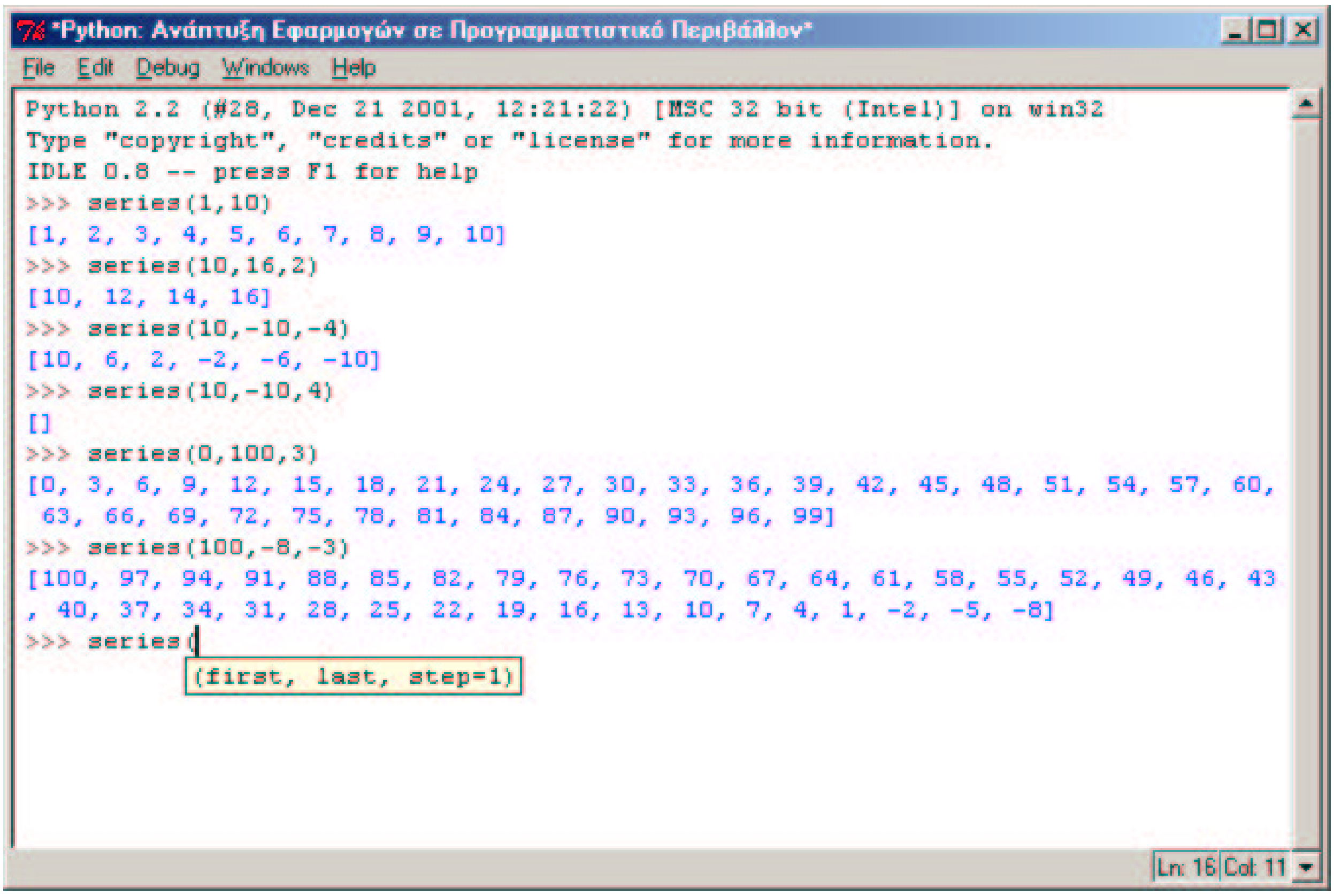

This picture is a tribute to the extensibility of the language, which may prove again useful for other educators, in the future. According to Harvey this characteristic should not be overlooked [HB84]:

> An extensible language is one in which user-defined procedures ' look like' primitive procedures. [...] Extensible languages are particularly valuable for teaching because a teacher can provide language extensions and teach them as if they were primitives.

Python's open development model and the available documentation supply good facilities for injecting in the language such special extensions, without having to reverse engineer the language or reengineer anything from scratch. Moreover, Python language has a standardized protocol with PEPs (Python Enhancement Proposals), that allows the mature outcomes of extension proposals like this, when they are deemed to be considerable improvement, to become part of the official version of the language. It remains to be seen though if this is really worthwhile to become a standard component or, it should rather be an ad-hoc function. Usually, the policy for such ideas is rather conservative for obvious reasons.

## *Recommendations on the DAPE course*

To begin with, the book has exceptionally rich content, so much, that most teachers will only be able to "consume" a subset of it in the class and leave the rest for the more interested students.

Among other qualities, it is a pleasant surprise to find detailed reference on algorithmic complexity in chapter five, and even complete chapters on Object Oriented Programming, Event-driven programming, Graphical Environments and User Interfaces chapters. They are way above the expectations for an introductory book in Programming, which had no prior one to follow.

### The pseudocode

The pseudocode is not runable by a computer system and this can be demonstrated, by the nature of errors found in the book: In page 86 there is a recursive program for binary search. The results of the called subfunctions are not passed to a variable, and are not stored anywhere. It is impossible to claim it is syntactically incorrect, since there is no formal grammar definition of the pseudocode.

Someone may consider writing a translator, interpreter or compiler for the pseudolanguage, since the language's keywords are specified semantically in pages 148-180. This would be substantial work and each technical path has its own problems to consider:

- a translator to another language (say Pascal, because this is the easiest) would be useful for programs that are correct, but not in the case of errors: The error messages will be often unintelligible, because they will only refer to the translated code (say, in Pascal for example). Such, a technique would violate the Hoare's principle of Security.
- interpreter or compiler: it is an effort of implementing another language environment. The complexity involved should not be underestimated and the final result is probably not going to be necessarily useful for a long time, since the content of the course may change at any moment. It is probably easier to generate a prototype for such a parser with the help of a functional language, like Haskell. Haskell has a special tool for such endeavours, named Happy, which is a parser-generator expecting input definitions in BNF.

### Existing languages in use: PASCAL & BASIC

There are a few items, which cannot be taught easily –if at all- with BASIC or PASCAL, while Python can make a real difference in this course:

- Tree data structures: Python can manage nested lists, so it' s possible to represent trees in them.
- Functions returning lists (tables). This relates with the flexibility to manage arbitrary data types. For instance, at page 91, there is a program for converting an amount in Euros, to the minimum possible number of coins. The algorithm seems to be an alternative of converting an amount to respective coins and notes. The proposed program returns only the quantity of coins but not their exact composition. It is believed that this happens because Pascal, where the pseudocode is based upon, cannot handle arbitrary arrays returned from functions. At least, the formal standard of Pascal by Wirth, does not allow this to happen!
- Object Oriented Programming. Teaching object orientation through any derivatives of Pascal, BASIC or C will be either a minor tool or have to use a proprietary, commercial environment.

# CONCLUSIONS AND RECOMMENDATIONS

This research attempted to deploy Python as a language for the needs of the 3rd class of the Greek Lyceum and, in particular, for the needs of the course:

"Development of Applications in a Programming Environment".

## Applicability conclusions

The deployment of the language was as extensive as what is the norm with the other languages. In any case, it was diverse and long enough for the teachers to deduce the first significant conclusions:

- Python is an easy language to learn with readily comprehensible syntax.
- It results in small length of code, which means small chance of errors.
- There are differences with Pascal, as well as the Pseudocode currently applied.
- Attention has to be given to the usage of some keywords, like the for…range() construct.
- It is noticeable improvement, in comparison with the existing languages in use.
- Currently, pseudocode is the first priority at school; this condition can prevent applying Python.

## Researcher's view

The comparison of Python with the other alternatives, gives the impression of a language that can well serve the students once they depart from the programming class for higher education training. From the technological standpoint, it lacks little compared to most other high level languages. Python is not a toy language, it concentrates considerable features and is a tool for serious work.

> *The long-term usefulness of a language comes from how well and how unobtrusively it supports the day-to-day work of programming, which consists not of writing new programs, but mostly reading and modifying existing ones* [RE00].

The latter observation is of particular importance to teachers, since a major part of their work is to study the programs of students and not to write code. The indentation structure of the language can greatly help teachers in their daily work. Evaluation of code written on paper is also simpler.

It is claimed that the reliability and validity of this research is analogous to the number of schools that have participated. It's relatability though, is much wider and covers areas and conditions well different than those originally proposed.

It is not confirming a conclusion for country-level decisions, however, it is formulating a strong message for any Computer Science Education decision-makers, as well as Scientists of other fields that are in need of a computer language:

*Python is an interesting language to look at and keep an eye on.*

## What comes next

The next step is to apply the language in a larger scale, in other classes, and perform further analysis of students programs. It would be great to do an experiment demonstrating examination scenarios. This could also be attempted as a comparative study with an existing language, as Pascal.

It is advisable to use the experience obtained in this research, in particular the newly proposed way of writing iteration constructs in Python with the function *series()*.

The research team that conducted this project will attempt to present the findings in consequent conferences and make the effort known. It is even considered to propose a slight modification of the Python language to include this extra function called series() as an extra component.

## As a final note

While considering computer languages as scientific literacy, let's remind of Knuth repeatedly quoting a statement, which was originally said by George Forsynthe [KD86, p. 9, 245]:

> *The most valuable acquisitions in a scientific or technical education are the general-purpose mental tools, which remain serviceable for a lifetime. I rate natural language and mathematics as the most important of these tools, and computer science as a third.*
> *[…] The learning of mathematics and computer science together has pedagogical advantages, for the basic concepts of each reinforce the learning of the other."*

It is unavoidable to endeavour in Computer Science, without learning a programming language.

*The language in which we express our ideas has a strong influence on our thought process*

Donald Knuth, 1974

# LIST OF REFERENCES

# WEB RESOURCES

# APPENDIX A – The questionnaires in Greek & English

# Interview questions

Fotis Georgatos
26-2-2001
UvA, Amstel Institute

Dear Research Participant,

I am a researcher in Mathematics and Science Education and Master Program in the Amstel Institute of University of Amsterdam, with a special interest in Computer Science Education.

The purpose of this interview is to gather feedback on the matters that have to do with applying a computer language in education, in particular Python in the higher secondary level (Lyceum).

The questionnaire-driven interview consists of two parts, one to be filled in before and one after teaching the language in a laboratory setting.

Name:____________________________
E-mail ___________________________
MALE O FEMALE O

Interview Dates: (Before)________
(After)________

**GENERAL INFORMATION ABOUT THE TEACHER**

1. Age. Which year did you finish high-school (Lyceum)? __________

2. What was your involvement with the educational system in a single sentence?
   What was the motivation to arrive on the chair of the "other" side of the desk?
______________________________________________________________________
______________________________________________________________________
______________________________________________________________________
______________________________________________________________________

## EDUCATIONAL EXPERIENCE & BACKGROUND

3. What was the major subject of your university diploma/degree?
________________________________________________________________________________
Please, mention any specialization courses (minor) relating to Computer Science Education:
________________________________________________________________________________
________________________________________________________________________________

4. Total years of education experience as a teacher: ______

5. Years of education experience teaching CS courses: ______

6. Which other courses are you teaching or you taught in the past?
____________________________________________________
____________________________________________________
____________________________________________________
____________________________________________________

7. What is your experience in programming? Have you ever been a professional programmer?
____________________________________________________
____________________________________________________
____________________________________________________
____________________________________________________

8. Please provide your current Teaching Levels in Computer Science (CS) related courses, along with the population of classes in each case.

Gymnasium A ____________
Gymnasium B ____________
Gymnasium C ____________
Lyceum A ____________
Lyceum B ____________
Lyceum C ____________

9. How would you describe your students and your school?
____________________________________________________
____________________________________________________
____________________________________________________
____________________________________________________

**TEACHING A COMPUTER LANGUAGE - BEFORE PYTHON COURSES**

10. Please, list your expectations from a computer language to be used for teaching children. What are or should be its characteristics?

____________________________________________________________
____________________________________________________________
____________________________________________________________
____________________________________________________________

11. Which is/are the language/s you are using currently, in CS courses?

____________________________________________________________
____________________________________________________________
____________________________________________________________
____________________________________________________________

12. Why did you choose the language you are using now?
Is that the experience of the teacher (you), of students or available tools? Etc.

____________________________________________________________
____________________________________________________________
____________________________________________________________
____________________________________________________________

13. Are there many language choices designed specifically with education in mind?
What is your opinion or experience with them?

____________________________________________________________
____________________________________________________________
____________________________________________________________
____________________________________________________________

14. Did you know anything about the computer language Python?
If yes, how have you been aware of it?
(Please ignore in this answer your current involvement in this project)

____________________________________________________________
____________________________________________________________
____________________________________________________________
____________________________________________________________

And a last question about the computing facilities at your school…

15. Is the equipment been adequate to teach a computer language in your school?
How many computers per student exist?

____________________________________________________________
____________________________________________________________
____________________________________________________________
____________________________________________________________

## TEACHING A COMPUTER LANGUAGE - AFTER PYTHON COURSES

16. What teaching materials did you depend on, for instructing with Python?
You may include course book as well as the educational technology.

____________________________________________________________
____________________________________________________________
____________________________________________________________
____________________________________________________________

As far as Python is concerned…

17. Which topics/concepts do you think would need more emphasis while teaching?
Would you rather avoid some topics? If yes, please explain what and why.

____________________________________________________________
____________________________________________________________
____________________________________________________________
____________________________________________________________

18. Comparing Python with a previous language(s), what are the pros and cons?

____________________________________________________________
____________________________________________________________
____________________________________________________________
____________________________________________________________

19. Did you have practical difficulties in using the language Python in the laboratory?

____________________________________________________________
____________________________________________________________
____________________________________________________________
____________________________________________________________

20. How applicable is Python as first computer language for teaching programming in higher secondary education? What is your overall opinion as an educator? Would you recommend it?

____________________________________________________________
____________________________________________________________
____________________________________________________________
____________________________________________________________

**Thank you for taking the time to participate in this research and interview!**

Yours Sincerely,
Fotis Georgatos

# Ερωτήσεις Συνέντευξης

Φώτης Γεωργάτος
26-2-2001
UvA, Amstel Institute

Αγαπητέ Συνεργάτη Ερευνητή,

Είμαι ένας ερευνητής στην Εκπαίδευση Θετικών Επιστημών (Mathematics and Science Education) που είναι πρόγραμμα επιπέδου Masters από το Amstel Institute στο Πανεπιστήμιο του Amsterdam, με ιδιαίτερο ενδιαφέρον στην Εκπαίδευση της Επιστήμης Η/Υ, γενικά της Πληροφορικής.

Ο σκοπός αυτής της συνέντευξης είναι να συγκεντρώσει την εμπειρία σε θέματα που έχουν να κάνουν με την εφαρμογή μιας γλώσσας προγραμματισμού στην εκπαίδευση, ειδικότερα της Python στην δευτεροβάθμια εκπαίδευση (Λύκειο).

Η από ερωτηματολόγιο οδηγούμενη συνέντευξη αποτελείται από δύο μέρη, ένα που πρέπει να συμπληρωθεί πριν και ένα που πρέπει να συμπληρωθεί μετά την διδασκαλία της γλώσσας στο Εργαστήριο Πληροφορικής.

Όνομα:____________________________
E-mail ____________________________
Άνδρας Ο Γυναίκα Ο

Ημερομηνίες Συνέντευξης: (Πριν)________
(Μετά)________

**ΓΕΝΙΚΕΣ ΠΛΗΡΟΦΟΡΙΕΣ ΓΙΑ ΤΟΝ ΚΑΘΗΓΗΤΗ**

1. Ηλικία. Πότε αποφοιτήσατε από το Λύκειο? __________

2. Τι οδήγησε στην ενασχόλησή σας με το εκπαιδευτικό σύστημα;
   Ποιος ήταν ο λόγος που σας έφερε στην "άλλη" μεριά της έδρας;

____________________________________________________________
____________________________________________________________
____________________________________________________________
____________________________________________________________

## ΕΚΠΑΙΔΕΥΤΙΚΗ ΕΜΠΕΙΡΙΑ ΚΑΙ ΥΠΟΒΑΘΡΟ

3. Ποιο ήταν το κύριο θέμα της διπλωματικής/ πτυχιακής σας εργασίας;
______________________________________________________________________________

Μήπως είχατε μαθήματα εξειδίκευσης που σχετίζονται με την Εκπαίδευση της Πληροφορικής;
______________________________________________________________________________
______________________________________________________________________________

4. Συνολικά χρόνια εκπαιδευτικής εμπειρίας ως καθηγητής: ______

5. Έτη εκπαιδευτικής εμπειρίας διδάσκοντας Πληροφορική: ______

6. Τι άλλα μαθήματα διδάσκετε ή διδάξατε στο παρελθόν?
__________________________________________________
__________________________________________________
__________________________________________________
__________________________________________________

7. Ποια είναι η εμπειρία σας στον προγραμματισμό; Υπήρξατε επαγγελματίας προγραμματιστής;
__________________________________________________
__________________________________________________
__________________________________________________
__________________________________________________

8. Παρακαλώ, καταγράψτε τα επίπεδα των τάξεων στις οποίες διδάσκετε πληροφορική, καθώς και τον πληθυσμό των σχετικών τάξεων.

Γυμνάσιο Α ____________
Γυμνάσιο Β ____________
Γυμνάσιο Γ ____________
Λύκειο Α ____________
Λύκειο Β ____________
Λύκειο C ____________

9. Πώς θα περιγράφατε τους μαθητές σας και το σχολείο σας;
__________________________________________________
__________________________________________________
__________________________________________________
__________________________________________________

**ΔΙΔΑΣΚΟΝΤΑΣ ΜΙΑ ΓΛΩΣΣΑ ΠΡΟΓΡΑΜΜΑΤΙΣΜΟΥ – ΠΡΙΝ ΤΗΝ PYTHON**

10. Παρακαλώ, καταγράψτε τις προσδοκίες σας από μια γλώσσα προγραμματισμού που πρόκειται να χρησιμοποιηθεί για την διδασκαλία νέων. Ποια θα έπρεπε να είναι τα χαρακτηριστικά της;

____________________________________________________________
____________________________________________________________
____________________________________________________________
____________________________________________________________

11. Ποια /ποιές είναι η / οι γλώσσα / -ες που χρησιμοποιείται επί του παρόντος στην Πληροφορική;

____________________________________________________________
____________________________________________________________
____________________________________________________________
____________________________________________________________

12. Γιατί επιλέξατε την γλώσσα αυτή;
Είναι μήπως λόγω εμπειρίας του καθηγητή (εσάς), των μαθητών, ή των διαθέσιμων εργαλείων; κλπ

____________________________________________________________
____________________________________________________________
____________________________________________________________
____________________________________________________________

13. Υπάρχουν αρκετές γλώσσες σχεδιασμένες ειδικά με την εκπαίδευση κατά νου;
Ποια είναι η γνώμη ή εμπειρία σας με αυτές;

____________________________________________________________
____________________________________________________________
____________________________________________________________
____________________________________________________________

14. Ξέρετε τίποτα για την γλώσσα προγραμματισμού Python;
Εάν ναι, από πού είσαστε ενήμεροι;
(Παρακαλώ αγνοείστε την τρέχουσά σας εμπλοκή σε αυτήν την εργασία)

____________________________________________________________
____________________________________________________________
____________________________________________________________
____________________________________________________________

Και μια ακόμα ερώτηση σχετικά με τις υπολογιστικές εγκαταστάσεις στο σχολείο σας:

15. Είναι ο εξοπλισμός κατάλληλος για να διδάξετε μια γλώσσα προγραμματισμού στο σχολείο σας; Πόσοι υπολογιστές υπάρχουν ανά μαθητή;

____________________________________________________________
____________________________________________________________
____________________________________________________________
____________________________________________________________

**ΔΙΔΑΣΚΟΝΤΑΣ ΜΙΑ ΓΛΩΣΣΑ ΠΡΟΓΡΑΜΜΑΤΙΣΜΟΥ – ΜΕΤΑ ΤΗΝ PYTHON**

16. Πάνω σε τι διδακτικά εργαλεία βασιστήκατε, για την διδασκαλία της Python;
Μπορείτε να συμπεριλάβετε τα σχολικά εγχειρίδια ή την εκπαιδευτική τεχνολογία.
________________________________________________________________
________________________________________________________________
________________________________________________________________
________________________________________________________________

Όσον αφορά την Python τώρα…

17. Ποια θέματα / έννοιες νομίζετε ότι θα χρειάζονταν περισσότερη έμφαση κατά την διδασκαλία; Μήπως θα αποφεύγατε κάποια θέματα; Εάν ναι, παρακαλώ εξηγείστε τι και γιατί;
________________________________________________________________
________________________________________________________________
________________________________________________________________
________________________________________________________________

18. Συγκρίνοντας την Python με προηγούμενες γλώσσες, ποια είναι τα υπέρ και τα κατά;
________________________________________________________________
________________________________________________________________
________________________________________________________________
________________________________________________________________

19. Είχατε πρακτικές δυσκολίες στο να χρησιμοποιήσετε την Python στο εργαστήριο;
________________________________________________________________
________________________________________________________________
________________________________________________________________
________________________________________________________________

20. Πόσο κατάλληλη είναι η Python σαν πρώτη γλώσσα για την διδασκαλία προγραμματισμού στην ανώτερη δευτεροβάθμια εκπαίδευση, δηλαδή το Λύκειο; Ποια είναι η γενική σας γνώμη ως εκπαιδευτή; Θα την συνιστούσατε;
________________________________________________________________
________________________________________________________________
________________________________________________________________
________________________________________________________________

Ευχαριστώ για τον χρόνο που αφιερώσατε σε αυτήν την έρευνα και συνέντευξη!

Με εγκάρδιους χαιρετισμούς,
Φώτης Γεωργάτος

# APPENDIX B – The answered questionnaires & interview transcripts

# Interview based Questionnaire

Fotis Georgatos
20-5-2002
UvA, Amstel Institute

Name: *Skentos Kostas*
E-mail: *kskentos@otenet.gr*
MALE *X* FEMALE O

Interview Date: 1/5/2002

[…start of tape, with the missing words…] overcame the technical problems, we are now ready to conduct an interview. The subject of the interview is to see if Python was, indeed, an interesting language in its application in Lyceum, possibly, also in other levels of education. What has been done so far is some teaching in a class and we, together, are going now to fill in a questionnaire. The questionnaire is comprised from the following parts: in the beginning we fill in general information, then information that relates with the situation before Python is taugh, what happens until today, and in the last part is the experience once Python has been taugh. Let's start, so.

I am called Fotis Georgatos; let's say this to have it recorded, and…

*Skentos Kostas*

Today is 1st of May 2002.
[Q1] Let me ask your age, when did you finish Lyceum?

*1987*

[Q2] What led in your involvement with the educational system? How did you find yourself located at the "other" side of the desk?

*It has always triggered my interest, it has to do with children, which makes it even more interesting, and has a lot of free time, which I like.*

## EDUCATIONAL EXPERIENCE & BACKGROUND

[Q3] So, what was the major subject of your university diploma work?

*It was about ATM networks and their future application in daily life, what they could offer to us. It was in the sector of Electronic Automation, University of Athens, Department of Informatics.*

Malista. Therefore you have finished Informatics?

*First degree was in Physics and I have a postgraduate degree in Informatics.*

So, how much time was each one of them?

*Four and two.*

Did you have any specialization courses relating to Computer Science Education?

*I had specialization courses that relate to Education in general, not necessarily just Informatics, these were "Didactical Methods" and "Education I", "Education II". They were optional courses; anyone wanting to be involved with education has been choosing these courses.*

These, I assume, were during the first degree?

*Yes.*

[Q5] So, soon after, there comes a moment that you are a teacher. Until now, how many years you have been a teacher in total?

*I have been a teacher of Informatics for 4 years.*

Very well. This is about teaching experience exclusively in Informatics. Assuming you have worked with other courses as well, what are they? Can you give us a picture?

*In Lyceum during these 4 years, I have also taugh "Communication Technology" in 2nd Class of Lyceum. As a teacher, I have also worked in IEK[79] teaching "Electronics" courses and I have taugh, in individual courses, Physics.*

For each one of them, how much time did it…

*IEK was for about a year, "Communication Technology" for about 4 years.*
*Physics has been for 6-7 years, in the form of hourly individual courses.*

...While you were a student?

*Yes, while I was a student, and later, until I was assigned the teaching position.*

[79] This stands for Vocational Education, in Greece

[Q7] Also, what is your experience in programming? Programming, except from an educational subject, can also be means for production, let's say, in a job. Have you ever been a professional programmer? Have you worked in a company writing code?

*I have worked in a company and the language I have used mostly, both as a student as well as later professionally, is Pascal. Regarding how much code... not too much like many thousands of lines of code, maybe less than five thousand lines.*

But it is a number of the order of a few thousands; right?

*Yes.*

Excluding Pascal, have you also used other languages?

*I have used C, and LOGO.*

Have you worked with these professionally? or in education?

*It's mostly in education, which is when I was student and later in projects we had to do.*

By the way, your experience with C and LOGO, what is that?

*The one language is very powerful, as we have all seen, ...*

We are talking about C now?

*Yes, C.*
*LOGO is a more specialized language; it is for special applications and for this reason it has not interested me as much as other languages.*

Well, LOGO is often promoted as a good language for education, as an educational tool.

*I don't think so.*

I have not formed a final positive or negative idea over this matter. Do you have an opinion to offer?

*First, its applications are so specialised that I would not recommend it as an educational tool.*
*At least not me...*

I can possibly agree that I share this mentality: Since LOGO has not been used in practical applications, maybe its educational value is limited. It may be teaching logical thinking in a self-consistent way, which is good for an educational tool, but educational tools have also to be a useful skill for a student and LOGO shows not to be such. Is this what we are talking about?

*Yes, exactly.*

Let's proceed with the next question...
[Q8] Please, provide a report of the classes, in which you teach Programming, along with the population of these classes...what are they…

*Well, I teach Informatics in the 1st class of Lyceum to two groups of 22,23 people each; in the 2nd class to four groups of 20-22 people in each; and in the 3rd class of Lyceum we teach Programming to two groups of "direction" students with 18 students each.*

I will ask you to repeat the last part…You said that in the 3rd Lyceum there are two groups of?

*17-18 people for Programming, "netto".*

You mean, in the technological direction?

*Yes.*

The total population of the classes, is respectively…

*How many groups are per class?*

I am just asking the number of students per class…

*Oh, well... In the first class there are about 72 people, in the 2nd of Lyceum are about 85, which means that nearly everyone attends Informatics' course, and in the 3rd of Lyceum are also nearly 85 people.*

The 3rd of Lyceum, as we said, is two groups with 17 and 18 students, which follow the Technological direction, respectively, isn't it?

*Correct.*

Which means that in one of the two groups, the Python course occurred. Which one was that? Was it called Gamma1 or Gamma2?

*Gamma2. GammaTaf2. Gamma Technologikh 2.*

Nice. Could we make a comparison and say something relating to Gamma1?

*Yes, GammaTaf1 was not selected, first, because its level is lower and, second, the pupils show less interest for the courses. We could have difficulty, due to the limited available time, in being efficient.*

So, it was a practical choice given the schedule limitations?

*Exactly.*

OK, let's not forget that this was the 3rd of Lyceum, in a period that certain things have to be done!

[Q9] A description of the students and the school?

*The school is an open, free school, which maybe needs improvement in organisational matters, but it is a quite good school to work in, because you don't have too many "directives".*

You had mentioned in the past the word self-governed. Do you want to comment on this?

*Yes. Self-organised and self-governed. Because we don't have a permanent director, every teacher is free to organise his course and his behaviour at school as he thinks it is the best. The same applies for the students.*

So, this is something that you find positive, right?

*Of course, both for children as well as teachers, because everybody works as he thinks is best.*

A more personal question: Do you think that other schools have something to "envy" and "copy" from this way of functioning?

*...Maybe yes. But it is so tight the connection in Lyceum with the examinations, that it's like a double-sided knife: if you let students and teachers too free, they possibly not be efficient later in the examinations, and may say something like... we don't study. don't do...that... you get the idea*

Good. But you do prefer it as a working environment…

*As a working environment it is ideal.*

## TEACHING A COMPUTER LANGUAGE - BEFORE PYTHON COURSES

Let's proceed with the next group of questions, which refer to what happens until today, covering the period that someone does not know Python, so I kindly request to answer similarly to these questions.

[Q10] What are the expectations for a computer language that is going to be used for teaching young people? Mostly for the Lyceum, but also for lower ages… what should be its characteristics?

*What we could say is what we are all saying… "to be easy to learn it"; well, this can be a little imprecise. The commands have to resemble the words that are in use, like Pascal, for example, that children have been learning in the past.*

So, we may comment over an approach of the "artificial" language close to the natural language…?

*Precisely…*
*…the opposite of what happens with C; which is unacceptable; we could never teach it to children, they would not understand what is the meaning in C.*

Please comment this about C a little further, maybe you have an example here…

*Yes, the commands it uses makes children often not to understand what they correspond to, in contrast to Pascal, that they have been learning up to now, where they know that **write** is the well-known γράφω (Engl. write) or **read** will ask something from us, or **print** will print something, which they understand in this way. A language that will use commands, more or less, further away from the natural language we are using, will be a difficult language.*

Now by the way, I recall from C++, and this applies for Java as well, that someone has to type "**cout** <<" in order to print something on the screen; this is exactly what you are mentioning now, isn't it?

*Yes, exactly…*

...it does not resemble natural language…

*…anyone… he does not learn it.*

[Q11] Which are the languages you are using at this moment, in CS courses? Maybe you have also used others in the past…

*The language I am using currently in Informatics is Pascal, both because in the textbook there is the Pascal example just next to the pseudocode, as well as, its commands resemble a lot the commands of the pseudocode that we use in the book. Formerly, we had also used LOGO mainly for younger children and children at Gymnasium (primary high school).*

What ages were they?
Which classes of the Gymnasium?

*2nd and 3rd of Gymnasium which means about 12-13 years old.*

What did you think of applying LOGO with these children?

*They were not very impressed... initially they were happy that they could draw on a computer but after a while the enthusiasm faded away, because it required a lot of commands that they could already do themselves with a drawing program, simply.*

So, could we summarize this like that "LOGO is initially attractive thanks to its graphical environment, but the opportunities are exhausted quickly…"

*Particularly for young children that have no great programming abilities, the things they can do are finishing quickly and their enthusiasm fades and its over.*

Let's see the next topic…

[Q12] You said that until now you are using Pascal. What is the reason that Pascal is the language you are using? Maybe you have a certain environment that you are suggesting to the children. That would be also good to hear about.

*Mostly Pascal, because the way the book is written, actually its commands, generally, the pseudocode and the language, the pseudolanguage that the book is using, is structured according to Pascal.*

It is quite obvious that this happens…

*Yes, so, this helps to tell them, "Look a language that you can use to write the respective program is Pascal and here is its code. That's it".*

Are there other reasons that promote the use of Pascal in such a case? As we said already, the pseudocode is Pascal-like, resembles Pascal like and the syntax of pseudocode is practically the same as Pascal. Are there other reasons that make Pascal a good language for education?

*Maybe its structure... generally...I could say that from the languages I have seen up to now, Pascal has been the one placing the student in a position to write good code directly without having too many things to consider. Of course, you have to do the declarations initially, but this something that all languages are doing.*

Declarations… so?

*So, variables, constants, arrays... I had to think what I am going to use later on and declare it in advance. This very thing, for the children it has been difficult to get used to this, and this has been the problem we faced so far in teaching Pascal.*

So, declaration of variables, that has to be done first thing in the code, is something that creates small troubles because students have to have thought in advance what they are going to use later in the code… *[Nodding positively, not written on tape ;-) ]*
So, this is a so-called "typed" language. Did you ever use non-strongly-typed languages?
*[pause]*

That they don't have to declare variables in advance? BASIC, for example, does not need declarations in the code and also Python as you may have seen…

*Python, yes. But..can we talk about Python, already…?*

Ehm, well… just discuss this a little!

*This is what I found as an advantage of Python, so far. It is not necessary to think in advance: "I have declared this, I have not declared this, constant, variable, integer etc… No. And this is an advantage.*

Did you ever write code in BASIC?

*A long time ago…*

Well, recalling from BASIC, it is not a typed language, you can work on the variables directly; although has been a problem in practice and for this reason later versions of the language encouraged to declare variables…

*Yes.*

I had something else to have a comment on…or a question:
Has it always been that you could declare variables in advance then go ahead, write the code and then everything magically worked or, did you have to come back and changes things?

*Many times. And for this reason many times when I wrote the code, either on paper or on the computer, I leave a few empty spaces in the beginning, because I know that I will always have to fill in in declarations a few things.*

This happens very often, right?

*Yes, this is what I say to the pupils as well: "First write the code, then look what you have used and carry on in declarations". Because always you will have to come back up there and add new things.*

So, actually the way the language is designed, enforces you to do at least two movements: first, to declare some initial variables and write the program, then correct this in a second step and bring the extra variables in declarations…

*Yes.*

I find this interesting… let's proceed though…

[Q13] There are language designed with education in mind. What is your opinion or experience with them? One that we already mentioned is LOGO…

*The disadvantage of LOGO that I observed is that it did not attract the interest of the children for a long period. Maybe it was not only the language, but also the supplied manuals… in any case, it was consumed very quickly.*

PASCAL is also an educational language, for which we have already discussed…

*PASCAL, yes, we already mentioned its limitations as well.*

Excluding LOGO and PASCAL, there can be other educational languages. Do you wish to add something to this list and comment on? There has been discussion over PROLOG…

*PROLOG, I find no reason to use it as a language in education. It maybe be following a logic reasoning and do this in a consistent way, but its applications are so specialised that make it me, according to me, a non-school language.*

A non-school language? Well, because it is not a skill for the students later on?

*Yes.*

The next question is…
[Q14] Do you know anything about the computer language Python? If yes, where from? Please ignore your involvement in this project, consider that this is asked a few months ago, maybe even before you knew me at all…

*Python, I didn't know about it. Maybe only once, as a reference on the Internet, for a programmer that said he was using Python; I didn't what kind of language it was… anything at all.*

[Q15] Is the equipment of your school adequate to teach a computer language?

*Yes. The school laboratory is very well equipped at this moment, comparing to what is the situation today; particularly for teaching programming there can be one student for computer, which is very good.*

What we saw in class recently, was that every student had its own computer, because the Technological Direction is a subset of the whole class and therefore it is possible for each student to have his own machine.

*Yes.*

I also saw that you have some equipment for networking… a router…

*Through the router, all the computers are connected with each other and they also have connection to the Internet. The information, that pupils have access on, is really quite much.*

Nice.

[noise while changing pages and looking notes]

I trust a lot this tape recorder… I think it is not necessary to have a break to check for it and we can continue…

*It's OK, I'm looking at the [VU meter] needle moving continuously…*

## TEACHING A COMPUTER LANGUAGE - AFTER PYTHON COURSES

This is about an evaluation of what has happened and what conclusion, possibly, exists.

[Q16] What teaching materials were used for instructing Python? Maybe books, or materials or something…

*The book on which we relied for teaching is the school textbook, which teaches programming in a pseudolanguage.*

So, the pseudolanguage, the pseudocode is the didactic tool, which can be found in the book for Informatics "Development of Applications in a Programming Environment".

17. Which topics/concepts do you think would need more emphasis while teaching? From what has been done so far, would you emphasize on something? Also, would you rather avoid a topic? Please answer to this and explain…

*I would avoid topics which although are included in the book are not examined in the end, because students would not have learned these, so learning such topics that they have not studied earlier from a book would be more difficult, and possibly even confuse them.*

*Furthermore, if I was to emphasise on something… do you mean what part of Python?*

Not only about this language. More interesting could be which topics from programming languages in general. There things which are examined at the end of the year and must have been taught. What is this study material? What are the subjects?

*The material is the structures: sequential, iteration and conditionals and arrays. What is not examined is lists, stacks and queues and everything else from then on.*

Variables? I assume it is so fundamental that…

*Oh yes, this is out of discussion: variables, declarations and so are coming before everything else. I mean, when we are talking about a programming language we must know these three structures of sequence, condition and iteration, in order to able to write a program based on those structures.*

Do you recall from the Python courses what has been taught? I mean, what is your impression of the students now that they have used Python while trying to learn these structures? Did they understand them?

*Yes. They have understood them and are able with what they know to write a program, while before that they were learning only pseudocode they were wondering: "OK, what we learn looks like a program, but can it be applied in a computer?". Now they see that what they know does help and using what they already know can write a program on the computer.*

Did you have the same result with Pascal?

*On the long term, yes; but they were having a hard time with the language types. It was difficult for them that they had to begin in a certain way, declare variables, constants, arrays… what kind of array…does it have integers or…*

Reals?

*Yes, reals.*

They are not exactly "reals", as we often call them in Informatics, and it is more correct to say "floating point" numbers…

*In the book, it is called as integers and reals…*

That's how it is taught at the school?

*Yes, the translation in Greek is just like this… "akeraioi kai pragmatikoi". So, it is enforcing to use the same terminology… you get the idea. You may know that it is not like this, BUT, when you go at school and say this to him, he will reply "mister, where is this said over here?" [hands as if showing a book]*

I get the idea… there are also the exams at the end of the year, they will write "floating point" and the examiner will come and say …
*…No…*
…it is wrong and so on. It requires some attention, indeed.

[Q18] Comparing Python with previous languages, and in this case Pascal, when do you benefit and when do you lose something? What are the pros and cons?

*So, with Pascal… its code is smaller, for sure, it is easier to write; it resembles a lot the pseudocode that pupils learn and they can write code immediately; it does not need the initial declarations which helps greatly to get to the point; so many will write the code at once that they have already thought over and have written in pseudocode. Yes.. that's it. Add anything else?*

Maybe something that was a problem and made it more difficult? This is also interesting.

*No.*

Well, I think there is a point, which has been discussed by the other teachers: Moreover, the way that the **for** command is implemented, has a parameter **range()**, which when is supplied the arguments 1 to 10 will not include the last step because the last step is the final and, as a result, this requires being familiar to it; the first time it is a little weird: if I say for example for i in range(1,10) it will takes values from 1 up to 9. This surprises initially, but I think it is just one of the things you encounter when using a new language, where few things are somewhat different and you have to…

*…Well, you say this once and eventually it becomes trivial for them.*

Actually, in this very school, of yours, I observed that this was not even discussed too much. Students saw it but no one considered it as difficult or weird. They saw it was different and worked with it; just proceeded writing programs considering this difference.

*Yes.*

My experience with Python, too, has been such: I thought that it is strange, why is it working so? But it is just a new language, which is simply a little differently defined, and then you learn it…

*…In fact.*

[Q19] Any practical difficulties encountered while using the language Python in the laboratory for Informatics, the room that the computers are located?

*Hm?…er…no… I think.*

I also believe that everything went optimally in this school case.

*In fact, I didn't recognise something that made the situation more difficult.*

Maybe it is even more interesting this: After the sequence of lessons that I was also present, at some point I left and then you had further discussion with the students using the language. Do you have anything to comment from that?

*Most students, and when I say the most I mean 70-80% of students, are positively placed for Python, and many are enthusiastic about it. The rest, which is the minority, believed that a programming language would give you the ability in a very short time to write programs like the ones they are already existing in their computers, which is something utopic.*

We had discussed about games, once…

*He, I think that they came in contact with reality…about a programming language. Learning a programming language does not mean that you will write in half an hour the program that took from somebody else half a year. I also think that it has been a negative influence that they are exposed to programs based on graphical user interfaces, buttons, such and so on, and they expected from a programming language something similar. They wanted to press a button on the screen and this was a kind of strange for them… Do I have to write code? Where's the button I am going to press now?*

That's true: the logic behind Python's design is that everything begins with code written as text of a line or a few, and everything that must be done graphically you have to do it through a library for it. Actually, it is not as hard as one might think initially to write graphical applications in Python, because there are tools, just like in Visual Basic, to built forms and connect later with Python code. Actually these are part of the CDROM, which we used for the installation of the language. Definitely, it's not as friendly as the Visual Basic environment that you can do drag'n'drop with the buttons.

*Even then, you still have to write code. In many cases, pupils are impatient and think that it is very easy to build applications. It is not like this. Actually, in order to write code, you must have prepared what you will build. Visual languages give you the illusion that you can build something very fast while it is not exactly like this.*

May impression over Visual Basic is also that: it will help you a lot to build easily an application with a few buttons, a form or two and these relate with each other in a simple manner. But when it comes to more specialized cases, then you cannot avoid learning well the environment and the language, and then you have to learn two things: the environment and the language.
Then, it is going to be more difficult in practice.

Last question…

[Q20] How applicable is Python as first computer language for teaching programming in higher secondary education? Let's speak about Greece, to be more precise and generalize later on.
What is your overall opinion as an educator? Do you recommend it to colleagues?

*I could recommend it with no reservation, for someone that has the time and needed laboratory hours to show to students a language and how they could write in a simple manner programs for their computer. This has as a prerequisite, that the course changes shape and its structure is not so much oriented towards the examinations. Until now, students are learning pseudocode, so that they can write such a program while being examined. If this condition were met, it would be an ideal language.*

So, what you are saying is that since up to now the school program defines that they have to go, at the end of the year, to write a pseudocode at the examinations, Python may deviate students from this aim, like any computer language could do, isn't it so?

*Yes, because it is needed at the same moment that the teacher that is going to correct the respective student examination sheets, to know the respective language. Since, at this moment teachers know only a few languages, which sums up to BASIC-PASCAL, having a student to write in a new language is probably undertaking a risk.*

I will put in a line BASIC, PASCAL and Python. Had you to choose one of these three, for this very case of the 3rd class of Lyceum, that students have examinations to attend at the end, which one would you choose?

*From what I have seen, Python, because it gives easier the same results as other languages.*

PASCAL though, gives you the advantage that it is closer to the pseudocode, isn't it so?

*On the other hand, we have to do with children, which can learn quite easily. They even have faster apprehension than we do, so they can easily understand a language and learn it much more quickly.*

I liked this! I will ask you to comment it. You said that children have faster apprehension than we do, didn't you?

*Sure! Why do children learn so much faster about computers than us? Because you tell them one, two, three…ten things and they will try to find out what you are doing. Teach a child about HTML. If you try to teach an adult, he will say "no way, what is this?".*

This is also what I believe, because children are in a younger age are in a better position to learn languages than adults are. I think this is an age that someone can learn a lot of languages quickly. And this applies both for natural as well as artificial languages.

Would you recommend to another teacher to have a look on this language named Python?

*Of course.*

You do believe it is something interesting?

*It is not only out of general interest, but why not learn something that will help?*
*…Since you are writing code anyway, - if you are not busy with programming, ok, whatever – but if you are programming why not learn this language that can ease your life a little?*

I don't know what other question to ask… we have exhausted the list of topics included in this questionnaire. Maybe… a generic picture… What have you seen after all this research, this procedure that took a little more than two months, counting from first discussion up to realisation; What remains after all this?

*There are things, which we may not have seen up to now, Python for example we didn't know what it was, which may prove to be useful at one point. The remaining question is: how could these be included in a programming course? Saying just that this is good? Yes, but..*
*It has to be included in the curriculum in order to help the student get busy with something creative, because the pseudocode, as it is currently, is inapplicable. You are writing a program that has no application in practice. This is what I think.*

So, the skill that the school tries to deliver to the student has no practical value. While with a tool like this, one can give knowledge with theoretical as well as practical value. If you were in the school in the students' shoes, and you can modify the Informatics' courses, you would include the programming courses, wouldn't you?

*Yes, but I would still keep it as optional course. Learn a programming language, once they have some knowledge with pseudocode, so that they know how to construct a program, then learn a programming language to apply my knowledge on it.*

So you wouldn't remove the pseudocode, right? You would leave it as is.

*Yes, I would leave it, because initially the children must be introduced smoothly in programming. Definitely not in the way that currently all is being done in pseudocode and afterwards it is applied in a language. Simply, let them read first the commands in Greek, see what they do, then tell them this is the real language, this corresponds to this, that corresponds to that and so on. This makes the transition smoother.*

So far, I have been hesitating about using the pseudocode…

*…Me as well…*

…Because in fact, you learn one extra language.

*For this reason, I would not keep it in the current form: I would use the pseudocode for an example or two, but not as much as: first teach in pseudocode, then in a real language, and do this all the time. Not like one-on-one. No. Mostly having the pseudocode as an introduction to a language. Just like they have been learning initially flow chart diagrams. Learn in one, two, three courses a few flow chart diagrams, here is the command, here is data input, here is printing, and once you understand how this is done with commands you forget about it. It introduces you to what is the flow of a program. This is what happens currently, albeit the book includes them until the end, we teach them only initially at the first few courses and then ignore them, that's it. It serves as an introduction only.*

If you were a student would you like to have Python in the curriculum?

*Yes… yes.*

There is some hesitation, or not?

*There is some hesitation, let me explain why: the students' program is already loaded enough, and every thing you add on it adds extra stress to them. What do I mean? Although you may like some courses, the fact alone that you are enforced to learn them, it gets difficult…*

I agree with this view –we are considering something else than Python now-, I also believe that when the school has a spirit of enforcement, then even trying to give the best skills to the students has the tension to fail. It is probably human nature this kind of behaviour…
*…Everyone in such conditions can act like this.*

So, Python could come in the school with another way, like an educational tool that students can work on with their own initiative and maybe teachers should do alike?

*Yes, maybe like this. By having students bringing their exercises done in Python, some small programs, and maybe at the optional courses to have more lessons in programming, unlike now that they work in Office or make WebPages in Frontpage etc.*

I think it has been a very successful interview – and the research that happened before -; I thank you for the time you have spent in all this; we will talk in the future. Maybe we are able to go together to the island of Rhodes[80] and present a result. We will see what comes next. Thank you very much.

*Me, too.*

---

[80] An educational conference will occur during late September 2002 in the island of Rhodes, which will likely be the first place to publicly present this research work, in Greece.

# Interview questions

Fotis Georgatos
18-5-2002
UvA, Amstel Institute

Dear Research Participant,

I am a researcher in Mathematics and Science Education and Master Program in the Amstel Institute of University of Amsterdam, with a special interest in Computer Science Education (CSE).

The purpose of this interview is to gather feedback on the matters that have to do with applying a computer language in education, in particular Python in the higher secondary level (Lyceum).

The questionnaire-driven interview consists of two parts, one to be filled in regarding the period before teaching the language in a computer laboratory setting and one for the period after.

Name: *Vagenas Nikos*
E-mail: *vagenas@di.uoa.gr*
MALE *X* FEMALE O

Interview Dates: (Reg. Before) *26/4/2002*
(Reg. After) *26/4/2002*

## GENERAL INFORMATION ABOUT THE TEACHER

1. Age. Which year did you finish high-school (Lyceum)? *1987*

2. What was your involvement with the educational system in a single sentence?
What was the motivation to arrive on the chair of the "other" side of the desk?

*Love for teaching and the possibility of learning through it. Possibility of being in contact with children, accompanied by a spirit to explore young people.*

## EDUCATIONAL EXPERIENCE & BACKGROUND

3. What was the major subject of your university diploma/degree?

*"Optical amplifiers doped with Ervio"*

Please, mention any specialization courses (minor) relating to Computer Science Education:

- *PATES / SELETE*

4. Total years of education experience as a teacher: *3 + 3 (in private courses)*

5. Years of education experience teaching CS courses: *6*

6. Which other courses are you teaching or you taught in the past?

*Except Informatics, another 5 years experience in teaching physics, chemistry & mathematics.*

7. What is your experience in programming? Have you ever been a professional programmer?

*Experience in writing programs in Pascal, C, Basic, Fortran.*

8. Please provide your current Teaching Levels in Computer Science (CS) related courses, along with the population of classes in each case.

Gymnasium A ___________
Gymnasium B ___________
Gymnasium C ___________
Lyceum A *60 (=30+30)*
Lyceum B *90 (=30+30+30)*
Lyceum C *90 (=30+30+30) (OBS: 15 students in the Technological Direction)*

9. How would you describe your students and your school?

*It is a public school, with probably average supply of equipment, specialised personnel with willingness to experiment. Students with willingness to explore.*

## TEACHING A COMPUTER LANGUAGE - BEFORE PYTHON COURSES

10. Please, list your expectations from a computer language to be used for teaching children. What are or should be its characteristics?

- *Easy in learning the grammar and syntax*
- *Ability of rapid creation of programs (simple as well as complex) with ease.*
- *Ability for easy error recognition and correction*

11. Which is/are the language/s you are using currently, in CS courses?

- *Pascal*
- *C*
- *Basic, Visual Basic*

12. Why did you choose the language you are using now?
Is that the experience of the teacher (you), of students or available tools? Etc.

- *Experience in teaching and excellent knowledge*
- *Language tools (debugging of logical and syntactical errors)*
- *Directly relevant with the respective book's approach*

13. Are there many language choices designed specifically with education in mind?
What is your opinion or experience with them?

- *LOGO – ( relevant, a toy language)*
- *BASIC – (a little relevant)*
- *PASCAL – (relevant)*

14. Did you know anything about the computer language Python?
If yes, how have you been aware of it?
(Please ignore in this answer your current involvement in this project)

*Having seen the name in bookcovers and on Internet. No knowledge about it.*

And a last question about the computing facilities at your school…

15. Is the equipment been adequate to teach a computer language in your school?
How many computers per student exist?

*There is still no dedicated computer laboratory (it will be ready for the following year 2002-2003). There is a small network of 5 computers, one being equipped with a projector. There is a ratio of 6 students per computer system.*

## TEACHING A COMPUTER LANGUAGE - AFTER PYTHON COURSES

16. What teaching materials did you depend on, for instructing with Python?
You may include course book as well as the educational technology.

- *Schoolbooks "Development of Applications in a Programming Environment" and "Computer Applications and Informatics" for basic knowledge in programming principles*
- *Python's tutorial which is included in the language's documentation. More precisely, I used these parts that relate with the basic programming principles.*

As far as Python is concerned…

17. Which topics/concepts do you think would need more emphasis while teaching?
Would you rather avoid some topics? If yes, please explain what and why.

- *Covered the basic programming structures (sequence, conditionals, iteration)*
- *The difference between of iteration command **for**, because it is approached with different way than Pascal*
- *The **range()** command*

18. Comparing Python with a previous language(s), what are the pros and cons?

*PROS: Simple in its syntax, therefore also in learning and program development. Smaller code, which means faster development. Easy keywords to remember, small chance of errors.*
*CONS: Difference with the existing approach used by the teaching books of "Development of Applications in a Programming Environment"*

19. Did you have practical difficulties in using the language Python in the laboratory?

*No practical problem. Easy installation in the network of computers.*

20. How applicable is Python as first computer language for teaching programming in higher secondary education? What is your overall opinion as an educator? Would you recommend it?

*Suitable language, because programs are developed in a fast and easy way. There is though non-direct correspondence between way of writing in pseudocode ("Development of Applications in a Programming Environment") and command syntax under Python.*

### Thank you for taking the time to participate in this research and interview!

Yours Sincerely,
Fotis Georgatos

# Interview questions

Fotis Georgatos
18-5-2002
UvA, Amstel Institute

Dear Research Participant,

I am a researcher in Mathematics and Science Education and Master Program in the Amstel Institute of University of Amsterdam, with a special interest in Computer Science Education (CSE).

The purpose of this interview is to gather feedback on the matters that have to do with applying a computer language in education, in particular Python in the higher secondary level (Lyceum).

The questionnaire-driven interview consists of two parts, one to be filled in regarding the period before teaching the language in a computer laboratory setting and one for the period after.

Name: *Voyiatzis Yiannis*
E-mail: *voyageri@otenet.gr*
MALE *X* FEMALE O

Interview Dates: (Reg. Before) *25/4/2002*
(Reg. After) *25/4/2002*

**GENERAL INFORMATION ABOUT THE TEACHER**

1. Age. Which year did you finish high-school (Lyceum)? *1986*

2. What was your involvement with the educational system in a single sentence?
   What was the motivation to arrive on the chair of the "other" side of the desk?

*It is a working environment near young people who have an intention to obtain knowledge.*

## EDUCATIONAL EXPERIENCE & BACKGROUND

3. What was the major subject of your university diploma/degree?

*The diploma subject was "Development of programs for text and string processing".*

Please, mention any specialization courses (minor) relating to Computer Science Education:

*After graduating in Informatics, I followed a postgraduate training in education, by SELETE.*

4. Total years of education experience as a teacher: *9*

5. Years of education experience teaching CS courses: *9*

6. Which other courses are you teaching or you taught in the past?

*Only Informatics*

7. What is your experience in programming? Have you ever been a professional programmer?

- *Co-operation with the software company Compusoft (12 months)*
- *Bank of Greece – IT department (18 months)*

8. Please provide your current Teaching Levels in Computer Science (CS) related courses, along with the population of classes in each case.

Gymnasium A ___________
Gymnasium B ___________
Gymnasium C ___________
Lyceum A *30+30*
Lyceum B *30+30*
Lyceum C *32 (15 students in the Technological Direction)*

9. How would you describe your students and your school?

*A high level regarding the average intelligence of students.*
*With high ambitions and possibilities.*

## TEACHING A COMPUTER LANGUAGE - BEFORE PYTHON COURSES

10. Please, list your expectations from a computer language to be used for teaching children. What are or should be its characteristics?

- *Easy to learn*
- *Efficient usage ( be able to produce results quickly)*
- *Easy error recognition, comprehensible by the students*

11. Which is/are the language/s you are using currently, in CS courses?

- *Pascal*
- *C*

12. Why did you choose the language you are using now?
Is that the experience of the teacher (you), of students or available tools? etc.

- *Experience*
- *Efficiency*
- *Available documentation*
- *Friendly environment*

13. Are there many language choices designed specifically with education in mind?
What is your opinion or experience with them?

- *LOGO – no experience*
- *PASCAL – it is being used widely; there are some problems initially, that relate with the need to declare variables and the syntax (var & begin-end)*

14. Did you know anything about the computer language Python?
If yes, how have you been aware of it?
(Please ignore in this answer your current involvement in this project)

*No*

And a last question about the computing facilities at your school…

15. Is the equipment been adequate to teach a computer language in your school?
How many computers per student exist?

*The Lyceum has not its own computer laboratory, therefore it is being used the laboratory of the collocated Gymnasium. The equipment is adequate (15 PCs), and corresponds one student per computer, excluding the cases that technical problems occur and students have to work in pairs.*

## TEACHING A COMPUTER LANGUAGE - AFTER PYTHON COURSES

16. What teaching materials did you depend on, for instructing with Python?
You may include course book as well as the educational technology.

*The students had already been instructed the basic principles of programming from the school textbook of the course "Development of Applications in a Programming Environment". For teaching Python I used the tutorial which is included in the language's documentation. More precisely, I used these parts that relate with the basic programming principles.*

As far as Python is concerned…

17. Which topics/concepts do you think would need more emphasis while teaching?
Would you rather avoid some topics? If yes, please explain what and why.

*What has been covered is the basic programming structures (sequence, conditionals, iteration, arrays). There should be some more explanation on the way the command **range()** works in Python. Furthermore, the way **for** works is different (a linear scale), while in Python is the browsing of a set. Also, a variable after a repetition loop has different values in Pascal and Python.*

18. Comparing Python with a previous language(s), what are the pros and cons?

*PROS: Its syntax is simple and compact in description. This results in fewer symbols, which means smaller code, which means decreased chance of an error occurring.*
*CONS: Its syntax is different than the design of the pseudocode, which is being instructed at school.*

19. Did you have practical difficulties in using the language Python in the laboratory?

*The installation was very easy; we had to reinstall though two computers later on (which of course has nothing to do with the language)*

20. How applicable is Python as first computer language for teaching programming in higher secondary education? What is your overall opinion as an educator? Would you recommend it?

*Generally speaking, Python is an easy language. It is designed for rapid development of programs. As such, it is suitable for use in education, but it is required to show attention in the 3rd class of the Greek Lyceum, due to the aforementioned differences with the instructed pseudocode.*

## Thank you for taking the time to participate in this research and interview!

Yours Sincerely,
Fotis Georgatos

# APPENDIX C – The programming examples of DAPE in Python

The programming examples, which are shown here are problems, mentioned in the school textbook. You may find each of them on the respective page, which is indicated in the comments.

```
Python 2.2 (#28, Dec 21 2001, 12:21:22) [MSC 32 bit (Intel)] on win32
Type "copyright", "credits" or "license" for more information.
IDLE 0.8 -- press F1 for help
>>> def Paradeigma_1():# DAPE, p.30, adding two integers
      a=int(raw_input())
      b=int(raw_input())
      c=a+b
      print c

>>> def Paradeigma_2():# DAPE, p.33, find the absolute value of an integer
      a=int(raw_input())
      if a<0:
            a=a*(-1)
      print a

>>> def Paradeigma_3():# DAPE, p. 34, if a<b find sum of a & b, else product
      a=int(raw_input())
      b=int(raw_input())
      if a<b:
            c=a+b
      else:
            c=a*b
      print c

>>> def Paradeigma_3():# DAPE, p.36, If input is 1,2 or 3 print 'A', 'B' or 'Γ'
      a=int(raw_input())
      if a==1:
            print 'A'
      elif a==2:
            print 'B'
      elif a==3:
            print 'Γ'
      else:
            print 'Αγνωστος'

>>> def Paradeigma_5():# DAPE, p.37, Ask age and print according messages
      print "Σε ποιά ηλικία άρχισες να μαθαίνεις προγραμματισμό;"
      age=int(raw_input())
      if age<0:
            print "Είπαμε ηλικία..."
      elif 0<=age<=5:
            print "Μάλλον τα παραλές!!"
      elif 5<=age<=60:
            print "Μπράβο"
      elif 60<=age<=100:
            print "Ποτέ δεν είναι αργά"
      elif age>100:
            print "Κάλλιο αργά παρά ποτέ"
```

```
>>> def Paradeigma_6():# DAPE, p. 38, Ask Weight and Height; comment it.
      varos=int(raw_input())
      ypsos=int(raw_input())
      if varos<80:
            if ypsos<1.70:
                  print "Ελαφρύς, κοντός"
            else:
                  print "Ελαφρύς, ψηλός"
      else:
            if ypso<1.70:
                  print "Βαρύς, κοντός"
            else:
                  print "Βαρύς, ψηλός"

>>> def Paradeigma_7():# DAPE, p.40, print 100 integers with a while loop
      i=1
      while i<=100:
            print i
            i=i+1

>>> def Paradeigma_8():# DAPE, p.41, demonstrate a while loop;
      x=int(raw_input())
      while x>0:
            print x
            x=int(raw_input())

>>> def Paradeigma_9():      # DAPE, p.42, demonstrate a repeat..until loop;
      return Paradeigma_8()  #  repeat..until does not exist in Python (!)

>>> def Paradeigma_10():     # DAPE, p.43, find the sum of numbers 1..100
      sum=0
      for i in range(1,101): # Oops, range() looks strange in this problem
            sum=sum+i
      print sum

>>> def Paradeigma_11():     # DAPE, p.43, find the sum of 2,4,6,...,98,100
      sum=0
      for i in range(2,101,2):
            sum=sum+i
      print sum

>>> def Pollaplasiasmos_ala_Rwsika(m1,m2):    # DAPE, p.48, multiplication
      p=0                                     # based on powers of 2
      while m2>0:
            if m2%2==1:
                  p=p+m1
            m1=m1*2                           # m1=m1<<1
            m2=m2/2                           # m2=m2>>1
      return p

>>> def Elax_Pinaka_original(table):          # DAPE, p.57, find the minimum
      min=table[0]                            # of an array; the book's way
      for i in range(1,100):
            if table[i]<min:
                  min=table[i]
      return min
```

```
>>> def Elax_Pinaka(table):              # DAPE, p.57, the minimum of an array
      min=table[0]                       # Python's "native" way
      for i in table[1:]:
            if i<min:
                  min=I
      return min

>>> def Athr_Pinaka_Original(m,n,table): # DAPE, p.58, find the sum of
      sum=0                              # a table's rows, columns and total
      for i in range(m): # range(1,m+1) is NOT correct; first element is #0
            row[i]=0
      for j in range(n): # same as above
            col[j]=0
      for i in range(m): # same as above
            for j in range(n):
                  sum=sum+table[i,j]
                  row[i]=row[i]+table[i,j]
                  col[j]=col[j]+table[i,j]
      return row,col,sum

>>> [0]*4 # A demonstration: this will initialize a vector-like table with nulls
[0, 0, 0, 0]

>>> def Athr_Pinaka(m,n,table):          # DAPE, p. 58, same as above
      sum=0                              # the Python's "native way"
      row=[0]*m # Initialize a vector-like matrix with m zeros
      col=[0]*n #      <<           <<        <<       n zeros
      for i in range(m):
            for j in range(n):
                  sum=sum+table[i][j]
                  row[i]=row[i]+table[i][j]
                  col[j]=col[j]+table[i][j]
      return row,col,sum

>>> Athr_Pinaka(2,2,[[1,4],[5,6]])
([5, 11], [6, 10], 16)

>>> def Fyssalida_Original(table,n):          # DAPE, p.68, Bubblesort, book
      for i in range(1,n):
            for j in range(n-1, i-1,-1):
                  if table[j-1]>table[j]:
                        table[j-1],table[j]=table[j],table[j-1]
      return table

>>> Fyssalida_Original([7,-34,6,10],4)
[-34, 6, 7, 10]

>>> def Fyssalida(table,n):              # DAPE, p.68, Bubblesort, Python's way
      for i in range(n-1,0,-1):
            for j in range(i):
                  if table[j]>table[j+1]:
                        table[j+1],table[j]=table[j],table[j+1]
      return table

>>> Fyssalida([7,-34,6,10],4)
[-34, 6, 7, 10]
```

```
>>> def Paragontiko(n):                 # DAPE, p.69, Factorial, version 1
      if n==0:
            product=1
      else:
            product=n*Paragontiko(n-1)
      return product

>>> def Paragontiko2(n):                # DAPE, p.70, Factorial, version 2
      product=1
      for i in range(2,n+1):
            product=product*i
      return product

>>> def Megistos_Koinos_Diaireths(x,y): # DAPE, p.71, Greatest Common Divisor
      if x<y:
            z=x
      else:
            z=y
      while ( x%z != 0 ) or ( y%z != 0 ):
            z=z-1
      return z

>>> def Eucleides(x,y):                 # DAPE, p.71, GCD, Euclides' algorithm
      z=y
      while z!=0:
            z=x%y
            x=y
            y=z
      return x

>>> def mkd(x,y):                       # DAPE, p.71, GCD, Euclides' algorithm
      if y==0:
            return x
      else:
            return mkd(y, x % y)        # RECURSIVELY

>>> mkd(30,4)
2     # cool

>>> mkd(300,27)
3     # cool*2

>>> mkd(300,30)
30    # cool*3

>>> def Fibonacci1(n):                  # DAPE, p.72, Fibonacci series, v. 1
      if n<=1:
            fib=n
      f0=0
      f1=1
      for i in range(2,n+1):
            fib=f0+f1
            f0=f1
            f1=fib
      return fib
```

```
>>> def Fibonacci2(n):                   # DAPE, p.72, Fibonacci series, v.2
      if n<=1:
            return n
      else:
            return Fibonacci2(n-1)+Fibonnacci2(n-2) # RECURSIVELY

>>> def Dyadikh_Anazhthsh(names,phones,onoma,arxi,telos): # DAPE, p.86
      # Binary search, as demonstration of "Divide and Conquer" technique
      meso=(arxi+telos)/2
      if onoma==names[meso]:
            Tel=phones[meso]
      else:
            if onoma<names[meso]:
                  Tel=Dyadikh_Anazhthsh(names,phones,onoma,arxi,meso-1)
            else:
                  Tel=Dyadikh_Anazhthsh(names,phones,onoma,meso+1,telos)
      return Tel

>>> def Dynamh1(a,b):                    # DAPE, p.89, Calculation of powers
      power=1
      for i in range(1,b+1):
            power=power*a
      return power

>>> Dynamh1(3,3)
27

>>> def Dynamh2(a,b):                    # DAPE, p.90, powers, based on 2
      # Initialize a table, actually a "dictionary", to store the variable a
      power={}
      power[1]=a
      i=1
      pow=1
      while pow<b:
            i=i+1
            pow=2*pow
            power[i]=power[i-1] * power[i-1]
      used=0
      result=1
      while used<b:
            if used<b:
                  if used+pow<=b:
                        result=result*power[i]
                        used=used+pow
                  pow=pow/2
                  i=i-1
      return result

>>> Dynamh2(3,3)
27

>>> Dynamh2(3,4)
81

>>> Dynamh2(4,4)
256
```

```
>>> def Nomismata(C,n,poso):             # DAPE, p.91, How many coins to use
      find=poso
      coins=0
      choice=n                           # actually n=len(C), as simple as that
      while (choice>0) and (find>0):
            if C[choice-1]<=find:        # remember the 0-based arrays
                  coins=coins+1
                  find=find-C[choice-1]  # remember...
            else:
                  choice=choice-1
      return coins

>>> EUROs=[1,2,5,10,20,50,100,500]       # Since January 2002.

>>> Nomismata(euros,8,1)
1

>>> Nomismata(euros,8,2)
1

>>> Nomismata(euros,8,3)
2

>>> Nomismata(EUROs,7,1000) # Consider only the first 7 coins, calculate
10

>>> Nomismata(euros,8,1000) # Consider all 8 coins, calculate
2

>>> Nomismata(euros,8,1001)
3

>>> Nomismata(euros,8,1002)
3

>>> Nomismata(euros,8,1003)
4

>>> Nomismata(euros,8,1004)
4

>>> Nomismata(euros,8,1005)
3
```

# APPENDIX D – Python's grammar in BNF

This chapter is based on Python's Reference Manual, as seen in the standard documentation of version 2.2.0. The BNF definition of Python is in a special format suitable for another program called yacc, which is a parser generator.

First, there is an exact copy of the documentation's section on Notation.

## Notation

The descriptions of lexical analysis and syntax use a modified BNF grammar notation. This uses the following style of definition:

```
name:           lc_letter (lc_letter | "_")*
lc_letter:      "a"..."z"
```

The first line says that a `name` is an `lc_letter` followed by a sequence of zero or more `lc_letter`s and underscores. An `lc_letter` in turn is any of the single characters "`a`" through "`z`". (This rule is actually adhered to for the names defined in lexical and grammar rules in this document.)

Each rule begins with a name (which is the name defined by the rule) and a colon. A vertical bar (`|`) is used to separate alternatives; it is the least binding operator in this notation. A star (`*`) means zero or more repetitions of the preceding item; likewise, a plus (`+`) means one or more repetitions, and a phrase enclosed in square brackets (`[ ]`) means zero or one occurrences (in other words, the enclosed phrase is optional). The `*` and `+` operators bind as tightly as possible; parentheses are used for grouping. Literal strings are enclosed in quotes. White space is only meaningful to separate tokens. Rules are normally contained on a single line; rules with many alternatives may be formatted alternatively with each line after the first beginning with a vertical bar.

In lexical definitions (as the example above), two more conventions are used: Two literal characters separated by three dots mean a choice of any single character in the given (inclusive) range of ASCII characters. A phrase between angular brackets (`<...>`) gives an informal description of the symbol defined; e.g., this could be used to describe the notion of `control character' if needed.

Even though the notation used is almost the same, there is a big difference between the meaning of lexical and syntactic definitions: a lexical definition operates on the individual characters of the input source, while a syntax definition operates on the stream of tokens generated by the lexical analysis. All uses of BNF in the next chapter (``Lexical Analysis'' ) are lexical definitions; uses in subsequent chapters are syntactic definitions.

```
identifier ::=
             (letter|"_") (letter | digit | "_")*

letter ::=
             lowercase | uppercase

lowercase ::=
             "a"..."z"

uppercase ::=
             "A"..."Z"

digit ::=
             "0"..."9"

stringliteral ::=
             [stringprefix](shortstring | longstring)

stringprefix ::=
             "r" | "u" | "ur" | "R" | "U" | "UR" | "Ur" | "uR"

shortstring ::=
             "'" shortstringitem* "'"
              | '"' shortstringitem* '"'

longstring ::=
             "'''" longstringitem* "'''"
              | '"""' longstringitem* '"""'

shortstringitem ::=
             shortstringchar | escapeseq

longstringitem ::=
             longstringchar | escapeseq

shortstringchar ::=
             <any ASCII character except "\" or newline or the quote>

longstringchar ::=
             <any ASCII character except "\">

escapeseq ::=
             "\" <any ASCII character>

longinteger ::=
             integer ("l" | "L")

integer ::=
             decimalinteger | octinteger | hexinteger

decimalinteger ::=
             nonzerodigit digit* | "0"
```

```
octinteger ::=
             "0" octdigit+

hexinteger ::=
             "0" ("x" | "X") hexdigit+

nonzerodigit ::=
             "1"..."9"

octdigit ::=
             "0"..."7"

hexdigit ::=
             digit | "a"..."f" | "A"..."F"

floatnumber ::=
             pointfloat | exponentfloat

pointfloat ::=
             [intpart] fraction | intpart "."

exponentfloat ::=
             (intpart | pointfloat)
              exponent

intpart ::=
             digit+

fraction ::=
             "." digit+

exponent ::=
             ("e" | "E") ["+" | "-"] digit+

imagnumber ::= (floatnumber | intpart) ("j" | "J")

atom ::=
             identifier | literal | enclosure

enclosure ::=
             parenth_form | list_display
              | dict_display | string_conversion

literal ::=
             stringliteral | integer
              | longinteger | floatnumber
              | imagnumber

parenth_form ::=
             "(" [expression_list] ")"

list_display ::=
             "[" [listmaker] "]"
```

```
listmaker ::=
             expression ( list_for
              | ( "," expression)* [","] )

list_iter ::=
             list_for | list_if

list_for ::=
             "for" expression_list "in" testlist
              [list_iter]

list_if ::=
             "if" test [list_iter]

dict_display ::=
             "" [key_datum_list] ""

key_datum_list ::=
             key_datum ("," key_datum)* [","]

key_datum ::=
             expression ":" expression

string_conversion ::=
             "`" expression_list "`"

primary ::=
             atom | attributeref
              | subscription | slicing | call

attributeref ::=
             primary "." identifier

subscription ::=
             primary "[" expression_list "]"

slicing ::=
             simple_slicing | extended_slicing

simple_slicing ::=
             primary "[" short_slice "]"

extended_slicing ::=
             primary "[" slice_list "]"

slice_list ::=
             slice_item ("," slice_item)* [","]

slice_item ::=
             expression | proper_slice | ellipsis

proper_slice ::=
             short_slice | long_slice
```

```
short_slice ::=
             [lower_bound] ":" [upper_bound]

long_slice ::=
             short_slice ":" [stride]

lower_bound ::=
             expression

upper_bound ::=
             expression

stride ::=
             expression

ellipsis ::=
             "..."

call ::=
             primary "(" [argument_list [","]] ")"

argument_list ::=
             positional_arguments ["," keyword_arguments
             ["," "*" expression ["," "**" expression]]]
              | keyword_arguments ["," "*" expression
            ["," "**" expression]]
            | "*" expression ["," "**" expression]
            | "**" expression


positional_arguments ::=
             expression ("," expression)*

keyword_arguments ::=
             keyword_item ("," keyword_item)*

keyword_item ::=
             identifier "=" expression

power ::=
             primary ["**" u_expr]

u_expr ::=
             power | "-" u_expr
              | "+" u_expr | "~" u_expr

m_expr ::=
             u_expr | m_expr "*" u_expr
              | m_expr "/" u_expr
              | m_expr "\%" u_expr

a_expr ::=
             m_expr | aexpr "+" m_expr
              aexpr "-" m_expr
```

```
shift_expr ::=
             a_expr
              | shift_expr ( "<<" | ">>" ) a_expr

and_expr ::=
             shift_expr | and_expr "\;SPMamp;" shift_expr

xor_expr ::=
             and_expr | xor_expr "\textasciicircum" and_expr

or_expr ::=
             xor_expr | or_expr "|" xor_expr

comparison ::=
             or_expr ( comp_operator or_expr )*

comp_operator ::=
             "<" | ">" | "==" | ">=" | "<=" | "<>" | "!="
              | "is" ["not"] | ["not"] "in"

expression ::=
             or_test | lambda_form

or_test ::=
             and_test | or_test "or" and_test

and_test ::=
             not_test | and_test "and" not_test

not_test ::=
             comparison | "not" not_test

lambda_form ::=
             "lambda" [parameter_list]: expression

expression_list ::=
             expression ( "," expression )* [","]

simple_stmt ::=
             expression_stmt
              | assert_stmt
              | assignment_stmt
              | augmented_assignment_stmt
              | pass_stmt
              | del_stmt
              | print_stmt
              | return_stmt
              | yield_stmt
              | raise_stmt
              | break_stmt
              | continue_stmt
              | import_stmt
              | global_stmt
              | exec_stmt
```

```
expression_stmt ::=
             expression_list

assert_statement ::=
             "assert" expression ["," expression]

assignment_stmt ::=
             (target_list "=")+ expression_list

target_list ::=
             target ("," target)* [","]

target ::=
             identifier
              | "(" target_list ")"
              | "[" target_list "]"
              | attributeref
              | subscription
              | slicing

augmented_assignment_stmt ::=
             target augop expression_list

augop ::=
             "+=" | "-=" | "*=" | "/=" | "\%=" | "**="
              | ">>=" | "<<=" | "\&=" | "\textasciicircum=" | "|="

target ::=
             identifier
              | "(" target_list ")"
              | "[" target_list "]"
              | attributeref
              | subscription
              | slicing

pass_stmt ::=
             "pass"

del_stmt ::=
             "del" target_list

print_stmt ::=
             "print" ( \optionalexpression ("," expression)* \optional","
                      | ">\code>" expression
                        \optional("," expression)+ \optional",")

return_stmt ::=
             "return" [expression_list]

yield_stmt ::=
             "yield" expression_list

raise_stmt ::=
             "raise" [expression ["," expression
              ["," expression]]]
```

```
break_stmt ::=
             "break"

continue_stmt ::=
             "continue"

import_stmt ::=
             "import" module ["as" name]
                ( "," module ["as" name] )*
              | "from" module "import" identifier
                ["as" name]
                ( "," identifier ["as" name] )*
              | "from" module "import" "*"

module ::=
             (identifier ".")* identifier

global_stmt ::=
             "global" identifier ("," identifier)*

exec_stmt ::=
             "exec" expression
              ["in" expression ["," expression]]

compound_stmt ::=
             if_stmt | while_stmt | for_stmt
              | try_stmt | funcdef | classdef

suite ::=
             stmt_list NEWLINE
              | NEWLINE INDENT statement+ DEDENT

statement ::=
             stmt_list NEWLINE | compound_stmt

stmt_list ::=
             simple_stmt (";" simple_stmt)* [";"]

if_stmt ::=
             "if" expression ":" suite
              ( "elif" expression ":" suite )*
              ["else" ":" suite]

while_stmt ::=
             "while" expression ":" suite
               ["else" ":" suite]

for_stmt ::=
             "for" target_list "in" expression_list
              ":" suite
              ["else" ":" suite]

try_stmt ::=
             try_exc_stmt | try_fin_stmt
```

```
try_exc_stmt ::=
             "try" ":" suite
              ("except" [expression ["," target]] ":"
              suite)+
              ["else" ":" suite]

try_fin_stmt ::=
             "try" ":" suite
              "finally" ":" suite

funcdef ::=
             "def" funcname "(" [parameter_list] ")"
              ":" suite

parameter_list ::=
             (defparameter ",")*
              ("*" identifier [, "**" identifier]
               | "**" identifier
               | defparameter [","])

defparameter ::=
             parameter ["=" expression]

sublist ::=
             parameter ("," parameter)* [","]

parameter ::=
             identifier | "(" sublist ")"

funcname ::=
             identifier

classdef ::=
             "class" classname [inheritance] ":"
              suite

inheritance ::=
             "(" [expression_list] ")"

classname ::=
             identifier

file_input ::=
             (NEWLINE | statement)*

interactive_input ::=
             [stmt_list] NEWLINE | compound_stmt NEWLINE

eval_input ::=
             expression_list NEWLINE*

input_input ::=
             expression_list NEWLINE
```

# APPENDIX E – Chomsky, Grammars & BNF notation

Noam Chomsky was probably, little aware while he was investigating natural languages in the 50s, that his under development theory of grammars would be readily applicable for the description of the upcoming domain of computer languages as well. Chomsky' s work on languages included a proposal on the classification of languages on the basis of grammar properties. He suggested four major language categories, each being a subset of the former, starting with *Universal Grammar*:

| Type (level) | Grammar Category | Notation or example | Computational Tool that recognizes this grammar |
|---|---|---|---|
| 0 | Universal Grammar | An infinite computer program | ? |
| 1 | Context Sensitive Grammar | A finite computer program | Compiler, interpreter |
| 2 | Context Free Grammar | BNF (or its subset LALR) | Yacc |
| 3 | Regular Grammar | Regular Expressions | lex, sed, awk |

The last two are constantly applied in computer science. In fact, during the development of the computer language ALGOL the meta-language BNF was created, which is precisely a Type-2 grammar. Backus and Naur worked independently of Chomsky. Following different paths ended up into a common indirect conclusion: in order to generate a language with unique syntactical meaning, a formalism describing strictly its production rules has to be used. **The rest of the information in this appendix comes mainly from the Internet; various and unknown sources.**

### *What is BNF notation?*

BNF is an acronym for "Backus-Naur Form". John Backus and Peter Naur introduced for the first time a formal notation to describe the syntax of a given language. This was for the description of the ALGOL 60 programming language. To be precise, most of BNF was introduced by Backus in a report presented at an earlier UNESCO conference on ALGOL 58. Few read the report, but when Peter Naur read it he was surprised at some of the differences he found between his and Backus' interpretation of ALGOL 58. He decided that for the successor to ALGOL, all participants of the first design had come to recognize some weaknesses, should be given in a similar form so that all participants should be aware of what they were agreeing to. He made a few modifications that are almost universally used and drew up on his own the BNF for ALGOL 60 at the meeting where it was designed. Depending on how you attribute presenting it to the world, it was either by Backus in 59 or Naur in 60. For more details on this period of programming languages, see the introduction to Backus' Turing award article in Communications of the ACM, Vol. 21, No. 8. August 1978.

Since then, almost every author of a new programming language used BNF to specify the syntax rules of the language, which in effect are Type-2 equivalent production rules. For example, this is a definition of an iteration statement in C [DR78, pp. 322]:

*iteration-statement*:
  while ( *expression* ) *statement*
  do *statement* while ( *expression* ) ;
  for ( [*expression*] ; [*expression*] ; [*expression*] ) *statement*

### *Is there a definition for BNF?*

The meta-symbols of BNF are:

| | |
|---|---|
| `::=` | meaning "is defined as" |
| `|` | meaning "or" |
| `< >` | angle brackets used to surround category names. |

The angle brackets distinguish syntax rules names (also called non-terminal symbols) from terminal symbols, which are written exactly as they are to be represented.
A BNF rule defining a nonterminal has the form:

| | |
|---|---|
| `nonterminal ::=` | sequence_of_alternatives consisting of strings of terminals, or nonterminals separated by the meta-symbol `|` |

For example, the BNF production for a mini-language is:

```
<program> ::=  program
                   <declaration_sequence>
               begin
                   <statements_sequence>
               end ;
```

This shows that a mini-language program consists of the keyword "program" followed by the declaration sequence, then the keyword "begin" and the statements sequence, finally the keyword "end" and a semicolon.

In fact, many authors have introduced some slight extensions of BNF for the ease of use:

- optional items are enclosed in meta-symbols [ and ], example:

```
     <if_statement> ::=  if <boolean_expression> then
                              <statement_sequence>
                         [ else
                              <statement_sequence> ]
                         end if ;
```

- repetitive items (zero or more times) are enclosed in meta-symbols { and }, example:

```
     <identifier> ::= <letter> { <letter> | <digit> }
```

This rule is equivalent to the recursive rule:

```
     <identifier> ::= <letter> |
                      <identifier> [ <letter> | <digit> ]
```

- terminals of only one character are surrounded by quotes (") to distinguish them from meta-symbols, example:

```
     <statement_sequence> ::= <statement> { ";" <statement> }
```

- In recent textbooks, terminal and non-terminal symbols are distinguished by using bold faces for terminals and suppressing < and > around non-terminals. This improves greatly the readability. The example then becomes:

```
     if_statement ::= if boolean_expression then
                          statement_sequence
                      [ else
                          statement_sequence ]
                      end if ";"
```

As a last example, maybe not the easiest to read!, here is the definition of BNF expressed in BNF:

```
syntax     ::=  { rule }
rule       ::=  identifier  "::="  expression
expression ::=  term { "|" term }
term       ::=  factor { factor }
factor     ::=  identifier |
                quoted_symbol |
                "("  expression  ")" |
                "["  expression  "]" |
                "{"  expression  "}"
identifier ::=  letter { letter | digit }
quoted_symbol ::= """ { any_character } """
```

BNF is not only important to describe syntax rules in books, but it is very commonly used, with variants, by tools for syntactical analysis. Read for example any book or manual page on LEX and YACC, the standard UNIX parser generators. If you have access to any Unix machine, you will probably find somewhere a chapter on the documentation of these tools.

Here is an example grammar of arithmetic expressions, which can also "understand" functions:

```
S    ::= E |  +E |  -E   # sentence
E    ::= T | E+T | E-T   # expression
T    ::= F | T*F | T/F   # term
F    ::= P | P^R         # factor
P    ::= i | n | R | (S) # primary
R    ::= i(L)            # function reference
L    ::= S | L, S        # argument list
```

For information on BNF, read M. Marcotty & H. Ledgard, The World of Programming Languages, Springer-Verlag, Berlin 1986., pages 41-50.

# APPENDIX F – Haskell, Functional Programming, Mathematics

The symbolism of λ has been proposed by the mathematician of logic, Alonzo Church, in "*The calculi of Lambda-Conversion*", Princeton University Press, New Jersey, 1941. The lambda calculus is the core of functional programming, of which a distinct member is the language Haskell. The following text, in italics, is from Haskell's website at **http://www.haskell.org**

λ

*About Haskell*

*Haskell is a computer programming language. In particular, it is a polymorphicly typed, lazy, purely functional language, quite different from most other programming languages. The language is named for Haskell Brooks Curry, whose work in mathematical logic serves as a foundation for functional languages. Haskell is based on lambda calculus, hence the lambda we use as a logo.*

*Why Use Haskell?*

*Writing large software systems that work is difficult and expensive. Maintaining those systems is even more difficult and expensive. Functional programming languages, such as Haskell, can make it easier and cheaper.*

*Haskell, a purely functional programming language, offers you:*
- *Substantially increased programmer productivity (Ericsson measured an improvement factor of between 9 and 25 in one set of experiments on telephony software).*
- *Shorter, clearer, and more maintainable code.*
- *Fewer errors, higher reliability.*
- *A smaller "semantic gap" between the programmer and the language.*
- *Shorter lead times.*

*Haskell is a wide-spectrum language, suitable for a variety of applications. It is particularly suitable for programs, which need to be highly modifiable and maintainable.*
*Much of a software product's life is spent in specification, design and maintenance, and not in programming. Functional languages are superb for writing specifications which can actually be executed (and hence tested and debugged). Such a specification then is the first prototype of the final program.*

*Functional programs are also relatively easy to maintain, because the code is shorter, clearer, and the rigorous control of side effects eliminates a huge class of unforeseen interactions.*

This is the algorithm QuickSort, expressed in Python. The brevity and elegance is "breathtaking":

```
>>> def qsort(L):                    # from Python Cookbook, ASPN # http://aspn.activestate.com
        if len(L)<=1: return L
        return qsort( [lt for lt in L[1:] if lt <L[0]] ) + [ L[0] ] + qsort( [ge for ge in L[1:] if ge>=L[0]] )

>>> qsort([4,5,76,7,8,3,345,34,23,23,1,4,5,54,0,54,-1])
[-1, 0, 1, 3, 4, 4, 5, 5, 7, 8, 23, 23, 34, 54, 54, 76, 345]
```

What is most noticeable in functional programming is that it takes mathematical symbolism to another level. To begin with, let's remind that throughout history both *symbolism (notation)* as well as *semantics* of mathematics have been under reconsideration over and over again:

- *Symbolism* has been reviewed every time a new concept arisen, like imaginary numbers, matrices, differential and integral calculations etc. It is the problem of how to represent a new concept on paper, next to the existing symbols and without using natural language. Think what happened with the introduction of functions like **sin**us, **cos**inus and operations like **mod**ulo.
- *Semantics* has been discussed when new fields were being explored and there was no assurance of what would be important to refer to later on. One of the most extensive discussions happened in the $16^{th}$ century involving Leibniz, Cramer, Laplace, about the appropriate definition for the determinant of a matrix. Gauss invented the term determinant, however his determinant was not the same as the one according to the current definition!

The fact that current mathematical symbolism can be further reviewed is little recognised, albeit there have been efforts towards this direction. A recent notable effort is the APL language defined by Kenneth Iverson in the early 60s for use in teaching. APL was not originally meant to be applied in computers and introduced many non-existing symbols; it was an advancement *per se*. APL was later implemented by IBM and belongs to the functional family of computer languages, the same family that includes LISP, Scheme, LOGO and Haskell.

One of the most important areas in which current symbolism could be challenged is in passing parameters to functions that are non-standard:

- matrices of imaginary numbers
- other functions (or matrices of, or functions that return functions !! )
- non-ordinary data types like trees, graphs etc.

To demonstrate an example, consider the meaning of the symbols $\prod$ and $\sum$ which are defined to correspond respectively to numerical operations * and +. If we assume a new numerical operation with transitive property, symbolised by ®, then we have no **formal** means to generate the new symbol $\Psi$ that will be the application of ® on a set. In fact, we may have to use natural language even for defining what ® means, albeit it may be describable with existing operations. Haskell and other functional languages solve this problem by being able to define operations with lists and functions and generalize over them. The aforementioned operations could be defined in Haskell as:

```
sum     = foldl (+) 0
product = foldl (*) 1
Ψ       = foldl (®) neutral_element
```

The Greatest Common Divisor, assuming a>b is written in Haskell as these two lines:

```
gcd(a,b) = if b==0 then a
                   else gcd(b, a mod b)
```

Functional programs are often shorter and more comprehensive than their imperative equivalents, due to their symbolism. Please read the following page for a comparison between Haskell and C in a certain programming case. You can find and download a Haskell98 interpreter, called Hugs, at: **http://www.haskell.org**

The following examples are found in the article "*Why functional programming matters*" by John Hughes [HJ89].

QuickSort in Haskell

```
qsort []     = []
qsort (x:xs) = qsort elts_lt_x ++ [x] ++ qsort elts_greq_x
                 where
                   elts_lt_x   = [y | y <- xs, y < x]
                   elts_greq_x = [y | y <- xs, y >= x]
```

QuickSort in C

```
qsort( a, lo, hi ) int a[], hi, lo;
{
  int h, l, p, t;

  if (lo < hi) {
    l = lo;
    h = hi;
    p = a[hi];
    do {
      while ((l < h) && (a[l] <= p))
          l = l+1;
      while ((h > l) && (a[h] >= p))
          h = h-1;
      if (l < h) {
          t = a[l];
          a[l] = a[h];
          a[h] = t;
      }
    } while (l < h);
    t = a[l];
    a[l] = a[hi];
    a[hi] = t;

    qsort( a, lo, l-1 );
    qsort( a, l+1, hi );
  }
}
```

Surely, the Haskell code looks cleaner and easier to comprehend. A technical note:

> *“It isn't all roses, of course. The C QuickSort uses an extremely ingenious technique, invented by Hoare, whereby it sorts the array in place; that is, without using any extra storage. As a result, it runs quickly, and in a small amount of memory. In contrast, the Haskell program allocates quite a lot of extra memory behind the scenes and runs rather slower than the C program. In effect, the C QuickSort does some very ingenious storage management, trading this algorithmic complexity for a reduction in run-time storage management costs. In applications where performance is required at any cost, or when the goal is detailed tuning of a low-level algorithm, an imperative language like C would probably be a better choice than Haskell, exactly because it provides more intimate control over the exact way in which the computation is carried out.”* [HJ89]

# APPENDIX G – Greatest Common Divisor, an ancient algorithm

The problem of the Greatest Common Divisor refers to the question of, given two positive integers that have a common divisor other than one, to find which one of their common divisors is the greatest[81]. It is often cited in Computer Science books and papers, proposing a method to calculate it and expressing this technique in a computer language's notation. It involves *sequential*, *iterational* and *conditional* actions, just like in imperative programming!

An algorithm for solving this problem, called briefly GCD, is often said to have been proposed first by Euclides, an ancient mathematician who lived in Sicily during the 3rd century BC. Euclides had stated his solution proposal along with a proof in ancient Greek, since at those times the only way to express such concepts was by means of natural language, This is how the solution algorithm and its proof looked like, as would be found in his book "Elements" (a translation in English follows):

136 ΣΤΟΙΧΕΙΩΝ ζ΄.

β΄.

**Δύο ἀριθμῶν δοθέντων μὴ πρώτων πρὸς ἀλλήλους τὸ μέγιστον αὐτῶν κοινὸν μέτρον εὑρεῖν.**

*Ἔστωσαν οἱ δοθέντες δύο ἀριθμοὶ μὴ πρῶτοι πρὸς ἀλλήλους οἱ ΑΒ, ΓΔ. δεῖ δὴ τῶν ΑΒ, ΓΔ τὸ μέγιστον κοινὸν μέτρον εὑρεῖν.*

*Εἰ μὲν οὖν ὁ ΓΔ τὸν ΑΒ μετρεῖ, μετρεῖ δὲ καὶ ἑαυτόν, ὁ ΓΔ ἄρα τῶν ΓΔ, ΑΒ κοινὸν μέτρον ἐστίν. καὶ φανερόν, ὅτι καὶ μέγιστον· οὐδεὶς γὰρ μείζων τοῦ ΓΔ τὸν ΓΔ μετρήσει.*

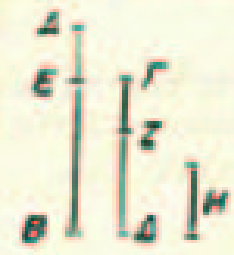

*Εἰ δὲ οὐ μετρεῖ ὁ ΓΔ τὸν ΑΒ, τῶν ΑΒ, ΓΔ ἀνθυφαιρουμένων ἀεὶ τοῦ ἐλάσσονος ἀπὸ τοῦ μείζονος λειφθήσεταί τις ἀριθμός, ὃς μετρήσει τὸν πρὸ ἑαυτοῦ. μονὰς μὲν γὰρ οὐ λειφθήσεται· εἰ δὲ μή, ἔσονται οἱ ΑΒ, ΓΔ πρῶτοι πρὸς ἀλλήλους· ὅπερ οὐχ ὑπόκειται. λειφθήσεταί τις ἄρα ἀριθμός, ὃς μετρήσει τὸν πρὸ ἑαυτοῦ. καὶ ὁ μὲν ΓΔ τὸν ΒΕ μετρῶν λειπέτω ἑαυτοῦ ἐλάσσονα τὸν ΕΑ, ὁ δὲ ΕΑ τὸν ΔΖ μετρῶν λειπέτω ἑαυτοῦ ἐλάσσονα τὸν ΖΓ, ὁ δὲ ΓΖ τὸν ΑΕ μετρείτω. ἐπεὶ οὖν ὁ ΓΖ τὸν ΑΕ μετρεῖ, ὁ δὲ ΑΕ τὸν ΔΖ μετρεῖ, καὶ ὁ ΓΖ ἄρα τὸν ΔΖ μετρήσει· μετρεῖ δὲ καὶ ἑαυτόν· καὶ ὅλον ἄρα τὸν ΓΔ μετρήσει. ὁ δὲ ΓΔ τὸν ΒΕ μετρεῖ· καὶ ὁ ΓΖ ἄρα τὸν ΒΕ μετρεῖ· μετρεῖ δὲ καὶ τὸν ΕΑ· καὶ ὅλον ἄρα τὸν ΒΑ μετρήσει· μετρεῖ δὲ καὶ τὸν ΓΔ· ὁ ΓΖ ἄρα τοὺς ΑΒ, ΓΔ μετρεῖ. ὁ ΓΖ ἄρα τῶν ΑΒ, ΓΔ κοινὸν μέτρον ἐστίν. λέγω δή, ὅτι καὶ μέγιστον. εἰ γὰρ μή ἐστιν ὁ ΓΖ τῶν ΑΒ, ΓΔ μέγιστον κοινὸν μέτρον, μετρήσει τις τοὺς ΑΒ, ΓΔ ἀριθμοὺς ἀριθμὸς μείζων ὢν τοῦ ΓΖ. μετρείτω, καὶ ἔστω ὁ Η. καὶ ἐπεὶ ὁ Η τὸν ΓΔ μετρεῖ, ὁ δὲ ΓΔ τὸν ΒΕ μετρεῖ, καὶ ὁ Η ἄρα τὸν ΒΕ μετρεῖ· μετρεῖ δὲ καὶ ὅλον τὸν ΒΑ· καὶ λοιπὸν ἄρα τὸν ΑΕ μετρήσει. ὁ δὲ ΑΕ τὸν ΔΖ μετρεῖ· καὶ ὁ Η ἄρα τὸν ΔΖ μετρήσει· μετρεῖ δὲ καὶ ὅλον τὸν ΔΓ· καὶ λοιπὸν ἄρα τὸν ΓΖ μετρήσει ὁ μείζων τὸν ἐλάσσονα· ὅπερ ἐστὶν ἀδύνατον· οὐκ ἄρα τοὺς ΑΒ, ΓΔ ἀριθμοὺς ἀριθμός τις μετρήσει μείζων ὢν τοῦ ΓΖ· ὁ ΓΖ ἄρα τῶν ΑΒ, ΓΔ μέγιστόν ἐστι κοινὸν μέτρον [ὅπερ ἔδει δεῖξαι].*

**Πόρισμα.**

*Ἐκ δὴ τούτου φανερόν, ὅτι ἐὰν ἀριθμὸς δύο ἀριθμοὺς μετρῇ, καὶ τὸ μέγιστον αὐτῶν κοινὸν μέτρον μετρήσει· ὅπερ ἔδει δεῖξαι.*

---

[81] I wish to dedicate this last chapter to my father, Michalis F. Georgatos. Not only for digging through his library to find this certain algorithm I asked him, but above all for the initial triggering of me in the importance of languages. This has been accompanied by years of effort with books, stories, poems and dedication.

**Problem:**

Given two numbers that are not prime to each other[82], let the greatest common divisor be found.

```
     A
     |
     |
     E     G
     |     |
     |     Z
     |     |
     |     |
     |     |     |
     |     |     |
     |     |     |
     B     D     | H
```

Assume two given numbers which are not prime to each other, named AB, GD.
So, if GD divides[83] AB and divides also itself, then GD is common divisor of AB, GD, and obvious that (it is) greatest as well; since no other greater than GD can divide GD.
Otherwise, when GD does not divide AB, the minimal is always mutually subtracted[84] from the maximal and a number will be derived, which is a divisor of itself.
A monad (=1) cannot be derived; otherwise, AB, GD are prime to each other; which was assumed not to be in the initial hypothesis.
Therefore, a number will be derived which will divide the one before it.
And while GD divides BE and leaves as a modulo the smaller AE,
EA will divide DZ and leave as a modulo the lesser ZG, while GZ will divide AE.
Because GZ divides AE, AE divides DZ, so will GZ divide DZ; and also itself; and so will also divide the whole GD. GD will divide BE and GZ divides AE; and so it can divide the whole of BA.
It also divides GD; so GZ can divide AB, GD. GD is, therefore, a common divisor of AB, GD.
I claim that this is the greatest as well.
Because, if GZ is not the greatest common divisor of AB, GD, a greater number than GZ will divide AB, GD.
Let it be so and call this H. And because H can divide GD and GD can divide BE, so does H divide BE; and it also divides the whole of BA; and so it divides AE. AE divides DZ; and H divides DZ.; and the whole DG; and therefore divides GZ, the greatest the smaller; which is impossible;
So, no number can divide AB, GD that is greater than GZ; therefore GZ is the greatest common divisor of AB, CD {**which is what had to be proved**}

The solution for the Greatest Common Divisor in Python is, assuming two integers x,y, x>y:

```
>>> def mkd(x,y):
      if y==0:    return x
      else:       return mkd(y, x % y)
```

Note: the solution is **not** a proof. It is provided for reader’s reflection on the features of languages.

---

[82] Two numbers are prime to each other when they don' t have a common divisor, excluding 1. For example, (12, 13), (14, 17), (16, 35). Note that these numbers are not necessarily primes; hence they have divisors, but not common ones.
[83] The original verb is "μετρείν" which can either be translated in the English language as "divide", or "measure". Similarly, the word "metron" is translated as "divisor" or noun "measure".
[84] This is a way to express recursion. Euclides explained this algorithm in an earlier paragraph of “Elements”, too.